\documentclass[aip,graphicx]{revtex4-1}
\usepackage{amsmath}
\usepackage{graphicx}
\usepackage{subcaption}
\usepackage[export]{adjustbox}
\usepackage{braket}
\usepackage{wrapfig}
\usepackage{mhchem}
\usepackage{dcolumn}
\usepackage{multirow} 
\usepackage{bm}
\usepackage[table, dvipsnames]{xcolor}
\DeclareUnicodeCharacter{2212}{-}
\draft 
\usepackage[most]{tcolorbox}
\usepackage{tabularx}
\usepackage{float}
\usepackage{titlesec}
\usepackage{subcaption} 

\usepackage{caption}
\DeclareCaptionJustification{justified}{\justifying}
\usepackage{ragged2e}

\usepackage{booktabs}
\usepackage{multirow}
\usepackage{siunitx}
\usepackage{tabularray}

\newenvironment{acknowledgement}
{\section*{Acknowledgement}}
{}
\newenvironment{data availability statement}
{\section*{Data Availability Statement}}
{}

\begin{document}
\title{Reaction-Level Consistency within the Variational Quantum Eigensolver: Homodesmotic Ring Strain Energies of Cyclic Hydrocarbons}

\author{Lisa Roy}
\affiliation{Quantum Information and Computation Lab, Department of Chemistry,
 Indian Institute of Technology Jodhpur, Rajasthan, India, 342030 
}%
\author{Maitreyee Sarkar}
\affiliation{Quantum Information and Computation Lab, Department of Chemistry,
 Indian Institute of Technology Jodhpur, Rajasthan, India, 342030 
}%
\author{Mitali Tewari}
 \affiliation{Quantum Information and Computation Lab, Department of Chemistry,
 Indian Institute of Technology Jodhpur, Rajasthan, India, 342030 
}%
 \author{Atul Kumar}
 \email{atulk@iitj.ac.in}
\affiliation{Quantum Information and Computation Lab, Department of Chemistry,
 Indian Institute of Technology Jodhpur, Rajasthan, India, 342030 
}%
 \author{Manikandan Paranjothy}
  \email{pmanikandan@iitj.ac.in}

\affiliation{Chemical Dynamics Research Group, Department of Chemistry,
 Indian Institute of Technology Jodhpur, Rajasthan, India, 342030 
}%

\begin{abstract}
Simulation of chemical reactions on quantum computing platforms using quantum–classical hybrid algorithms such as the Variational Quantum Eigensolver (VQE) is challenged by the need for a reaction-consistent treatment of electron correlation in reaction energy evaluations. In this work, we employ a previously reported symmetry-guided active space selection protocol to compute ring strain energies of cyclic hydrocarbons using homodesmotic reaction schemes. The protocol enforces symmetry consistency across all reactants and products by selecting active spaces that yield identical symmetry-matched fraction (SMF) values, thereby ensuring balanced correlation treatment at the reaction level. When multiple active spaces satisfy this criterion for a given molecule, larger active spaces often provide improved correlation treatment; however, smaller symmetry-consistent active spaces can also yield comparable agreement due to favorable error cancellation within the homodesmotic framework. Using this framework, ring strain energies were evaluated for a series of saturated and unsaturated cyclic hydrocarbons, ranging from cyclopropane to the structurally complex adamantane. The resulting energies achieve chemical accuracy relative to density functional theory (DFT) and remain in close agreement with coupled-cluster singles and doubles (CCSD) benchmarks. The systematic performance across increasing molecular complexity highlights the effectiveness of combining homodesmotic reaction design with symmetry-consistent VQE calculations. This approach, which enforces physically grounded consistency across reaction species, demonstrates clear potential for extending reaction-based quantum simulations to larger molecular systems and broader classes of chemical reactions. 

\end{abstract}

\maketitle


\section{\label{sec:level1} Introduction}

Due to the growing potential of quantum computers to simulate molecular systems with improved accuracy and more favorable scaling, quantum chemistry has emerged as a natural and important application of quantum computing.\cite{cao2019quantum, mcardle2020quantum, lanyon2010towards, claudino2022basics, kais2014introduction, fano2019quantum, kassal2011simulating, dirac1929quantum} On conventional computing platforms, the representation of the molecular wavefunction becomes exponentially complex as the number of particles increases,\cite{feynman2018simulating} severely restricting the predictive capabilities of established electronic-structure methods for large molecular systems. Quantum computers are expected to naturally encode quantum states, enabling more efficient representations that can scale polynomially in the fault-tolerant regime.\cite{aspuru2005simulated, armaos2020computational} \par

In computational chemistry, coupled cluster singles and doubles (CCSD) is a post Hartree–Fock method that systematically accounts for electronic correlation by incorporating single and double excitations from a reference Hartree–Fock state, and it is widely regarded as the “gold standard” for small to medium-sized molecular systems. However, its computational cost scales steeply with system size, making it challenging to apply to larger molecules. The integration of quantum algorithms with electronic-structure methods such as CCSD holds the potential to overcome these scaling limitations, opening pathways for accurate and scalable quantum simulations in chemistry. Among various quantum algorithms, the Variational Quantum Eigensolver (VQE)\cite{kandala2017hardware} has emerged as one of the most practical approaches in the noisy intermediate-scale quantum (NISQ) era.\cite{preskill2018quantum} VQE, a hybrid quantum–classical algorithm, uses parameterized quantum circuits (PQCs)\cite{ostaszewski2021structure, benedetti2019parameterized, whitfield2011simulation, lee2018generalized} to approximate the ground-state energy (GSE)\cite{lanyon2010towards} of molecular systems, effectively combining the expressive power of quantum hardware with the robustness of classical numerical optimization. \par

In understanding reactivity trends and the driving forces behind chemical reactions, strain concepts have played a central role in mechanistic organic chemistry.\cite{wiberg1986concept, bach2002effect, greenberg2013strained} The original strain concept originates from von Baeyer’s strain theory,\cite{wiberg1986concept} which proposed that the stability of cyclic organic molecules, particularly cycloalkanes, is governed by deviations of internal bond angles from the ideal tetrahedral angle of $109.5^\circ$. This concept was subsequently extended to include contributions from bond-length distortions, torsional strain, and non-bonded interactions\cite{wiberg1986concept} (Fig.~1). Accurate determination of molecular thermochemistry is therefore essential for quantifying such strain effects in cyclic systems. Experimentally, ring strain energies (RSEs) for cyclopropane and cyclobutane were determined by evaluating heats of formation derived from combustion data.\cite{wiberg1986concept, cox1970thermochemistry} In general, the excess energy obtained by comparing a cyclic molecule with an appropriate strain-free reference defines its ring strain energy.\cite{wiberg1986concept, khoury2004ring} \par

For many chemical systems, however, accurate thermochemical data are difficult to obtain, necessitating the use of theoretically motivated reaction schemes\cite{khoury2004ring,dudev1998ring, dill1979substituent}. While high-level ab initio methods provide reliable results, their computational cost becomes prohibitive for larger hydrocarbons. To address this challenge, systematic error-canceling hypothetical reaction schemes, such as isodesmic, homodesmotic, and hyperhomodesmotic reactions, have been developed. These schemes balance bond types, hybridization, and stereo-electronic effects, allowing chemically meaningful estimates of thermochemical quantities at modest levels of theory. Among these, homodesmotic reactions \cite{wheeler2012homodesmotic} have proven particularly effective for hydrocarbons, often achieving near chemical accuracy and offering valuable insight into stability trends. Homodesmotic reaction schemes, when combined with quantum chemical calculations, provide a powerful and computationally efficient framework for probing virtual properties such as ring strain energies, resonance energies, aromatic stabilization energies, and hyperconjugative effects in hydrocarbons. Such approaches enhance our understanding of the fundamental structural and energetic principles governing organic molecules.\par

In the present work, we integrate the Variational Quantum Eigensolver with homodesmotic reaction schemes to predict the ring strain energies of several cyclic hydrocarbons. By applying VQE to compute the enthalpies of reactants and products within carefully constructed homodesmotic reactions, reaction-level error cancellation can be maximized while employing quantum algorithms for electronic structure calculations. A key challenge in this context is the consistent selection of active spaces across all reaction species. To address this, we employ a previously introduced symmetry-guided active space selection strategy based on group-theoretical analysis,\cite{sarkar2025quantum} ensuring a balanced and reaction-consistent treatment of electron correlation. The generality of this framework is demonstrated across both saturated and unsaturated cyclic hydrocarbons, including structurally complex systems, while also providing physical insight into the role of intrinsic molecular symmetry in reaction-consistent quantum simulations. This synergy between VQE, symmetry considerations, and homodesmotic reactions provides a systematic framework for evaluating ring strain energies and exploring virtual thermodynamic properties of hydrocarbons within a quantum computational paradigm. To the best of our knowledge, this work represents the first systematic application of the Variational Quantum Eigensolver to homodesmotic reaction schemes for evaluating ring strain energies of cyclic hydrocarbons, including structurally complex molecules, within a symmetry-consistent active space framework.
\begin{figure}[H]
    \centering
    \includegraphics[width=0.75\textwidth]{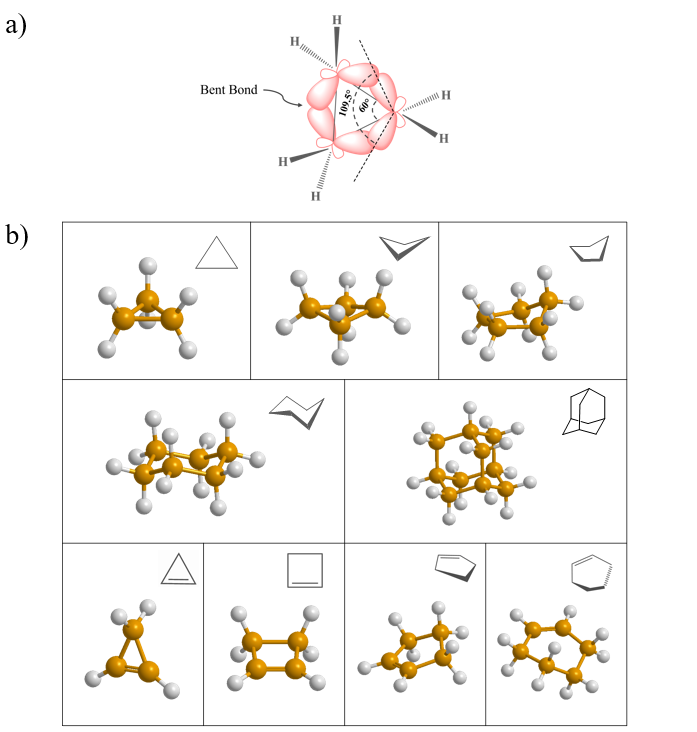}
    \caption{Illustration of ring strain in cyclic hydrocarbons. The top panel shows the bent-bond geometry in cyclopropane, highlighting the deviation from the ideal tetrahedral bond angle and the associated angular strain. The remaining panels depict the representative equilibrium conformations of cyclopropane, cyclobutane (butterfly), cyclopentane (envelope), cyclohexane (chair), and adamantane.}
\end{figure}


\section{\label{sec:level1} Theoretical Overview}

\subsection{Variational Quantum Eigensolver Algorithm}

VQE is a hybrid quantum-classical algorithm for estimating molecular ground-state energies (GSEs), based on the variational principle. 

\begin{equation}
    E(\vec{\theta}) = \langle \psi(\vec{\theta}) | H | \psi(\vec{\theta}) \rangle \geq E_0
\end{equation}
\par
Where $H$ is the molecular Hamiltonian and $ | \psi(\vec{\theta}) \rangle$ is the parameterized trial wavefunction, and $E_0$ is the exact GSE \cite{alteg2022study}. 
The Hamiltonian is expressed in qubit form (via first and second quantization, and Jordan-Wigner or Bravyi-Kitaev transformation) as a linear combination of Pauli operators. 
\begin{equation}
    H = \sum_i h_i P_i , \quad P_i \in \{I, X, Y, Z\}^{\otimes n}
\end{equation}

So the expectation value is measured as,
\begin{equation}
    E(\vec{\theta}) = \sum_i h_i \langle \psi(\vec{\theta}) | P_i | \psi(\vec{\theta}) \rangle
\end{equation}

The choice of ansatz is central to the success of VQE, because it determines how well the wavefunction can approximate the true ground state. A too-simple ansatz might fail to capture essential electronic correlation, while an overly complex one might introduce excessive parameters and circuit depth that are impractical on NISQ devices. This ansatz serves as the bridge between quantum hardware limitations and the accuracy of molecular simulations. \par
Chemistry-inspired ansatz, such as Unitary coupled cluster (UCC Ansatz) \cite{grimsley2019trotterized}, are important because they incorporate known physical structure (excitations from a Hartree-Fock reference), ensuring the trial wavefunction explores a meaningful subspace. The mathematical form of the UCCSD ansatz is as follows, encoding $T(\vec{\theta})$ as the excitation operator on a reference Hartree-Fock state. \par

\begin{equation}
    |\psi(\vec{\theta})\rangle = e^{T(\vec{\theta}) - T^\dagger(\vec{\theta})} |\phi_{\text{HF}}\rangle
\end{equation} \par
The parameters are optimized by a classical optimizer \cite{singh2023benchmarking}: 
\begin{equation}
    \vec{\theta}^* = \arg \min_{\vec{\theta}} E(\vec{\theta})
\end{equation}
\par
The final minimized value $E(\vec{\theta}^*)$ is taken as the best approximation of the ground state.


\subsection{Homodesmotic Reaction}

Error-balanced hypothetical reaction schemes have long played a central role in computational chemistry for obtaining reliable thermochemical quantities through systematic cancellation of correlation and basis-set errors\cite{hehre1970molecular, hehre1976ab, pople1971molecular, wheeler2012homodesmotic}. Such approaches remain particularly valuable for properties such as ring strain energies, where direct computation of absolute energies can be challenging even with advanced electronic structure methods. Among these schemes, isodesmic reactions were introduced to ensure conservation of bond types between reactants and products, thereby reducing sensitivity to the level of electronic structure theory employed\cite{pople1971molecular}. Bond separation reactions, a special subclass of isodesmic reactions, further refine this idea by decomposing molecules into fragments with matched bond environments, enabling enhanced error cancellation in thermochemical predictions\cite{wheeler2012homodesmotic}. Building on these concepts, homodesmotic reactions were formulated to achieve an even higher degree of balance by simultaneously conserving bond types, hybridization patterns, and the local bonding environments of atoms in both reactants and products\cite{george1976homodesmotic, wheeler2012homodesmotic}. As a result, homodesmotic reactions provide a more consistent treatment of electron correlation effects and have proven especially effective for hydrocarbons.\par

Homodesmotic reaction schemes have been successfully employed to evaluate ring strain energies and stabilization energies in cyclic hydrocarbons, yielding results that often approach chemical accuracy when combined with quantum chemical calculations\cite{george1975alternative, george1976homodesmotic, wheeler2012homodesmotic}. By carefully balancing the bonding environments on both sides of the reaction, these schemes minimize systematic errors and isolate the energetic contribution associated with molecular strain.\par

In the present work, homodesmotic reactions were considered for evaluating ring strain energies within a quantum computational framework. By computing the enthalpies of reactants and products using the Variational Quantum Eigensolver, the reaction-based formulation maximizes error cancellation while mitigating limitations arising from finite active
spaces and circuit depth. This synergy is particularly advantageous for near-term quantum simulations, where absolute energy calculations remain resource intensive. Based on two well-established sets of homodesmotic reaction schemes proposed by Khoury et al.\cite{khoury2004ring}and Wheeler\cite{wheeler2012homodesmotic}, we evaluate the ring strain energies of five cyclic hydrocarbons - cyclopropane, cyclobutane, cyclopentane, cyclohexane, and adamantane - using VQE-based quantum simulations. Additionally, few cyclic alkenes were also considered as shown below. The corresponding reaction schemes and expressions for ring strain energy (RSE) are summarized below.


\begin{tcolorbox}[colback=white!5!white, colframe=violet!80!black, title=Set I,before skip=5pt, after skip=5pt]
\begin{center}
\includegraphics[width=\textwidth]{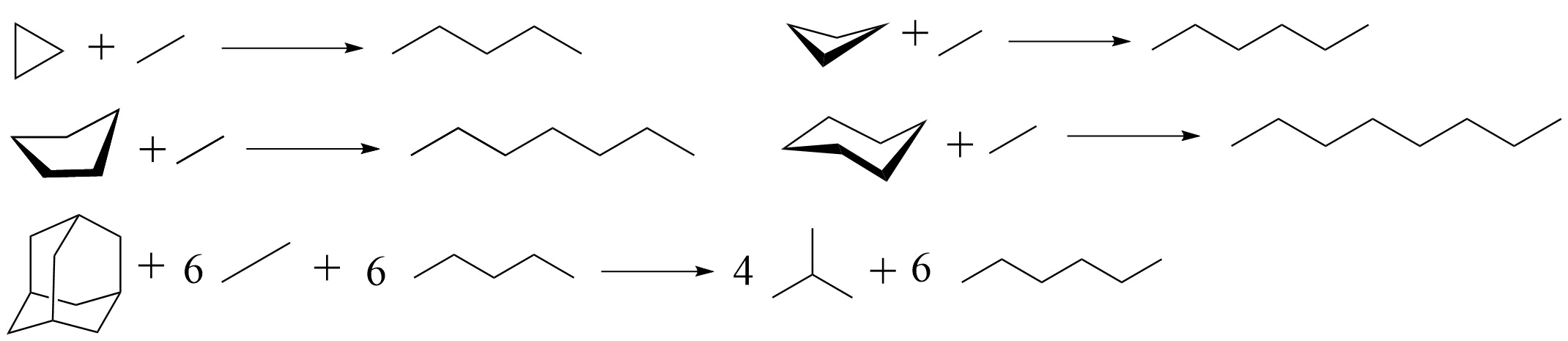}
\end{center}

\par\vspace{1pt}
\noindent\textcolor{violet}{\rule{\linewidth}{0.80pt}}
\vspace{6pt}

\includegraphics[width=\textwidth]{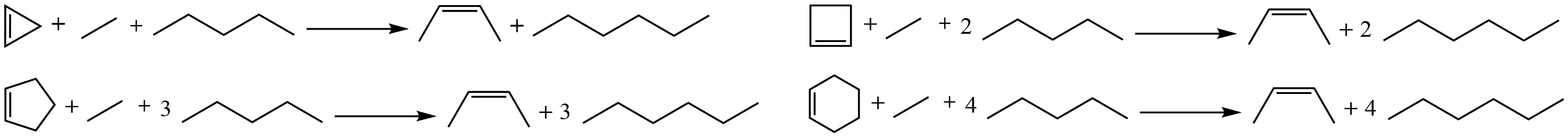}

\par\vspace{6 pt}
\noindent\textcolor{violet}{\rule{\linewidth}{0.8pt}}
\vspace{6pt}

This set of Homodesmotic reactions has been obtained from the work done by Khoury and his coworkers\cite{khoury2004ring}, and the ring strain energy is represented by:
\begin{center}
\includegraphics[width=0.4\textwidth]{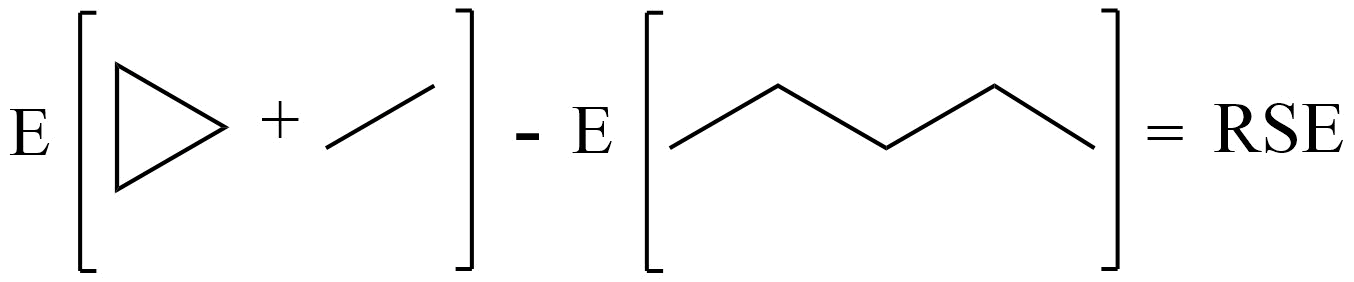}
\end{center}
Similarly, we can calculate the RSE for other rings by following their respective reaction schemes.
\end{tcolorbox}


\begin{tcolorbox}[colback=white!5!white, colframe=violet!80!black, title=Set II,before skip=5pt, after skip=5pt]
\begin{center}
\includegraphics[width=\textwidth]{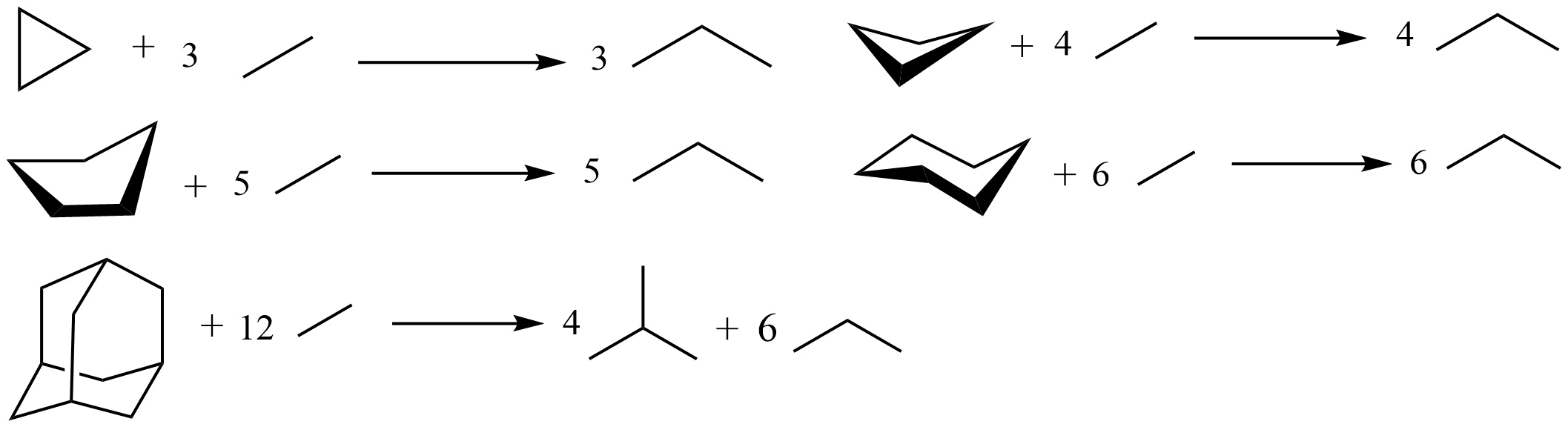}
\end{center}
\vspace{6pt}
\noindent\textcolor{violet}{\rule{\linewidth}{0.8pt}}
\vspace{6pt}

\includegraphics[width=\textwidth]{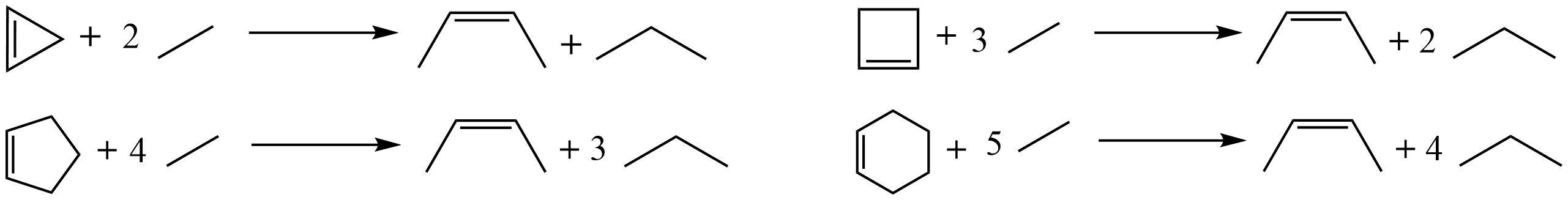}

\par\vspace{6pt}
\noindent\textcolor{violet}{\rule{\linewidth}{0.8pt}}
\vspace{6pt}
This set of Homodesmotic reactions has been obtained from the paper done by Wheeler\cite{wheeler2012homodesmotic}, and the ring strain energy is represented by.
\begin{center}
\includegraphics[width=0.4\textwidth]{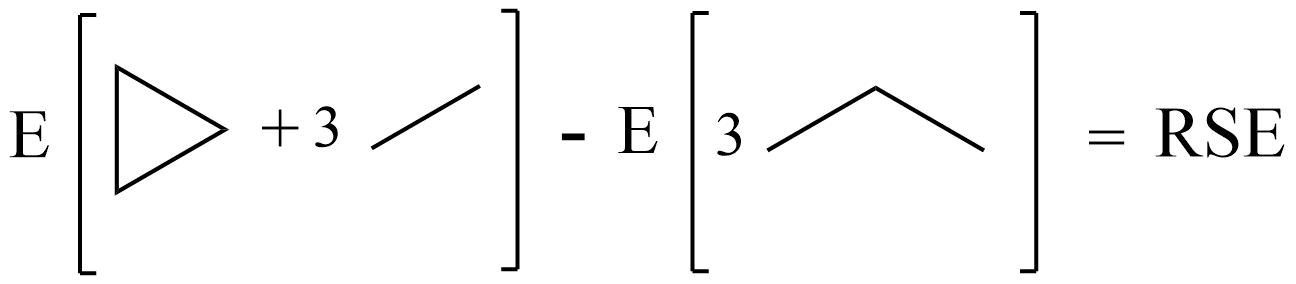}
\end{center}
Similarly, we can calculate the RSE for other rings by following their respective reaction schemes.
\end{tcolorbox}


\section{ Computational Methodology}

\subsection{Computational Chemistry}

Quantum chemical calculations for all reactants and products were performed using the NWChem software package. Geometry optimizations and ground-state energy calculations were carried out within density functional theory using the B3LYP exchange–correlation functional in conjunction with the 6-31G* basis set. These optimized geometries and corresponding
electronic energies were used to evaluate the ring strain energies for each cyclic hydrocarbon through the homodesmotic reaction schemes described in Section II. In addition to density functional theory, coupled-cluster calculations with single and double excitations (CCSD) were performed to obtain reference ring strain energies for comparison. CCSD
energies were computed using single point calculations using the density functional theory optimized geometries. All quantum chemical calculations were carried out on conventional computing platforms. 

For the symmetry-guided active space selection, NWChem package was employed to determine the irreducible representations of the molecular orbitals by exploiting the highest applicable abelian point group symmetry for each reactant and product involved in the homodesmotic reactions. The highest applicable abelian point group was employed to ensure unambiguous orbital symmetry labeling and consistent symmetry matching across reactants and products, which is essential for constructing reaction-consistent active spaces. These symmetry labels form the basis for constructing symmetry-consistent active spaces across reaction species, as described in the following subsection.


\subsection{Quantum Computing and Group theory}

The accuracy of Variational Quantum Eigensolver (VQE) calculations is strongly influenced by the choice of active space used to represent the molecular electronic structure. In reaction energy calculations, this choice becomes particularly critical, as an imbalanced treatment of electron correlation between reactants and products can lead to significant systematic errors. To ensure a consistent and symmetry-balanced description across all species involved in a reaction, we employ a group-theoretic strategy for active space selection, following and extending our previously developed framework for reaction energy prediction. 

For each molecular species, multiple candidate active spaces were constructed, ranging from a minimal (2, 2) active space to larger active spaces up to (10, 10). Each active space defines a truncated Hilbert space within which the VQE calculation is performed.  For a given active space, the Hartree–Fock reference state is expressed as a single Slater determinant,

\begin{equation}
|\Psi_0\rangle = |\phi_1 \bar{\phi}_1 \, \phi_2 \bar{\phi}_2 \cdots \phi_n \bar{\phi}_n \rangle
\end{equation}

where $|\phi\rangle$ and $|\bar{\phi}\rangle_i$ denote spin-up and spin-down molecular orbitals, respectively. The irreducible representation of each molecular orbital was determined using density functional theory calculations in NWChem, employing the highest applicable abelian point group symmetry for each molecule. While the use of non-Abelian point-group symmetry would be desirable in principle, current quantum algorithms and fermion-to-qubit mappings predominantly exploit Abelian symmetries, making Abelian symmetry adaptation the practical choice in VQE implementations. Accordingly, all symmetry analyses in this work are carried out within the highest-order Abelian point group compatible with each molecular structure.


\begin{figure}[H]
    \centering
    \includegraphics[width=0.8\textwidth]{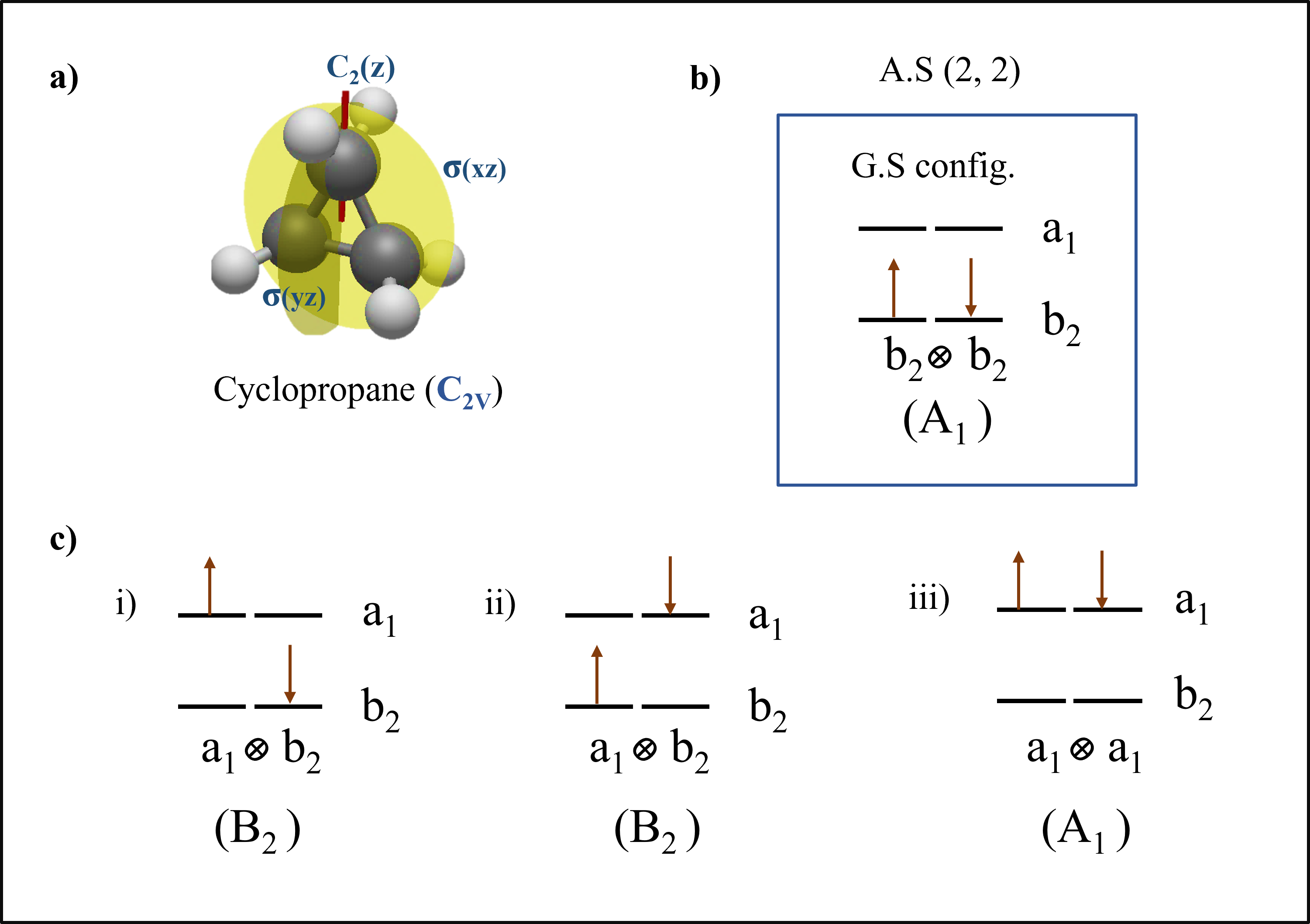}
    \caption{Illustration of symmetry classification of excitations for cyclopropane in a (2, 2) active space under $C_{2v}$ point group symmetry.\newline 
    (a) Cyclopropane structure with sym. elements indicated.\newline 
    (b) Ground-state electronic config., transforming as the $A_{1}$ irrep.\newline 
    (c) Possible single excitations relative to the ground state: (i) and (ii) excitations transforming as $B_{2}$, and (iii) excitation transforming as $A_{1}$.}
\end{figure}

For every candidate active space, all possible single and double excited configurations relative to the Hartree–Fock reference state were enumerated. Each excited configuration was assigned an irreducible representation of the molecular point group, obtained from the direct product of the irreps of the molecular orbitals involved in the excitation. Based
on this classification, we previously have introduced in one of our works a quantitative metric, termed the Symmetry Matched Fraction (SMF)\cite{sarkar2025quantum}, defined as the fraction of excited configurations that transform according to the same irreducible representation as the reference state,

\[
\textit{SMF}= 
\frac{N_{\text{excitation with same irrep with the ref.~state}}}
     {N_{\text{total excitations}}}
\times 100
\]

The SMF provides a measure of how effectively a given active space preserves symmetry compatibility between the reference state and its associated excited configurations, thereby serving as a proxy for the extent to which symmetry-allowed correlation effects are retained within the truncated Hilbert space. As an illustrative example, Fig. 2 shows the symmetry
classification of excitations for cyclopropane in a (2, 2) active space under $C_{2v}$ symmetry. The Hartree–Fock reference configuration transforms as the $a_{1}$ irreducible representation. Within this active space, three excited configurations are possible, comprising two single excitations transforming as $b_{2}$ and one double excitation transforming as $a_{1}$. Since only one of the three excited configurations shares the ground-state symmetry, the SMF for
this active space is 33.3\%.

For reaction energy calculations, including ring strain energy evaluation, symmetry consistency across all reactants and products is essential. Accordingly, active spaces were selected such that the SMF values of the reactants and products matched exactly. In cases where multiple active spaces yielded the same SMF value for a given molecule, we observed that larger active spaces often provide improved correlation treatment. However, smaller symmetry-consistent active spaces can in several instances yield comparable or marginally improved agreement due to favorable error cancellation within the homodesmotic framework. This indicates that symmetry consistency is the primary requirement, while active space size provides additional flexibility rather than a strictly monotonic improvement in accuracy. Ground-state energy calculations were performed using the VQE algorithm with the unitary coupled-cluster singles and doubles (UCCSD) ansatz as implemented in Qiskit (version 1.0.0). The molecular electronic Hamiltonians were mapped to qubits using the parity mapping with symmetry-based tapering to reduce the qubit count. Optimization of the variational parameters was carried out using the simultaneous
perturbation stochastic approximation (SPSA) algorithm. All VQE simulations were performed using the statevector simulator backend.


\section{ Result and discussion}

Using the symmetry-guided active space selection strategy described above, we computed the ring strain energies (RSEs) of five cyclic hydrocarbons following two distinct sets of homodesmotic reaction schemes. In total, ten homodesmotic reactions were examined, and internally consistent trends were obtained for all systems considered, ranging from cyclopropane to the structurally more complex adamantane.

As outlined in the Computational Methodology, for each reactant and product we evaluated the symmetry-matched fraction (SMF) associated with multiple candidate active spaces. We found that more than one active space can yield identical SMF values (SI, Table 2-18). In such cases, the final choice of active space was guided by a balance between system size and chemical considerations. For smaller rings such as cyclopropane and cyclobutane, relatively compact active spaces were sufficient to capture the dominant correlation effects while maintaining chemical accuracy and minimal computational overhead. In contrast, for moderately sized and structurally complex systems, expansion of the active space often improved quantitative agreement, although in several cases smaller symmetry-consistent active spaces were already sufficient to achieve chemical accuracy.

The VQE-computed RSE values were bench marked against DFT and CCSD results, as shown in Figs. 3–7. In each bar plot, the third bar corresponds to RSE values obtained using a uniform minimal active space of (2, 2) for all reaction species, without enforcing any symmetry-based selection criteria. The subsequent bars represent results obtained using the symmetry-guided protocol, where common SMF values of 33.33 \% or 50 \% were employed to select symmetry consistent active spaces across all reactants and products.

For a deeper assessment of the robustness and the transferability of the symmetry-guided active space selection protocol, we further extended our analysis of ring strain energy (RSE) determination for unsaturated cyclic hydrocarbon systems. Involvement of $\pi$-electron correlation and symmetry sensitive electronic structure effects in these unsaturated cyclic hydrocarbons provide a more stringent evaluation of the SMF-based active space selection strategy within the Homodesmotic reaction schemes.


\begin{figure*}[htbp]
    \centering
    \begin{subfigure}{0.38\textwidth}
        \centering
        \includegraphics[width=\linewidth]{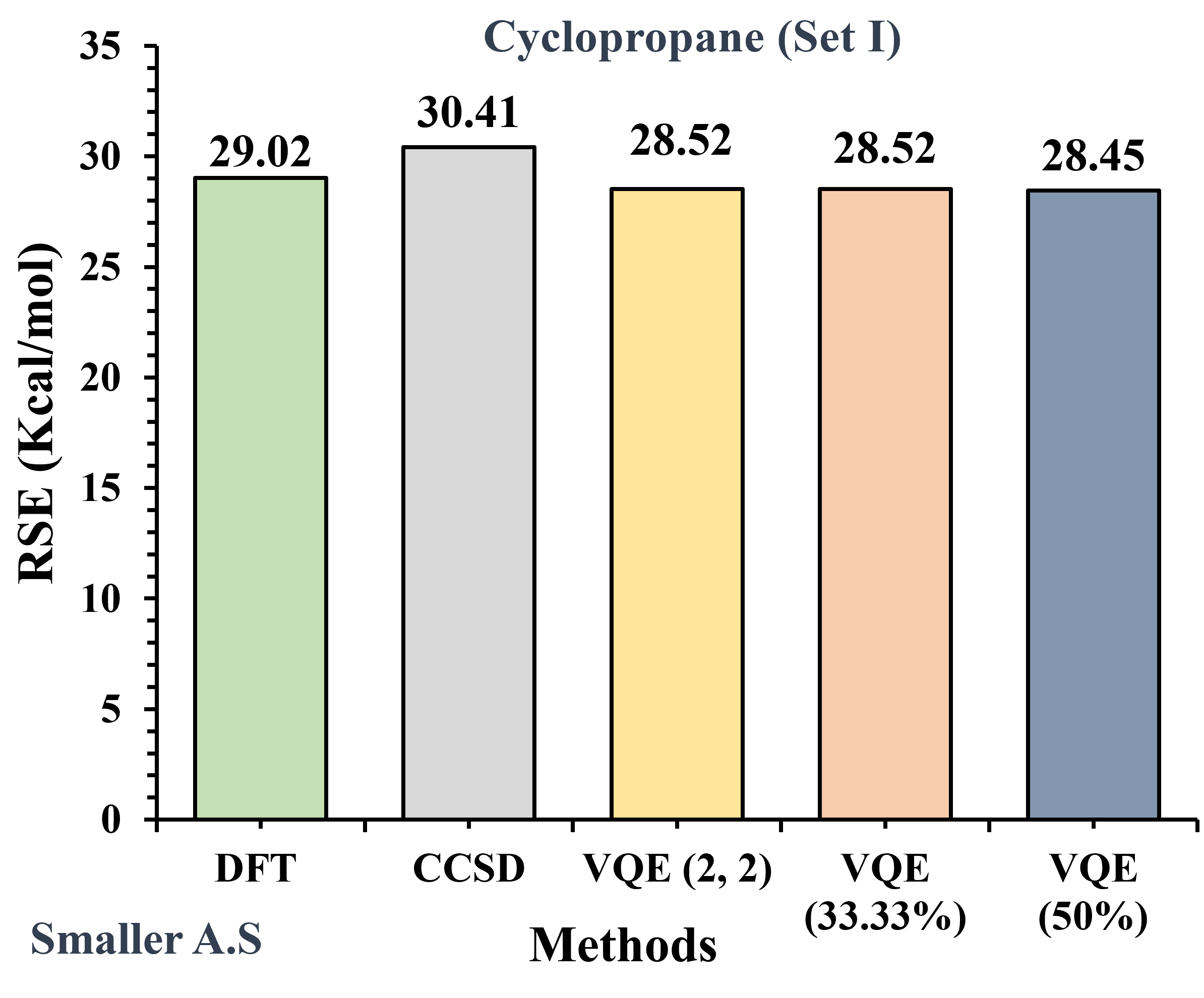}
        \caption{}
    \end{subfigure}
    \hspace{0.11\textwidth}
    \begin{subfigure}{0.38\textwidth}
        \centering
        \includegraphics[width=\linewidth]{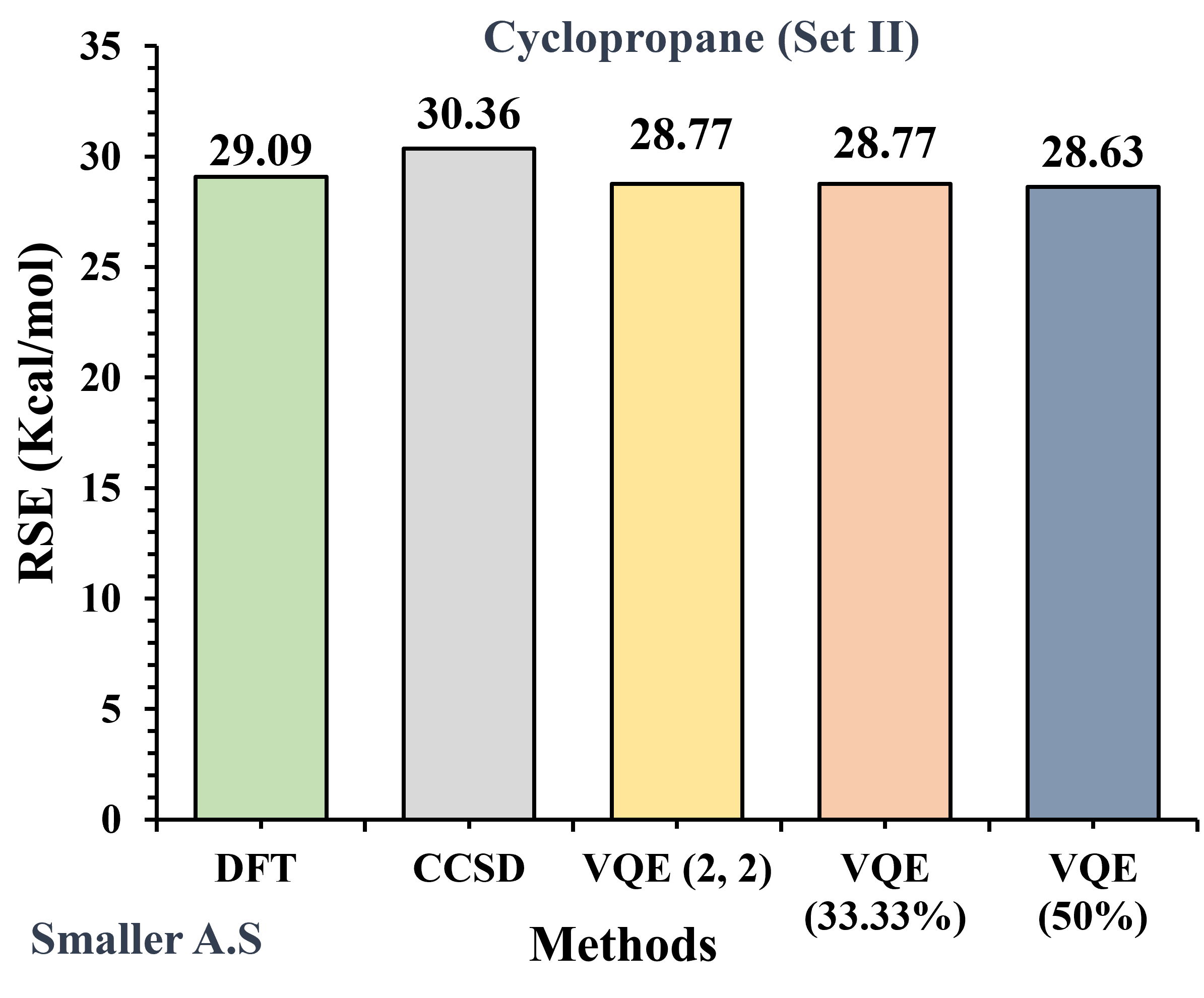}
        \caption{}
    \end{subfigure}

    \vspace{0.5cm} 

    \begin{subfigure}{0.38\textwidth}
        \centering
        \includegraphics[width=\linewidth]{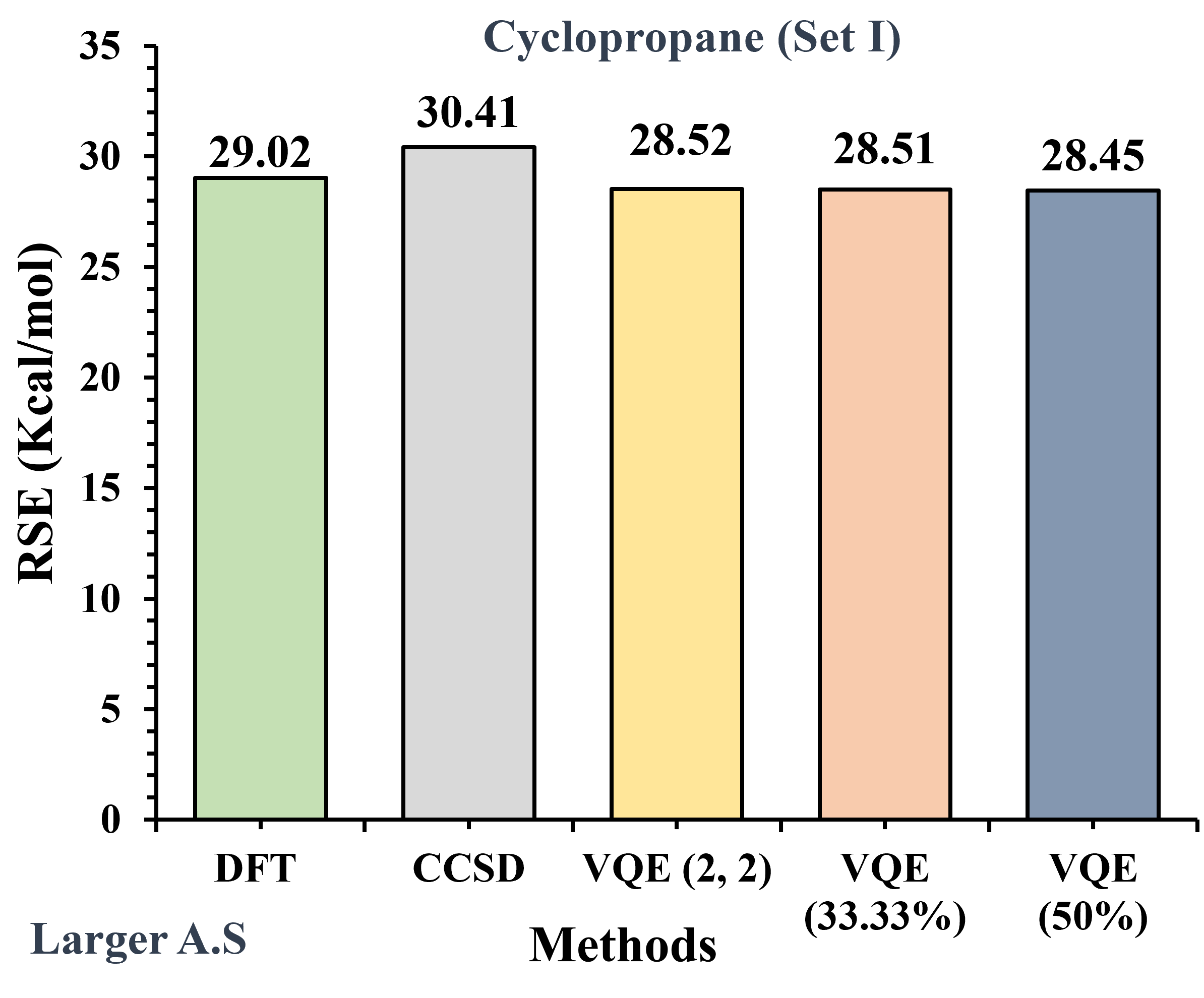}
        \caption{}
    \end{subfigure}
    \hspace{0.11\textwidth}
    \begin{subfigure}{0.38\textwidth}
        \centering
        \includegraphics[width=\linewidth]{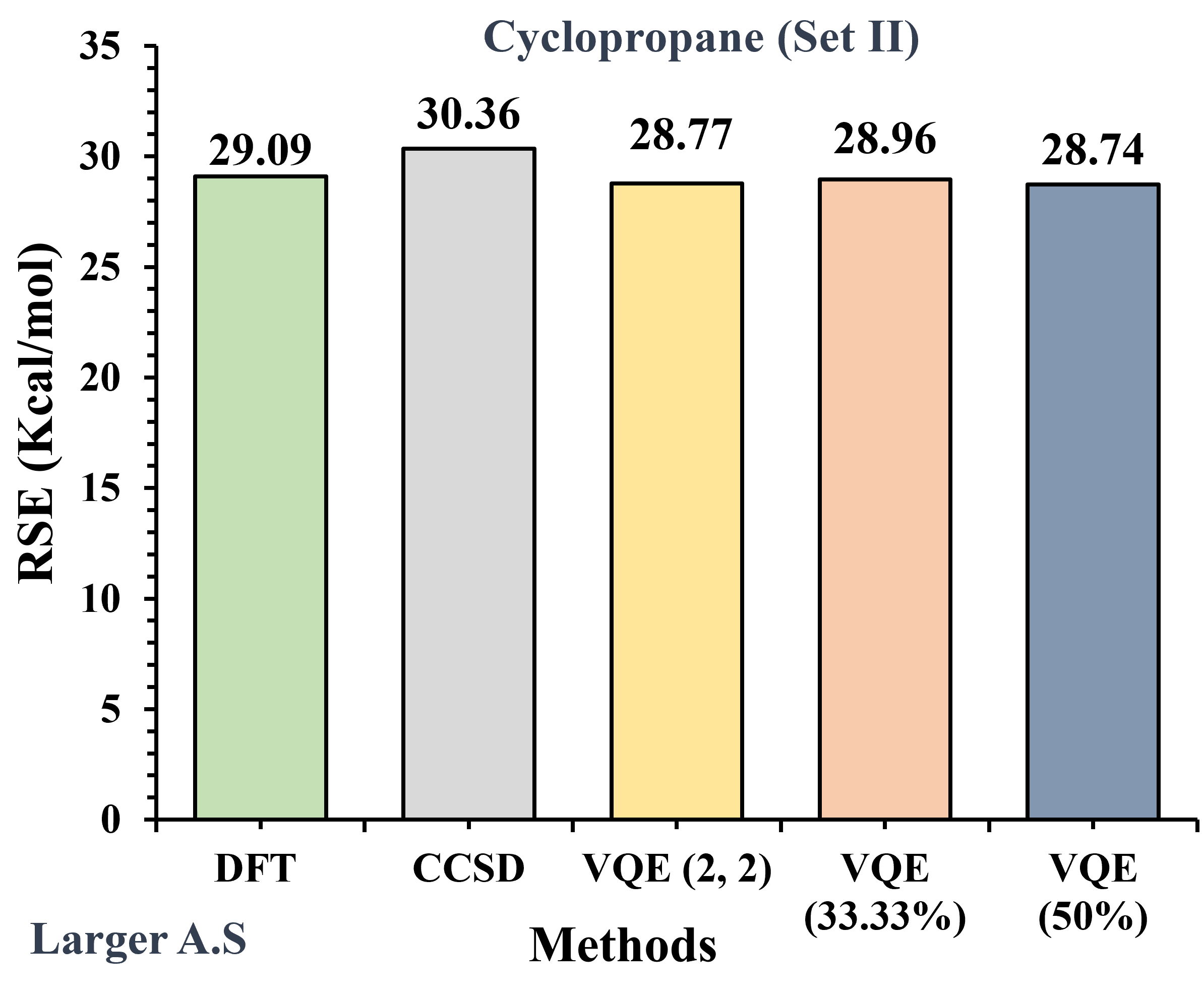}
        \caption{}
    \end{subfigure}

\caption{Bar plots comparing ring strain energies (RSEs) of cyclopropane obtained from VQE calculations using different active space selection strategies for Set I and Set II homodesmotic reaction schemes. Panels (a) and (b) correspond to Set I and Set II reactions, respectively, using the smallest symmetry-consistent active spaces, while panels (c) and (d) show results obtained with larger symmetry-consistent active spaces. In each panel, bars (from left to right) represent density functional theory (DFT), coupled-cluster singles and doubles (CCSD), VQE with a uniform (2, 2) active space without symmetry guidance, and VQE with the symmetry-guided active space selection protocol employing a matched SMF value of 33.33 \% across all reactants and products.}
\end{figure*}


\subsection{Saturated Cyclic Hydrocarbons}

\subsubsection{Cyclopropane}

The bar plots for cyclopropane are shown in Fig. 3. For this smallest ring system, the uniform (2, 2) active space and the symmetry-guided selection with an SMF match of 33.33 \% yield identical RSE values, as both approaches effectively correspond to the same active spaces for all reaction species in both Set I and Set II homodesmotic reactions. The resulting RSE values are within 0.3–0.5 kcal/mol of the DFT reference and within 1.5–1.8 kcal/mol of the CCSD values [Figs. 3(a,b)]. Upon increasing the active space size while maintaining the same SMF value, a modest improvement is observed for the Set II reaction scheme, with deviations reduced to approximately 0.1 kcal/mol relative to DFT and 1.3 kcal/mol relative to CCSD [Fig. 3(d)], whereas no significant change is seen for the Set I scheme [Fig. 3(c)].


\begin{figure*}[htbp]
    \centering
    \begin{subfigure}{0.38\textwidth}
        \centering
        \includegraphics[width=\linewidth]{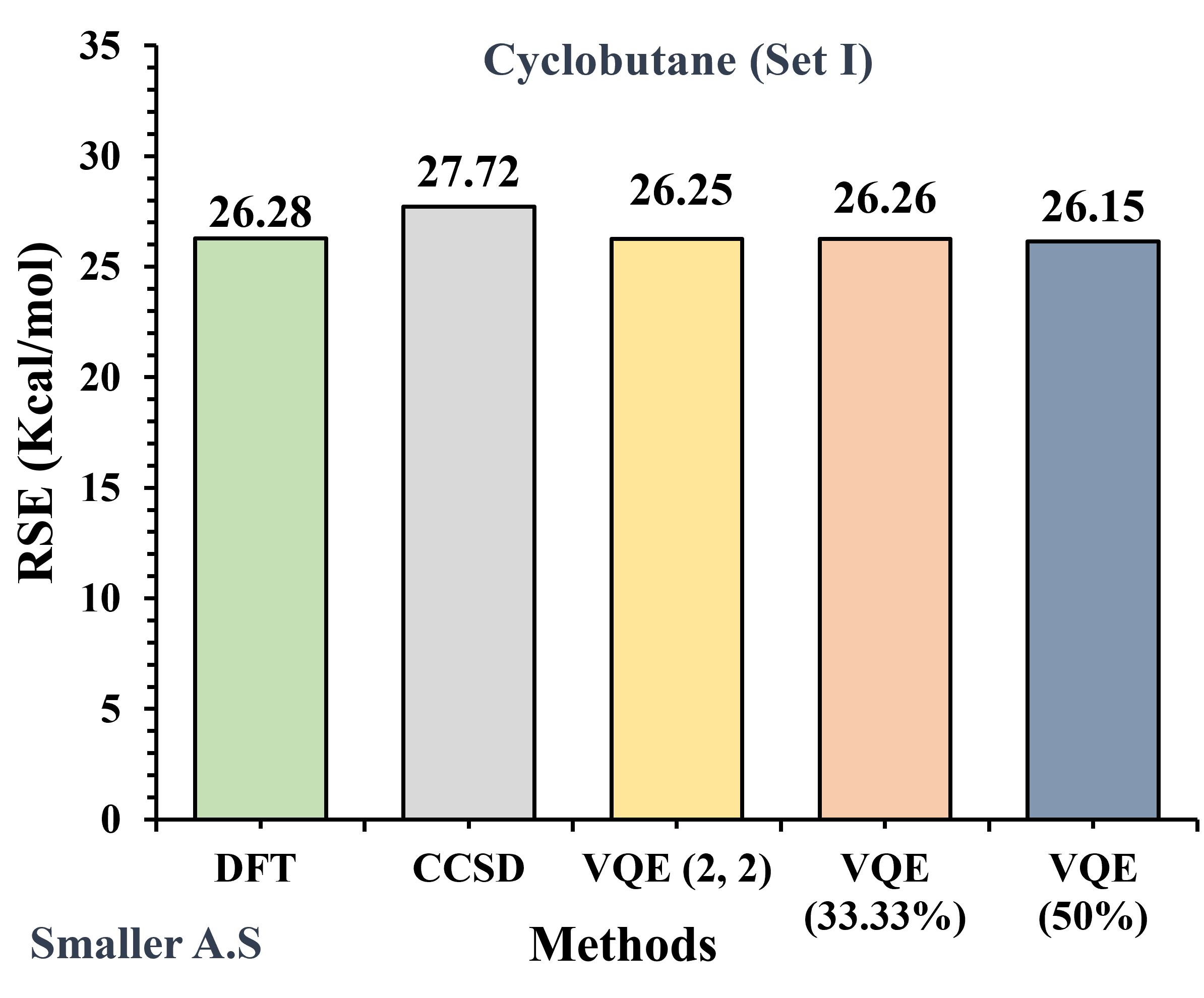}
        \caption{}
    \end{subfigure}
    \hspace{0.11\textwidth}
    \begin{subfigure}{0.38\textwidth}
        \centering
        \includegraphics[width=\linewidth]{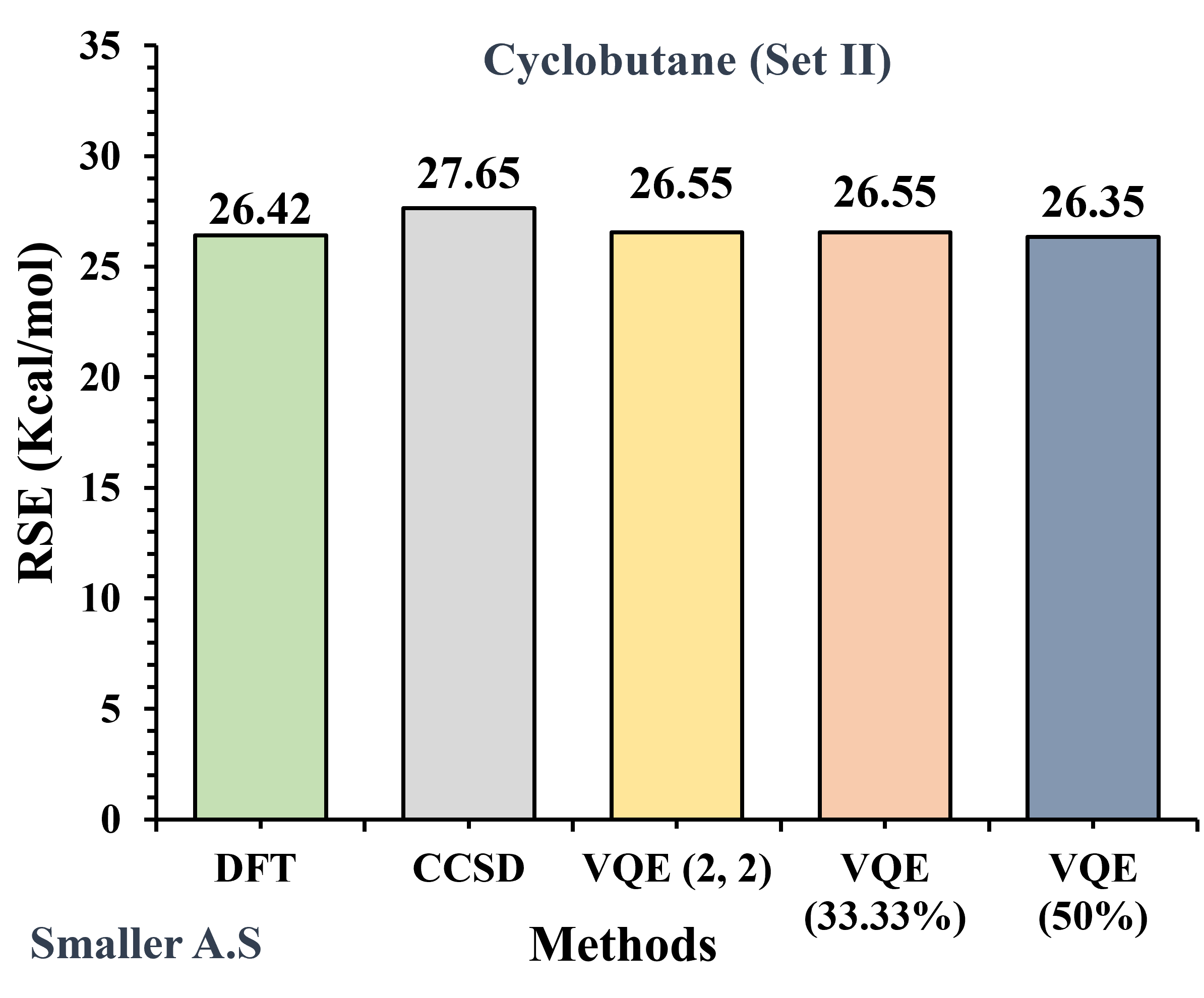}
        \caption{}
    \end{subfigure}

    \vspace{0.5cm} 

    \begin{subfigure}{0.38\textwidth}
        \centering
        \includegraphics[width=\linewidth]{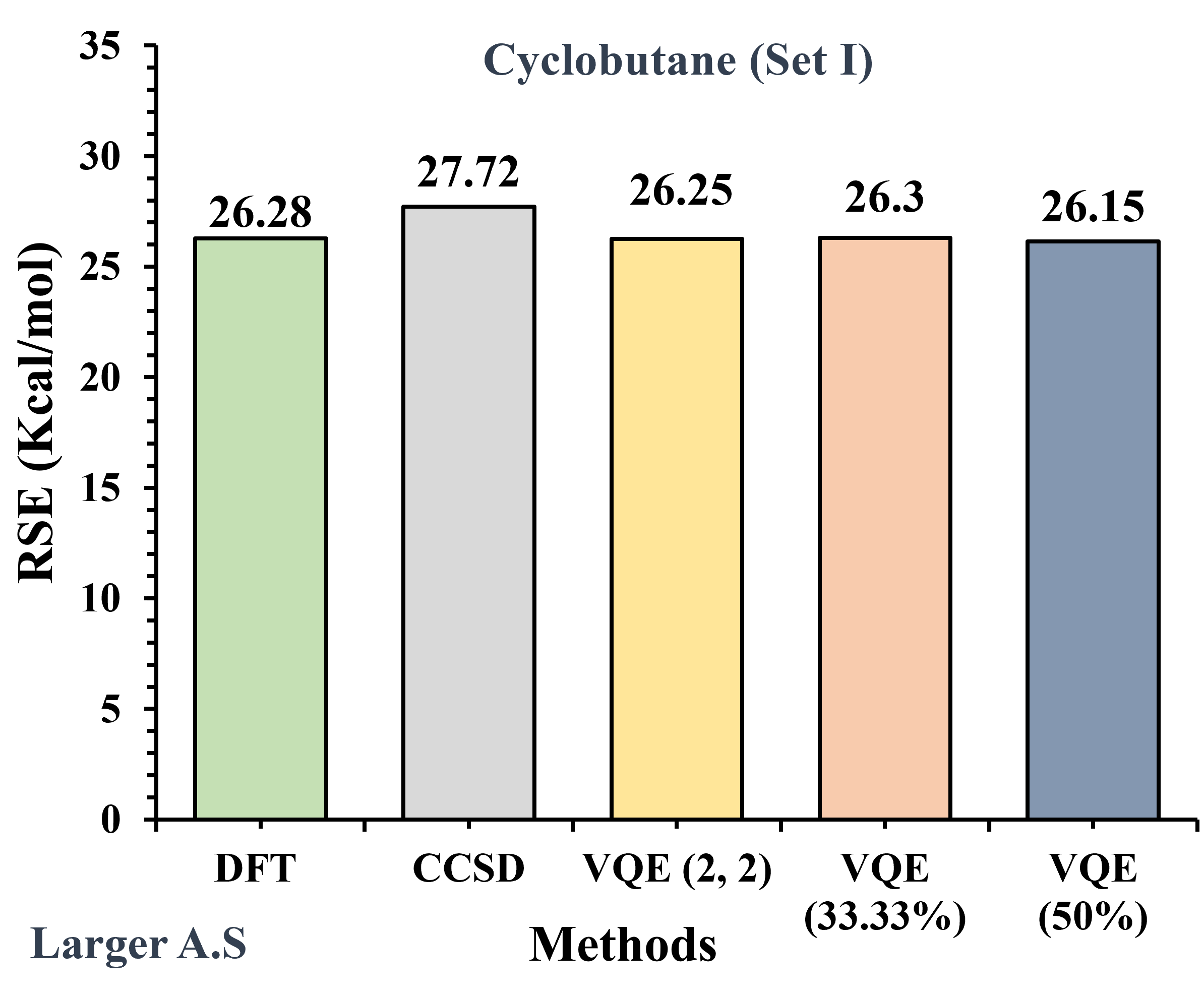}
        \caption{}
    \end{subfigure}
    \hspace{0.11\textwidth}
    \begin{subfigure}{0.38\textwidth}
        \centering
        \includegraphics[width=\linewidth]{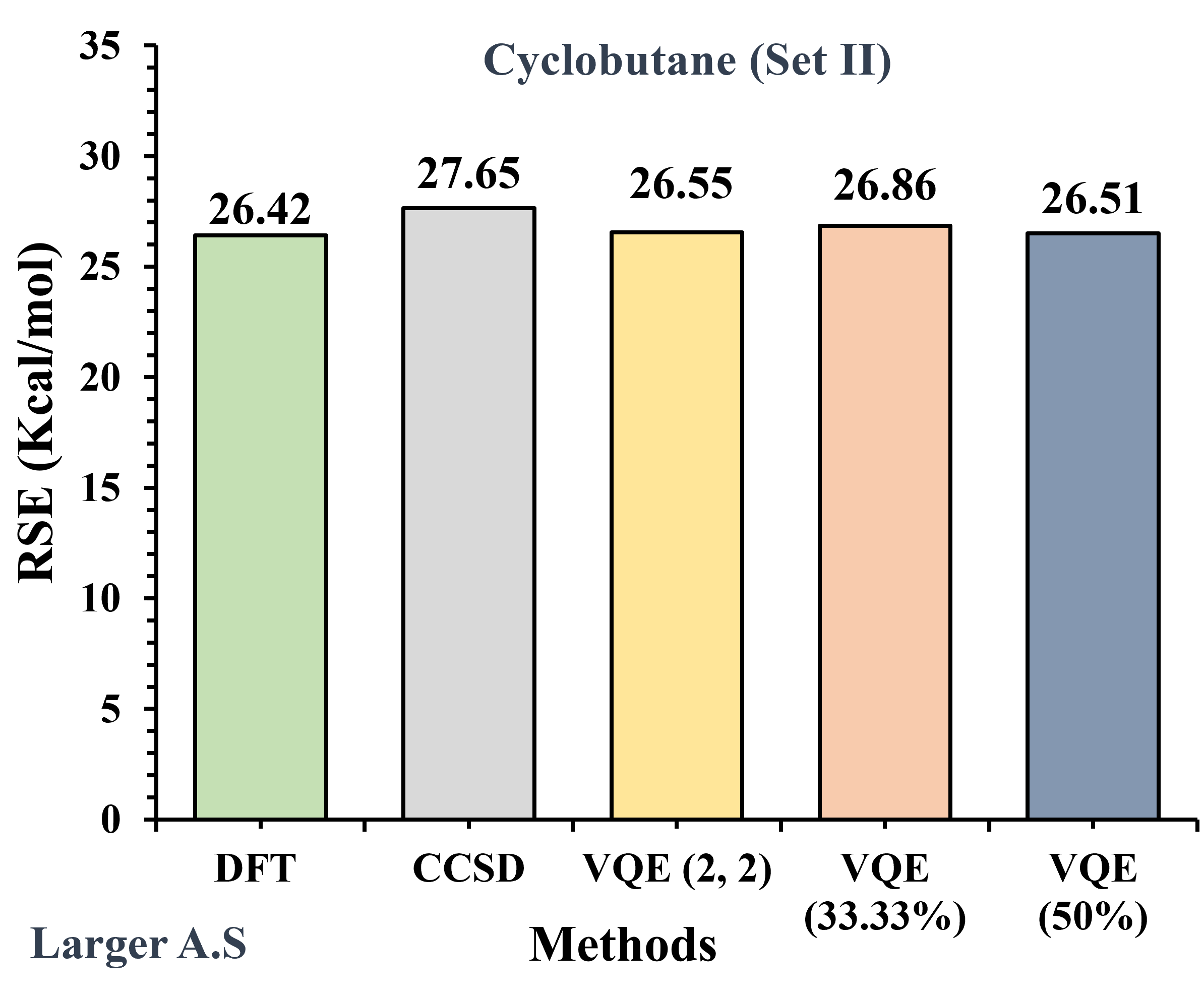}
        \caption{}
    \end{subfigure}
\caption{Bar plots for cyclobutane}
\end{figure*}

\subsubsection{Cyclobutane}
Cyclobutane exhibits trends closely analogous to those observed for cyclopropane (Fig. 4). Both the uniform (2, 2) active space and the symmetry-guided selection with an SMF of 33.33 \% produce RSE values with deviations of less than 0.02 kcal/mol relative to DFT and 1.08–1.45 kcal/mol relative to CCSD for both homodesmotic reaction sets [Figs. 4(a,b)]. Increasing the active space size while preserving the SMF condition leads to further improvement for the Set II reaction scheme, reducing the deviation to approximately 0.77 kcal/mol relative to CCSD [Fig. 4(d)], whereas Set I again shows minimal sensitivity to active space enlargement [Fig. 4(c)].


\begin{figure*}[htbp]
    \centering
    \begin{subfigure}{0.38\textwidth}
        \centering
        \includegraphics[width=\linewidth]{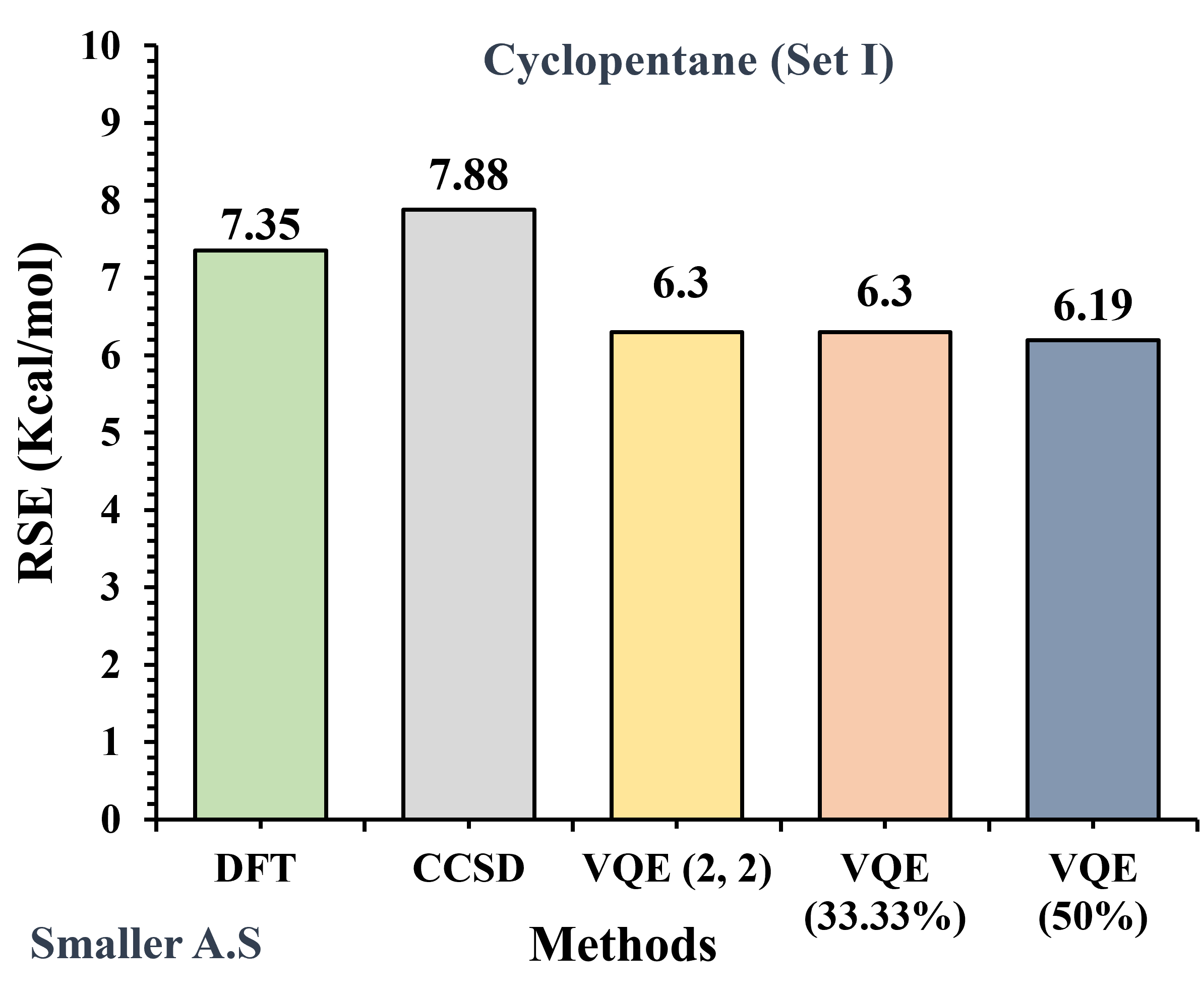}
        \caption{}
    \end{subfigure}
    \hspace{0.11\textwidth}
    \begin{subfigure}{0.38\textwidth}
        \centering
        \includegraphics[width=\linewidth]{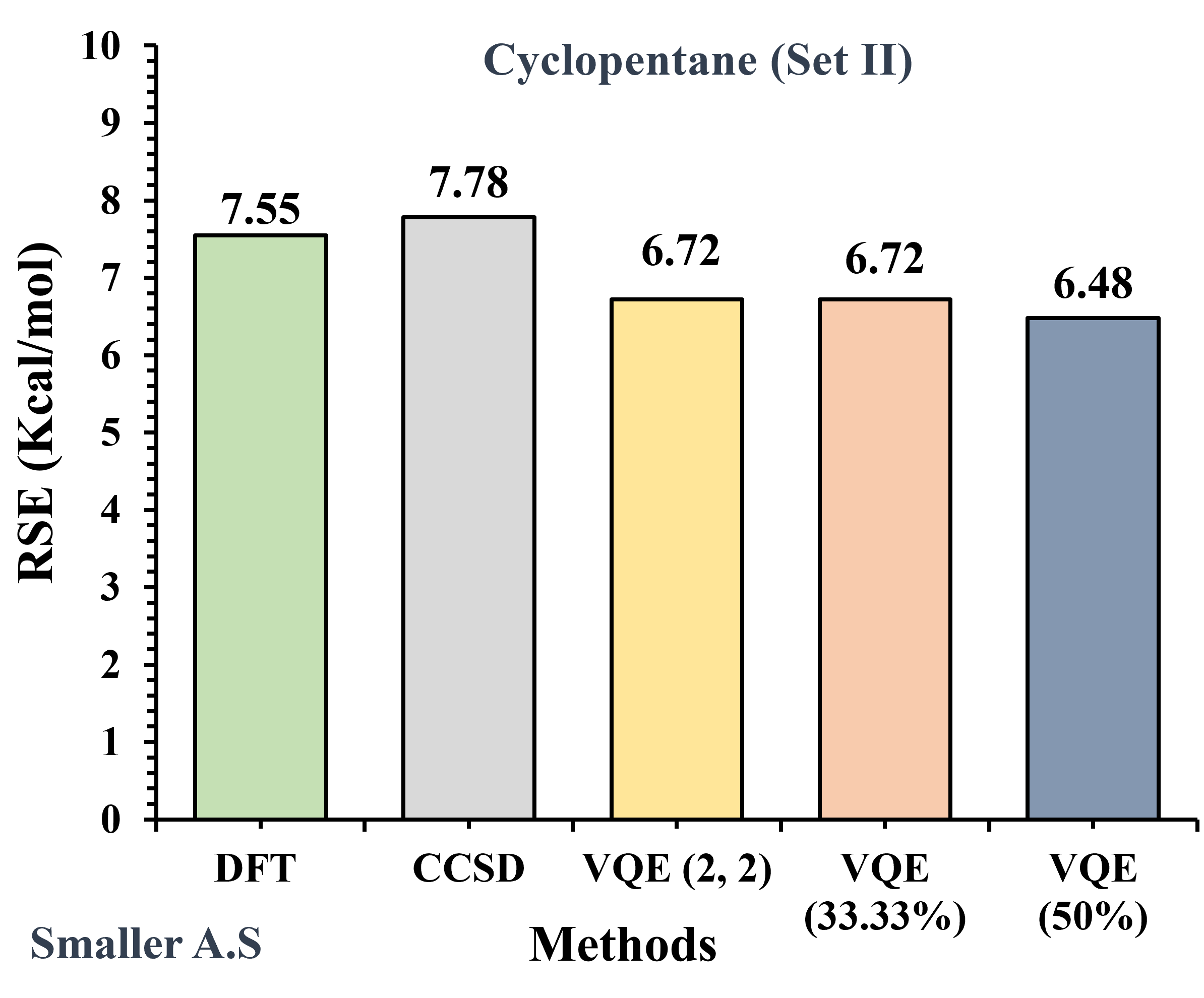}
        \caption{}
    \end{subfigure}

    \vspace{0.5cm} 

    \begin{subfigure}{0.38\textwidth}
        \centering
        \includegraphics[width=\linewidth]{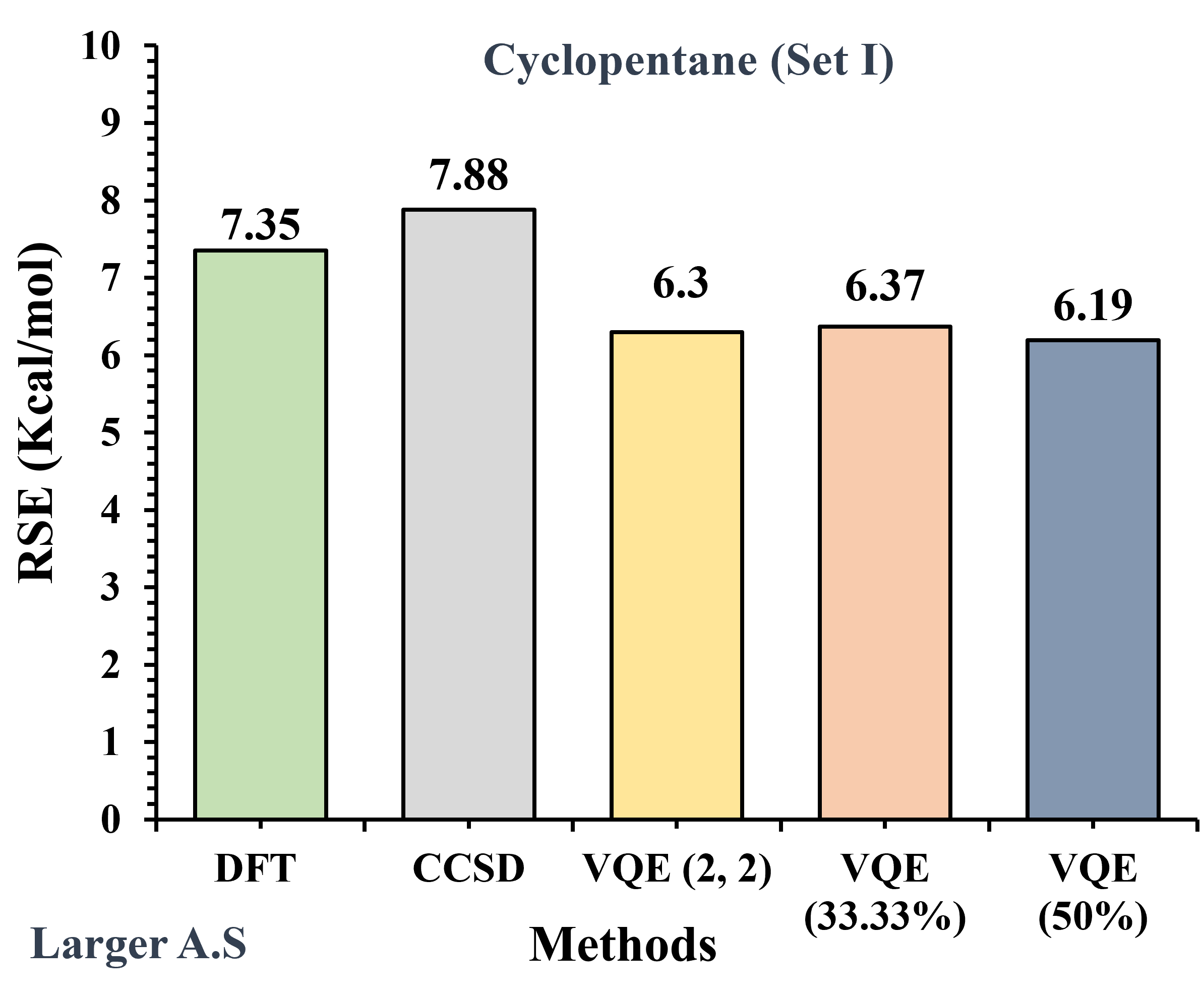}
        \caption{}
    \end{subfigure}
    \hspace{0.11\textwidth}
    \begin{subfigure}{0.38\textwidth}
        \centering
        \includegraphics[width=\linewidth]{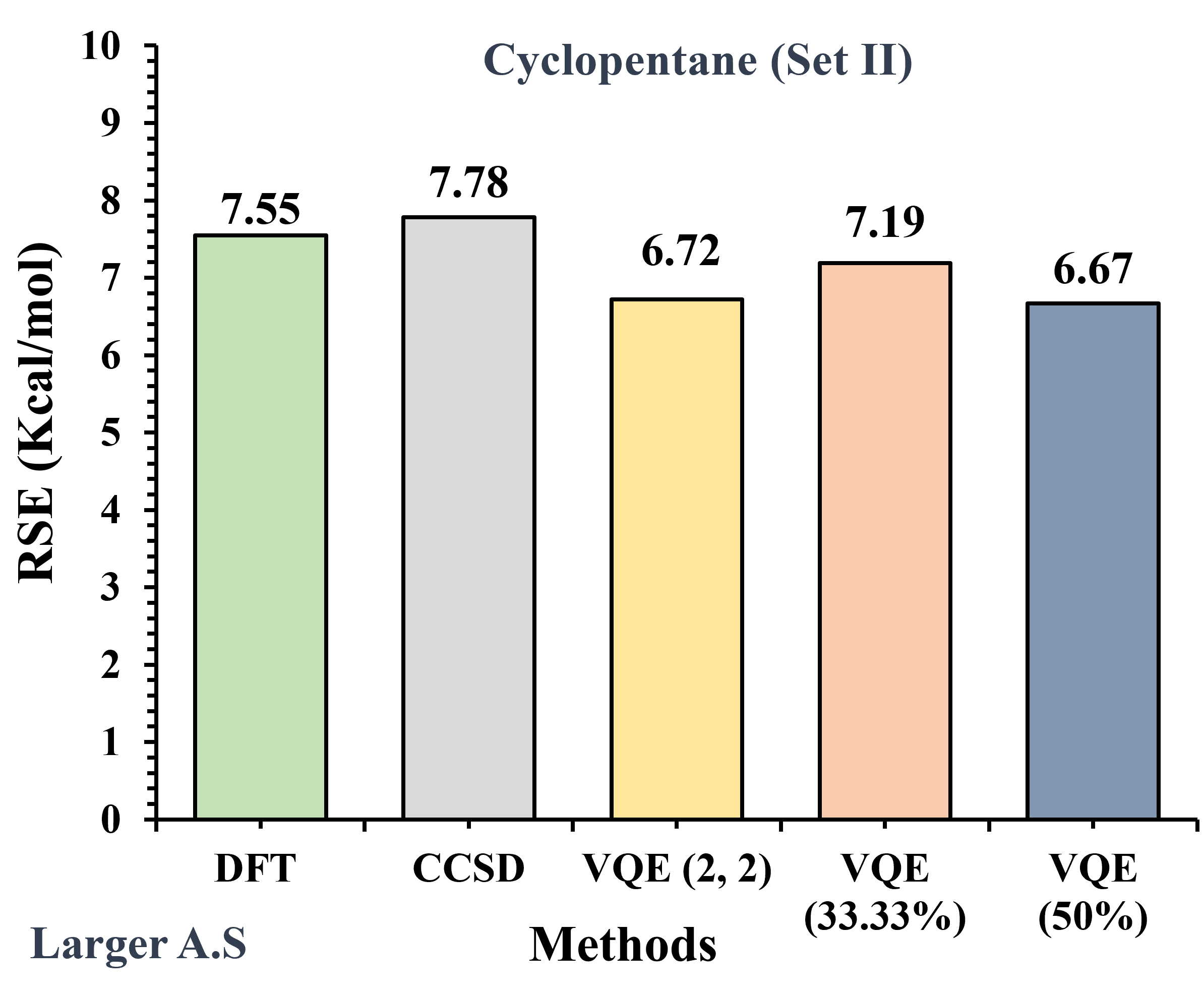}
        \caption{}
    \end{subfigure}
\caption{Bar plots for cyclopentane}
\end{figure*}

\subsubsection{Cyclopentane}

For cyclopentane, the uniform (2, 2) active space yields deviations of 0.8–1.05 kcal/mol relative to DFT and 1.04–1.5 kcal/mol relative to CCSD for both homodesmotic schemes (Fig. 5). Application of the symmetry-guided protocol with small active spaces does not lead to a substantial improvement [Figs. 5(a,b)]. However, upon increasing the active space size, a noticeable enhancement is obtained, particularly for the Set II reaction scheme. In this case, an SMF match of 33.33 \% yields better agreement than the 50 \% condition, highlighting the importance of both symmetry matching and sufficient active space flexibility [Figs. 5(c,d)].


\begin{figure*}[htbp]
    \centering
    \begin{subfigure}{0.38\textwidth}
        \centering
        \includegraphics[width=\linewidth]{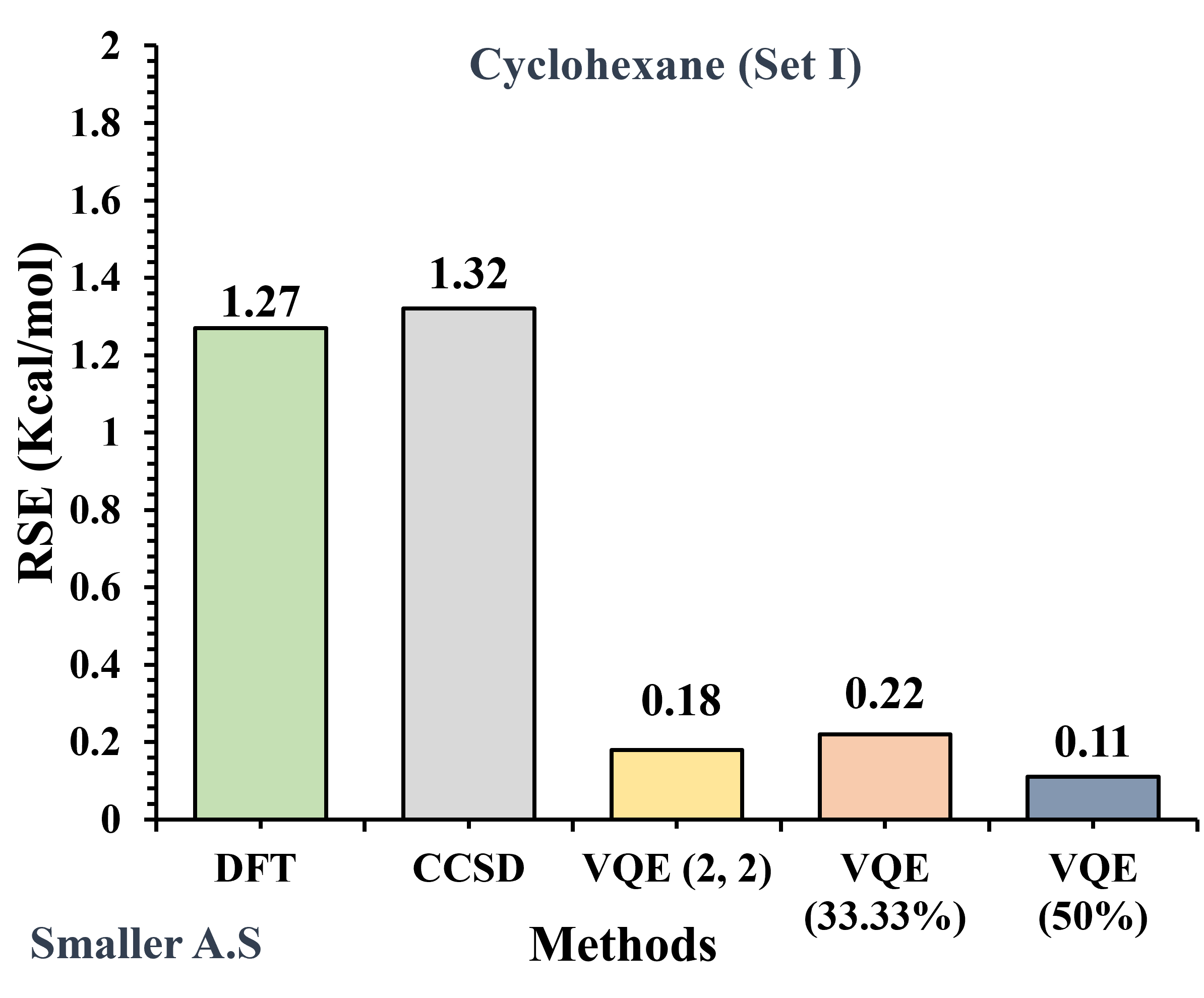}
        \caption{}
    \end{subfigure}
    \hspace{0.11\textwidth}
    \begin{subfigure}{0.38\textwidth}
        \centering
        \includegraphics[width=\linewidth]{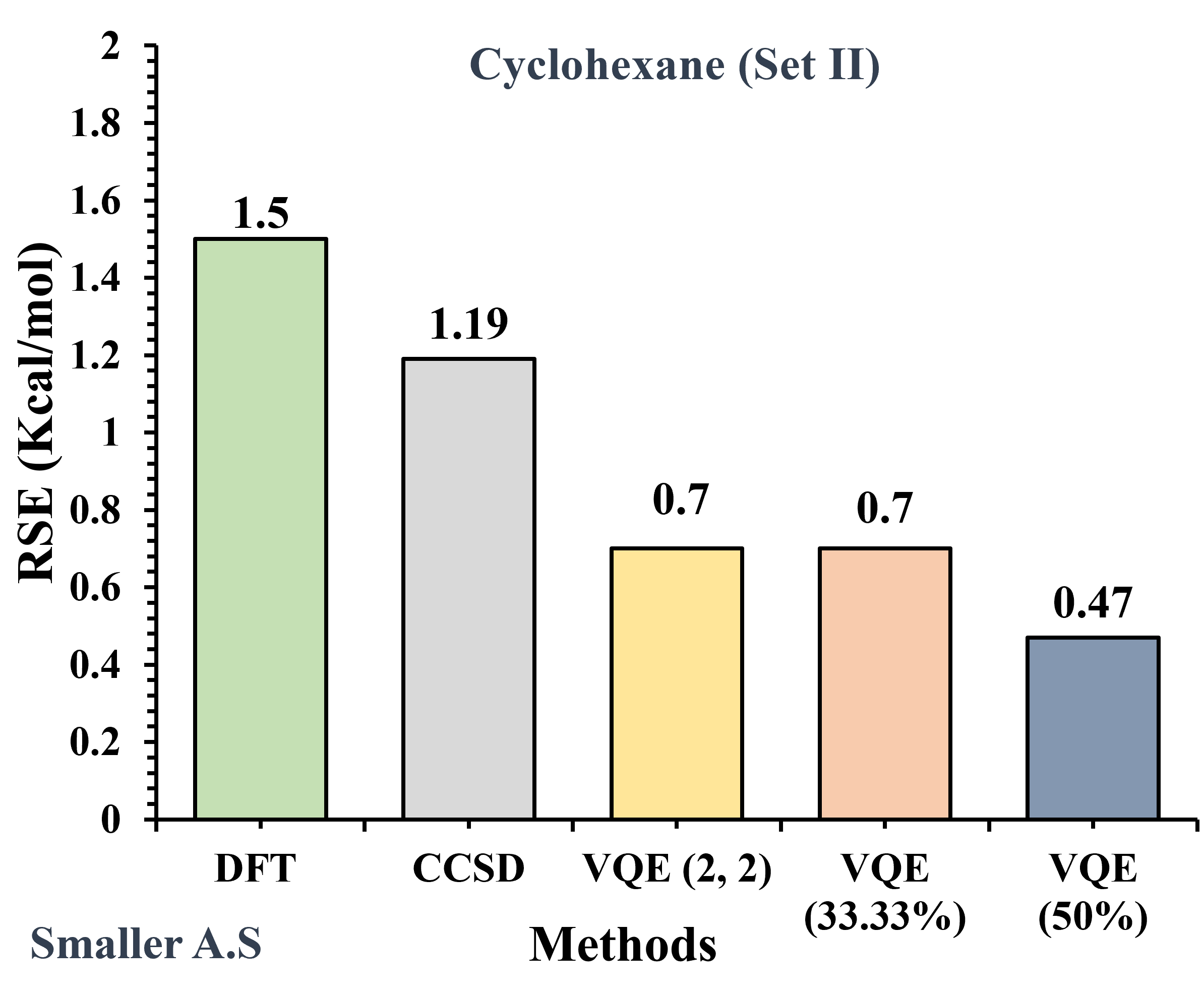}
        \caption{}
    \end{subfigure}

    \vspace{0.5cm} 

    \begin{subfigure}{0.38\textwidth}
        \centering
        \includegraphics[width=\linewidth]{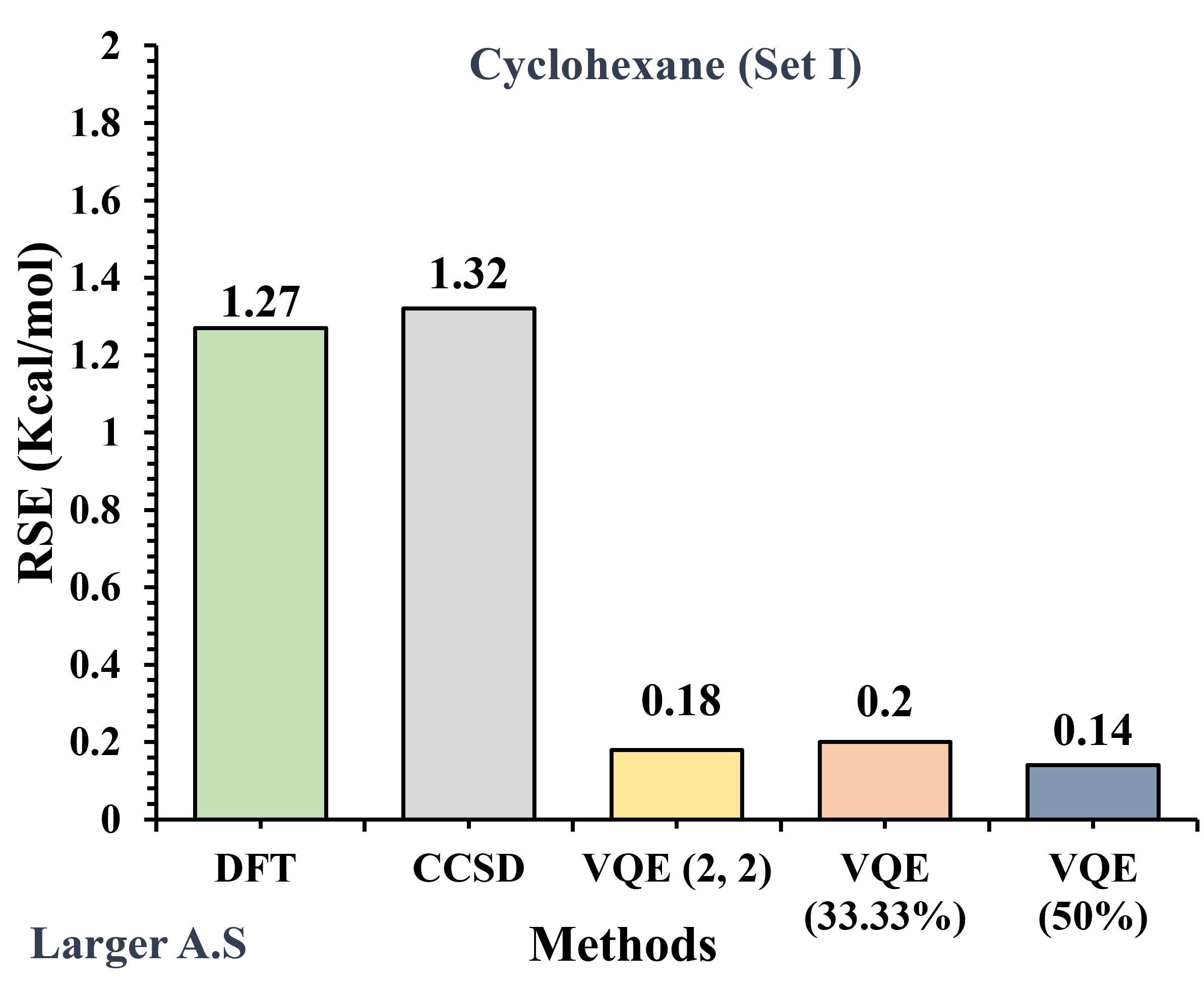}
        \caption{}
    \end{subfigure}
    \hspace{0.11\textwidth}
    \begin{subfigure}{0.38\textwidth}
        \centering
        \includegraphics[width=\linewidth]{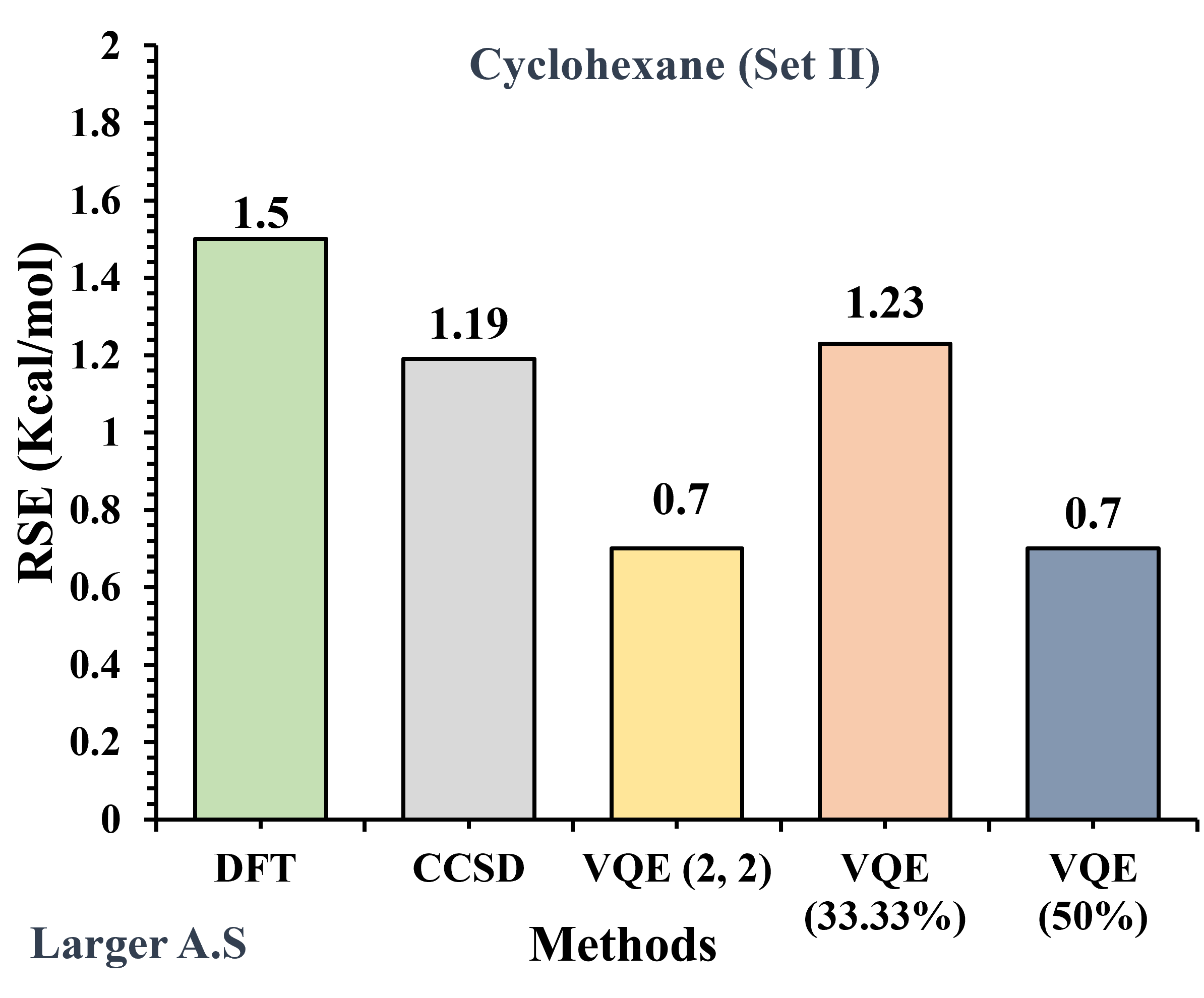}
        \caption{}
    \end{subfigure}
\caption{Bar plots for cyclohexane}
\end{figure*}

\subsubsection{Cyclohexane}

Cyclohexane follows a pattern similar to cyclopentane (Fig. 6). Using a uniform (2, 2) active space results in deviations of 0.8–1.09 kcal/mol with respect to DFT and 0.5–1.1 kcal/mol relative to CCSD. The symmetry-guided strategy does not significantly alter the results for the Set I reaction scheme, irrespective of active space size [Figs. 6(a,c)]. In contrast, for the Set II scheme, employing a larger symmetry-consistent active space with an SMF of 33.33 \% leads to a marked improvement in the predicted RSE [Figs. 6(b,d)].


\begin{figure*}[htbp]
    \centering
    \begin{subfigure}{0.38\textwidth}
        \centering
        \includegraphics[width=\linewidth]{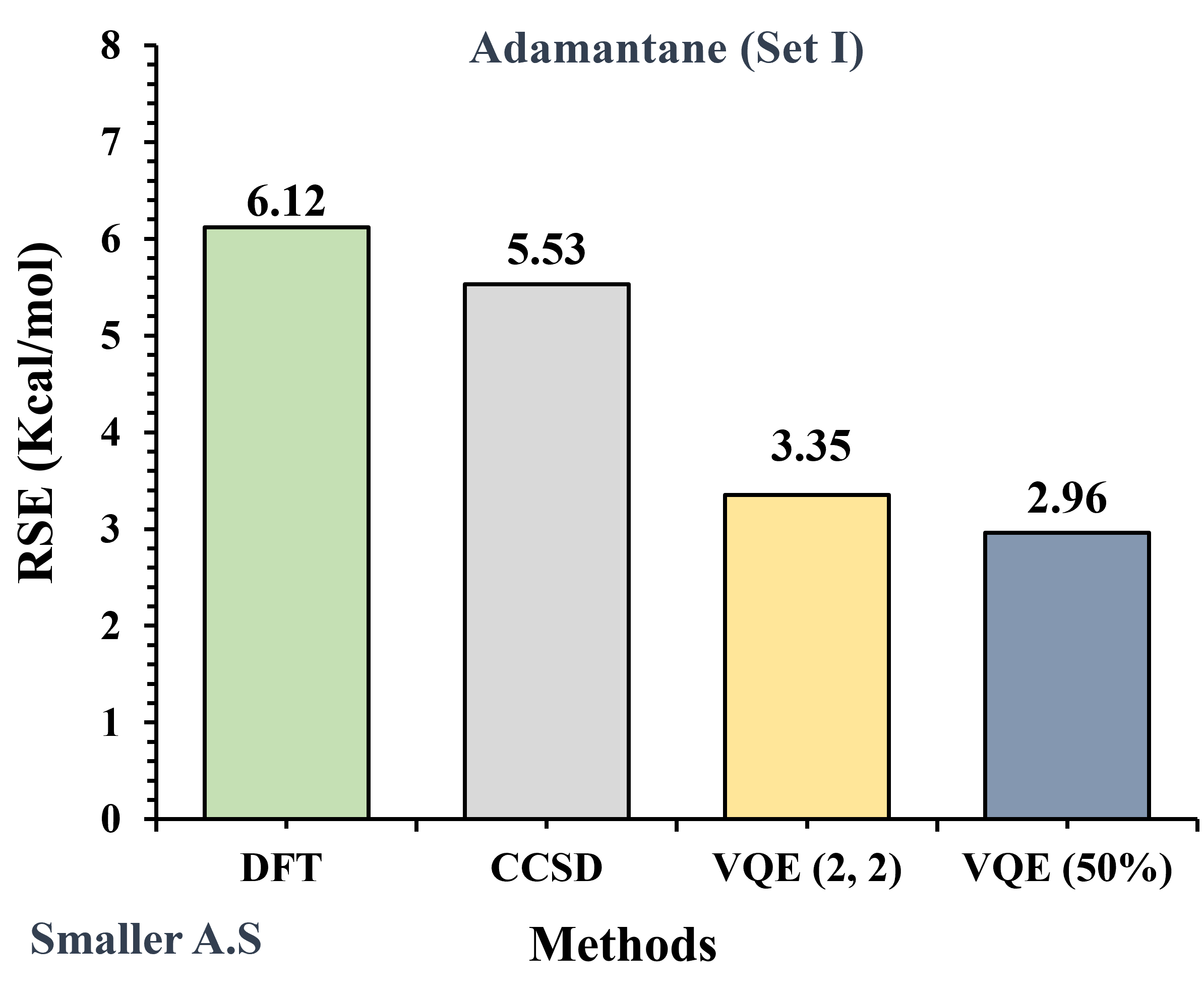}
        \caption{}
    \end{subfigure}
    \hspace{0.11\textwidth}
    \begin{subfigure}{0.38\textwidth}
        \centering
        \includegraphics[width=\linewidth]{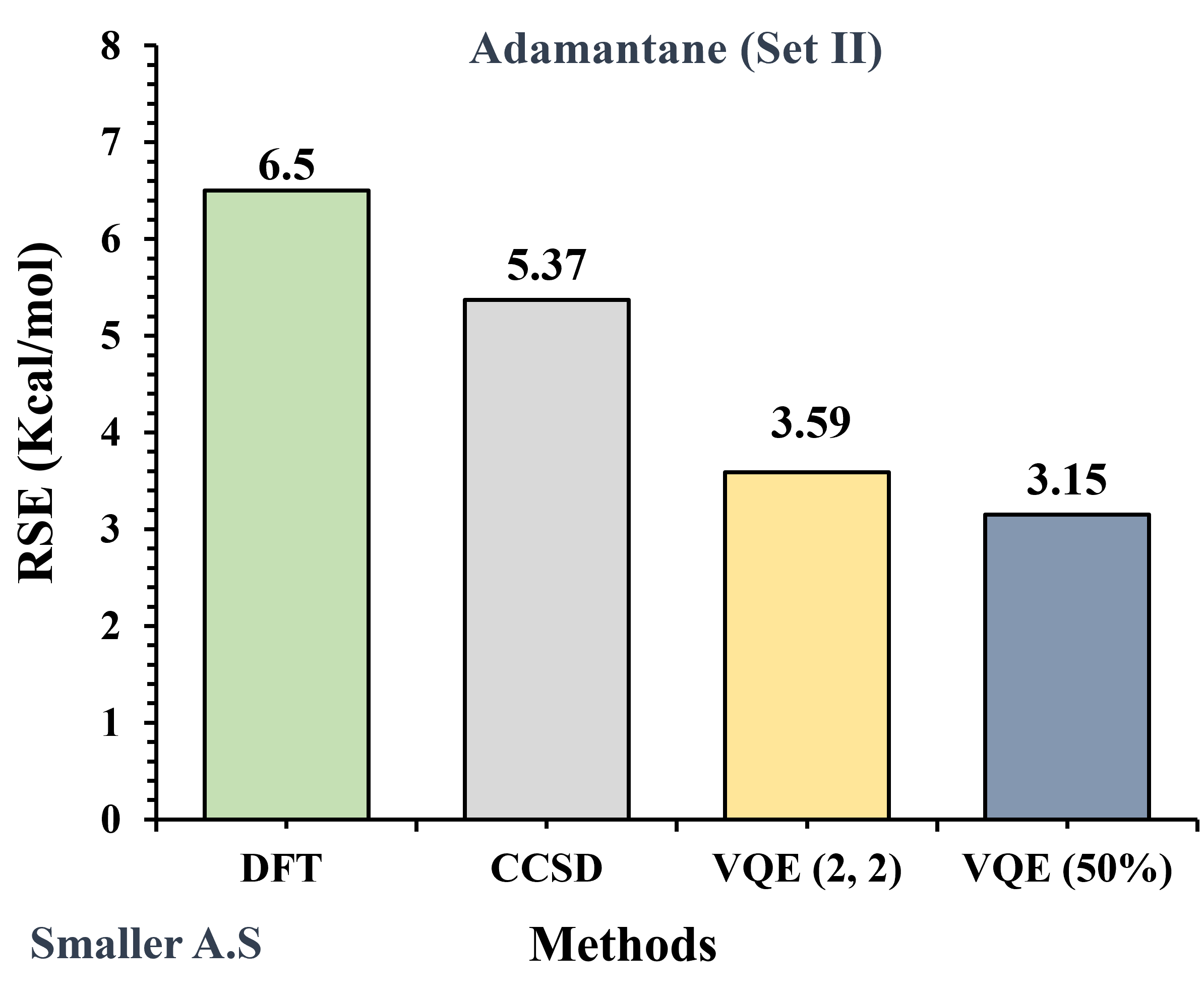}
        \caption{}
    \end{subfigure}

    \vspace{0.5cm} 

    \begin{subfigure}{0.38\textwidth}
        \centering
        \includegraphics[width=\linewidth]{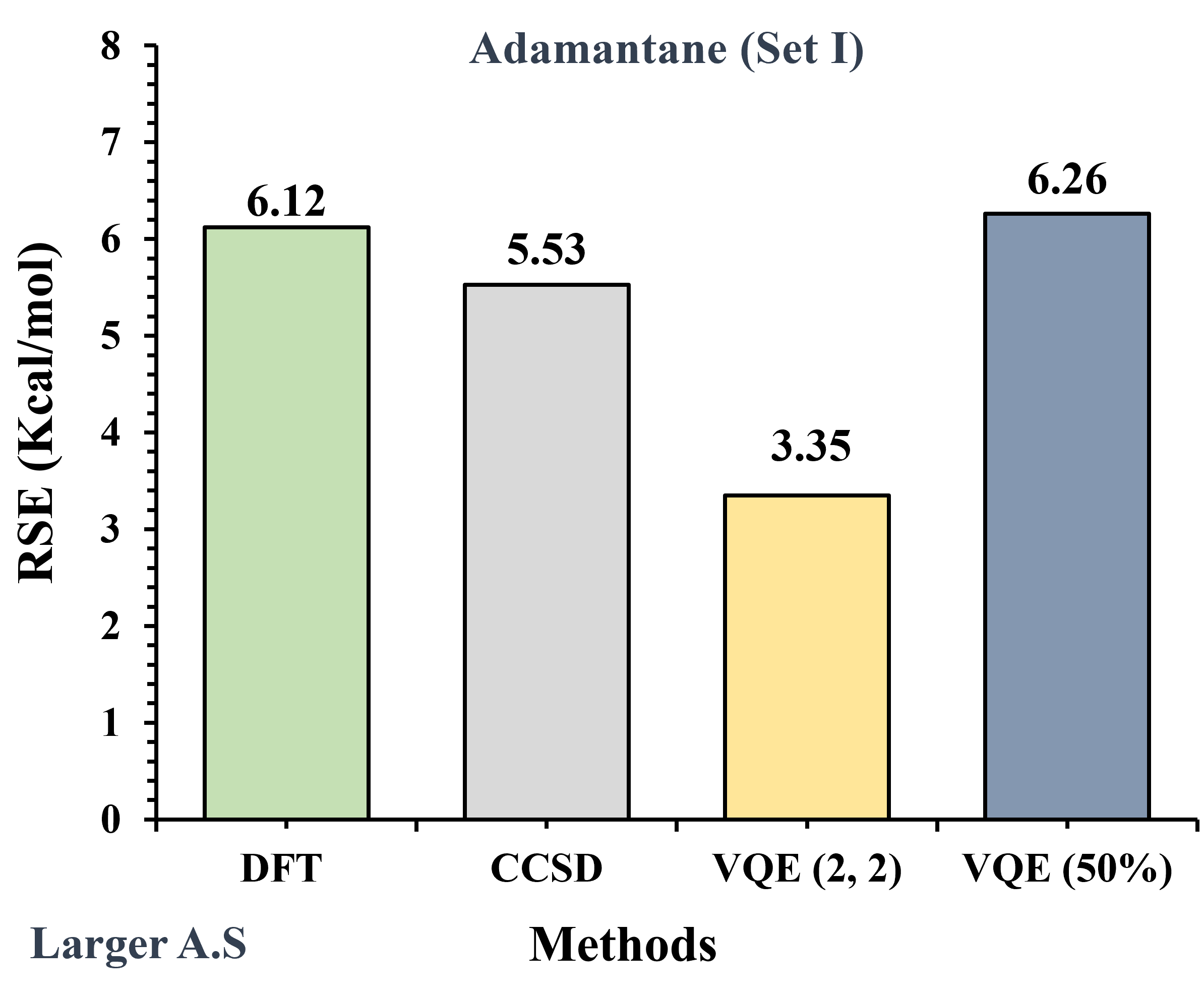}
        \caption{}
    \end{subfigure}
    \hspace{0.11\textwidth}
    \begin{subfigure}{0.38\textwidth}
        \centering
        \includegraphics[width=\linewidth]{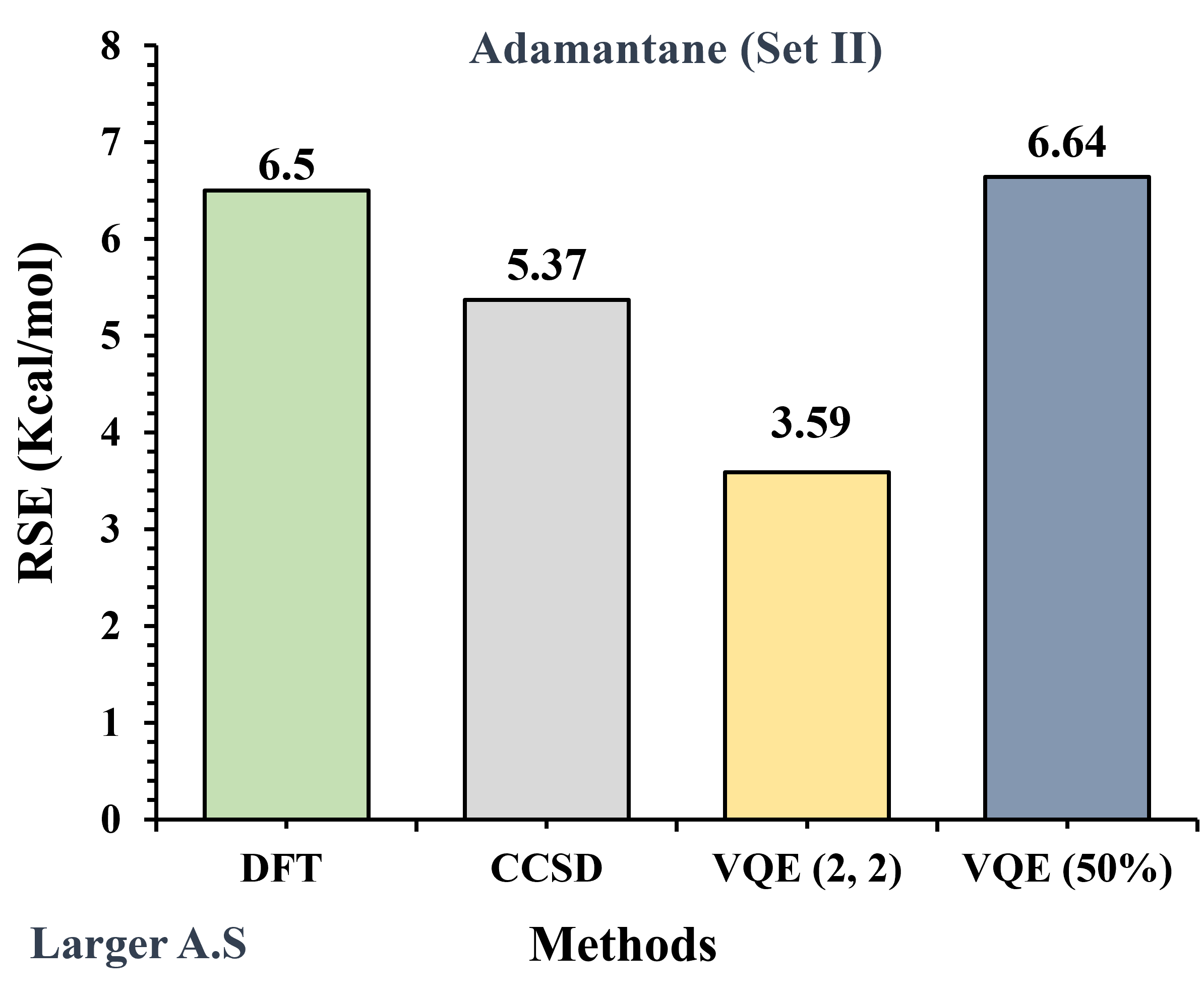}
        \caption{}
    \end{subfigure}
\caption{Bar plots for adamantane}
\end{figure*}

\subsubsection{Adamantane}

The most pronounced impact of the symmetry-guided active space selection strategy is observed for adamantane (Fig. 7). When a uniform (2, 2) active space is used, the RSE deviates by approximately 2.0–2.4 kcal/mol from DFT and by about 1.5 kcal/mol from CCSD. Upon implementing the symmetry-guided protocol with larger active spaces, the agreement improves substantially [Figs. 7(c,d)]. Unlike the smaller rings, a consistent SMF value of 33.33 \% could not be obtained for all reaction species involved in the homodesmotic schemes for adamantane. Consequently, the RSE was evaluated using a common SMF value of 50 \%. Despite the increased active space requirements, this symmetry-consistent approach yields excellent agreement with DFT and moderate agreement with CCSD, underscoring the applicability of the method to larger and more structurally complex systems.

Across the series from Cyclopropane to Cyclohexane, a systematic decrease in the reference RSE values is observed, reflecting the progressive release of angular strain as the ring size increases. Notably, the group-theory-inspired active space selection protocol within the VQE framework successfully reproduces this chemically expected trend using hypothetical reaction schemes. Error analyses for both homodesmotic sets (Fig. 8) further indicate that Set II reactions consistently yield closer agreement with both DFT and CCSD benchmarks. The consistently superior performance of the Set II homodesmotic reaction schemes can be attributed to their stricter balancing of local bonding patterns, hybridization states, and stereo electronic environments across reactants and products. This enhanced chemical balance promotes more effective cancellation of correlation energy contributions at the reaction level, which complements the symmetry-guided active space selection protocol employed in this work. Consequently, enforcing a common symmetry-matched fraction across reaction species is particularly effective for Set II schemes, leading to systematically improved accuracy, especially for larger and structurally more complex ring systems.


\begin{figure*}[htbp]
    \centering
    \begin{subfigure}{0.45\textwidth}
        \centering
        \includegraphics[width=\linewidth]{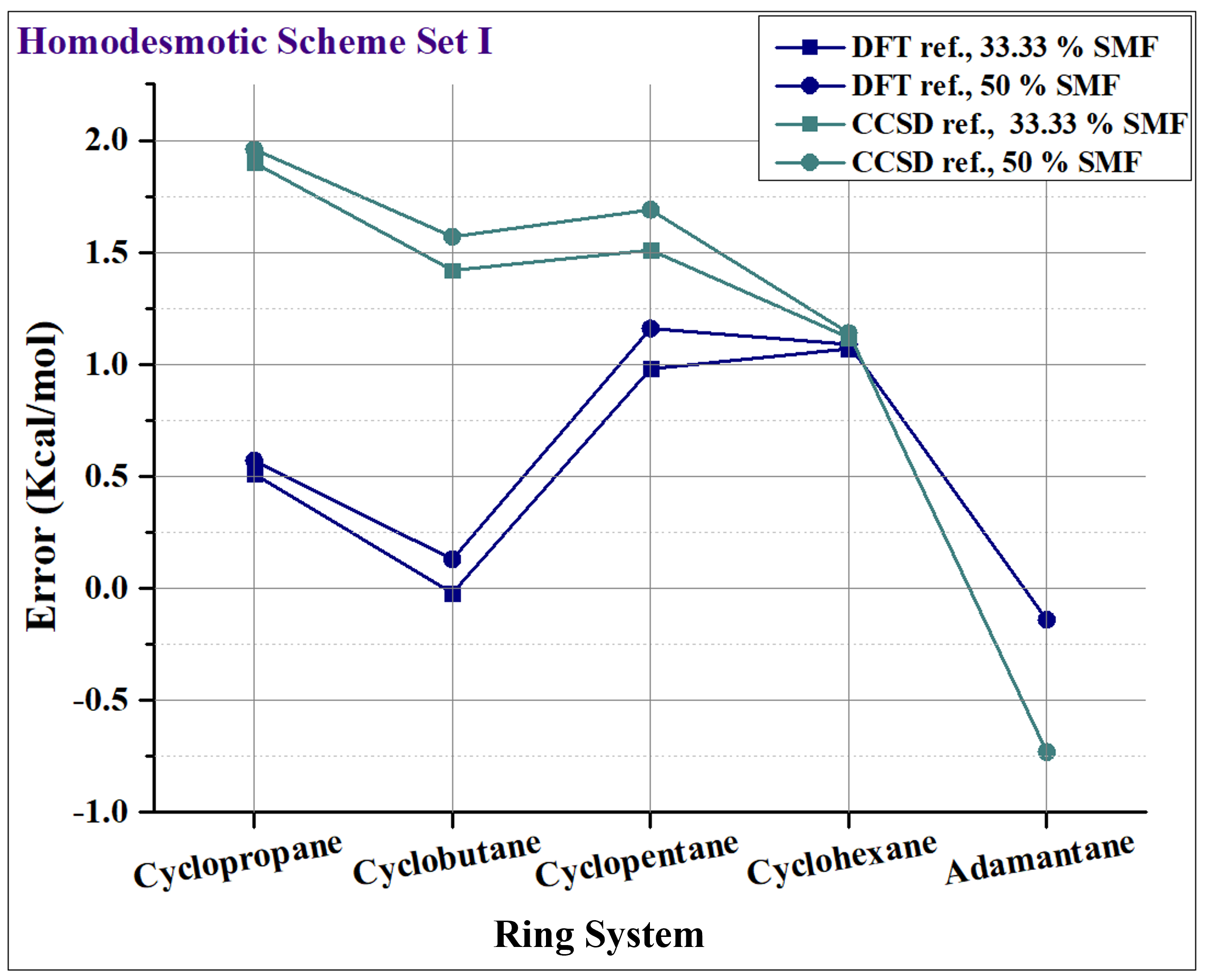}
        \caption{}
    \end{subfigure}
    \hspace{0.05\textwidth}
    \begin{subfigure}{0.45\textwidth}
        \centering
        \includegraphics[width=\linewidth]{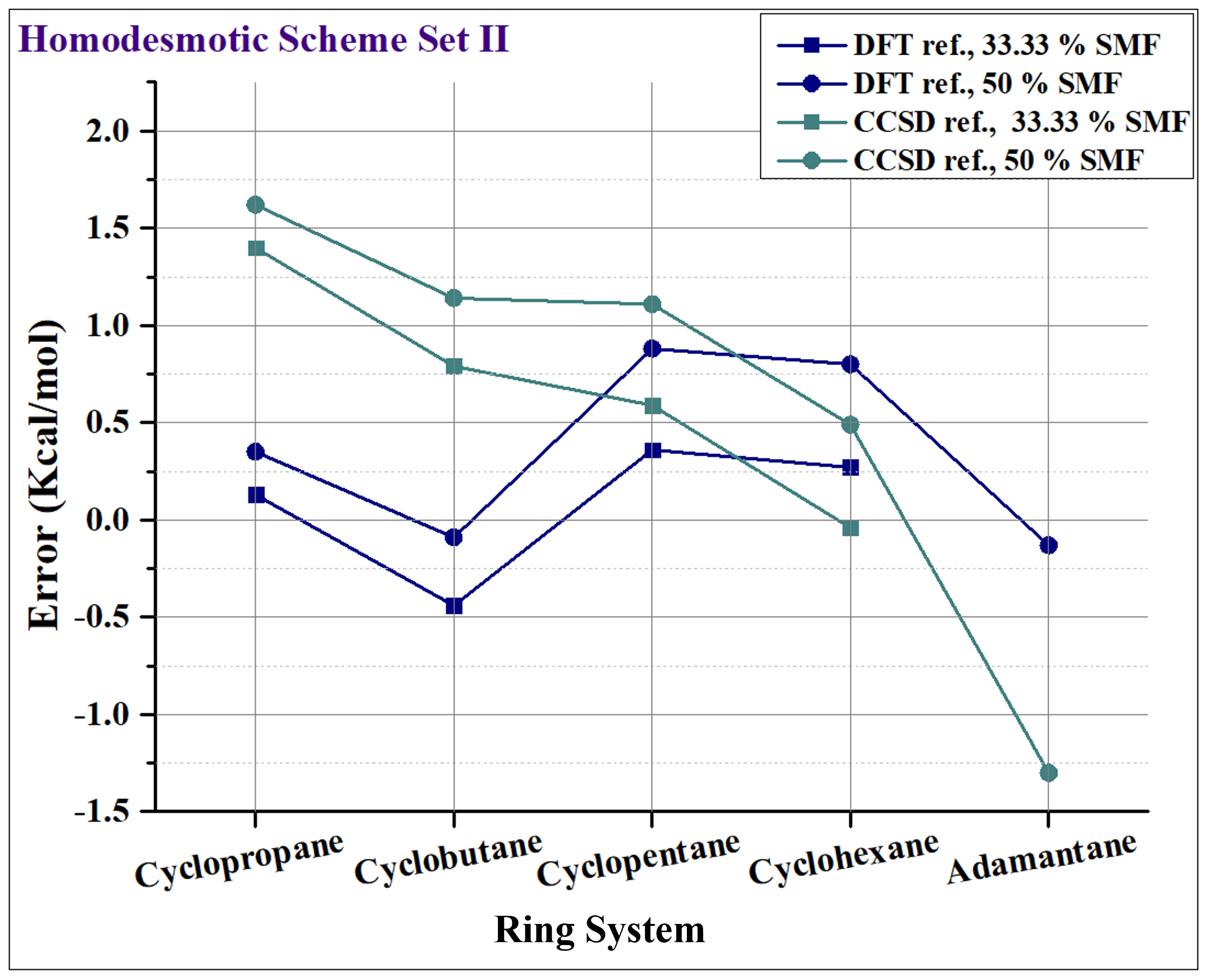}
        \caption{}
    \end{subfigure}

\caption{(a) and (b) Errors in ring strain energies (RSEs) obtained from VQE calculations for saturated cyclic hydrocarbons using Set I and Set II homodesmotic reaction schemes, respectively. Errors are reported relative to DFT and CCSD reference values for symmetry-matched fractions (SMF) of 33.33\% and 50\%. The horizontal axis denotes the ring system (cyclopropane through adamantane), while the vertical axis represents the deviation (kcal/mol) of the VQE results from the corresponding reference methods. The plots illustrate the dependence of VQE accuracy on both the homodesmotic reaction scheme and the choice of SMF-guided active space across different ring sizes.}
\end{figure*}

Overall, the symmetry-guided active space selection strategy produces RSE values within 0.2–1.0 kcal/mol of DFT and within approximately 1.8 kcal/mol of CCSD across all five systems, thereby approaching chemical accuracy. These results demonstrate that enforcing symmetry consistency across reactants and products is crucial for balanced correlation treatment in homodesmotic reaction-based VQE calculations. The systematic agreement of symmetry-guided VQE results with both density functional theory and coupled-cluster benchmarks across all five cyclic hydrocarbons indicates that enforcing symmetry consistency across reactants and products provides a chemically meaningful route to balanced
correlation effects in homodesmotic reaction schemes. For larger and structurally complex systems, such as adamantane, the transition from a uniform (2, 2) active space to symmetry matched active spaces improves the predicted ring strain energies by up to $\sim$2 kcal/mol, restoring chemical accuracy and underscoring the necessity of symmetry-consistent correlation treatment in homodesmotic VQE calculations. 


\begin{figure*}[htbp]
    \centering
    \begin{subfigure}{0.45\textwidth}
        \centering
        \includegraphics[width=\linewidth]{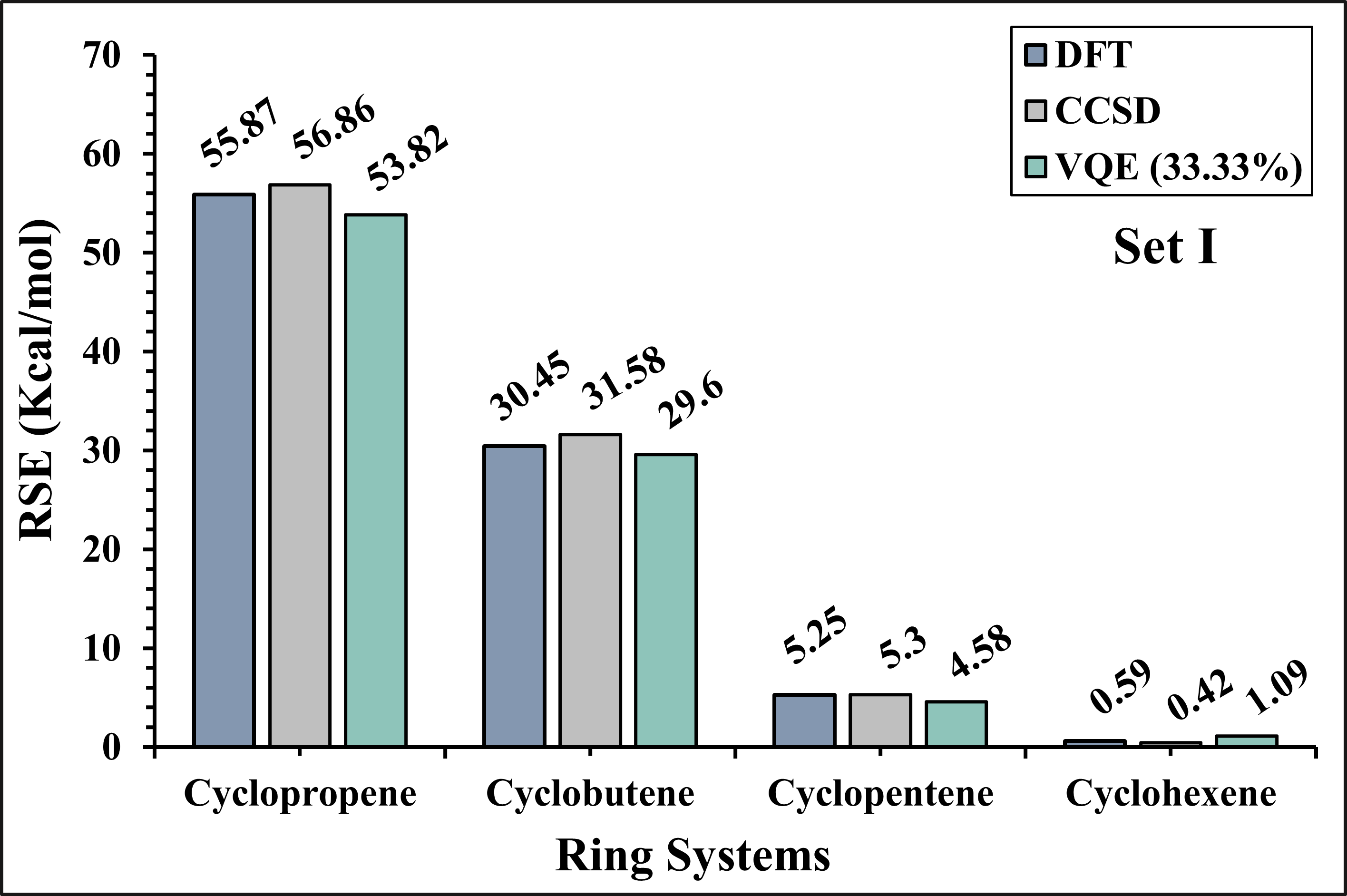}
        \caption{}
    \end{subfigure}
    \hspace{0.05\textwidth}
    \begin{subfigure}{0.45\textwidth}
        \centering
        \includegraphics[width=\linewidth]{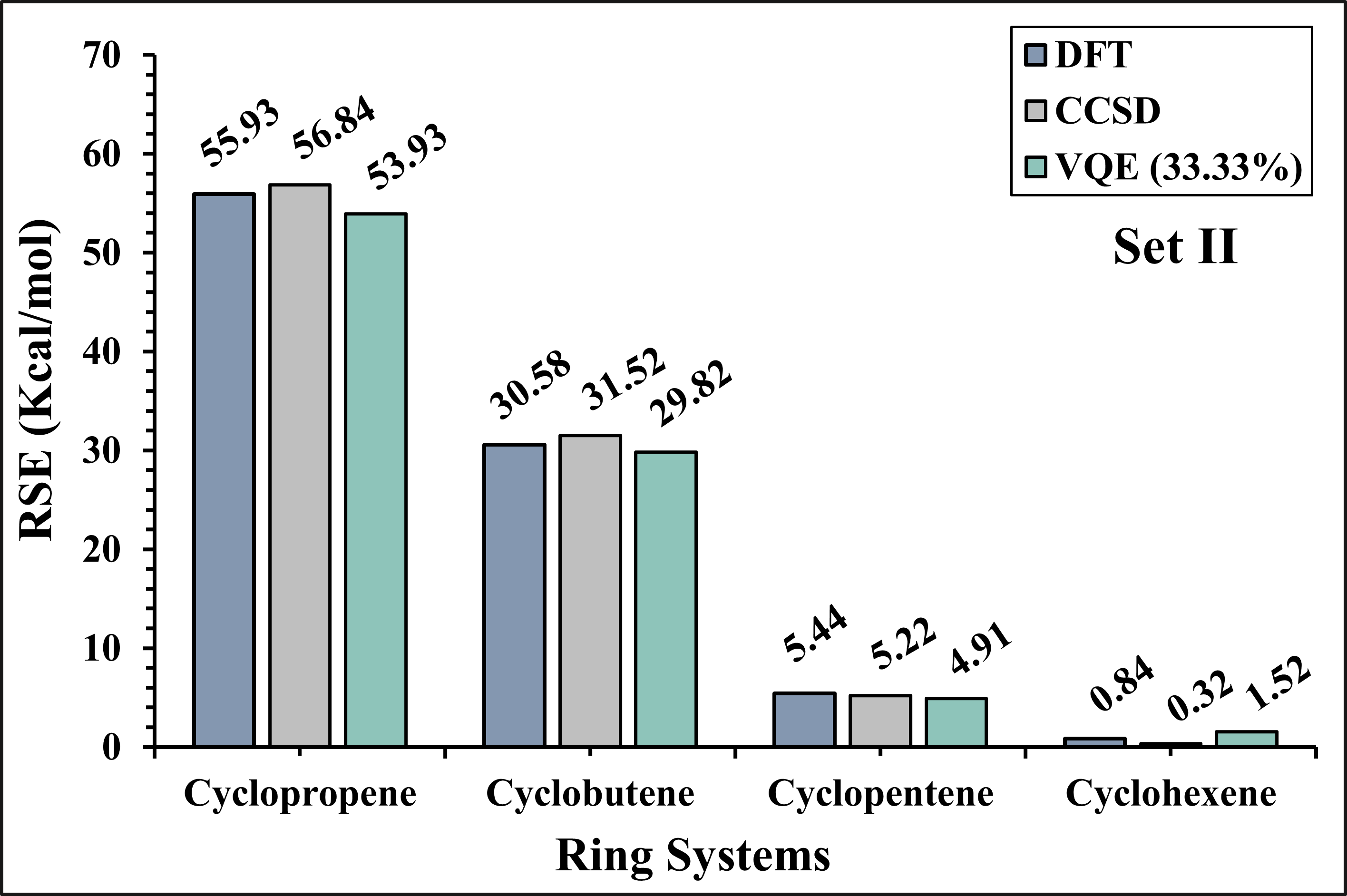}
        \caption{}
    \end{subfigure}

    \vspace{0.5cm} 

    \begin{subfigure}{0.45\textwidth}
        \centering
        \includegraphics[width=\linewidth]{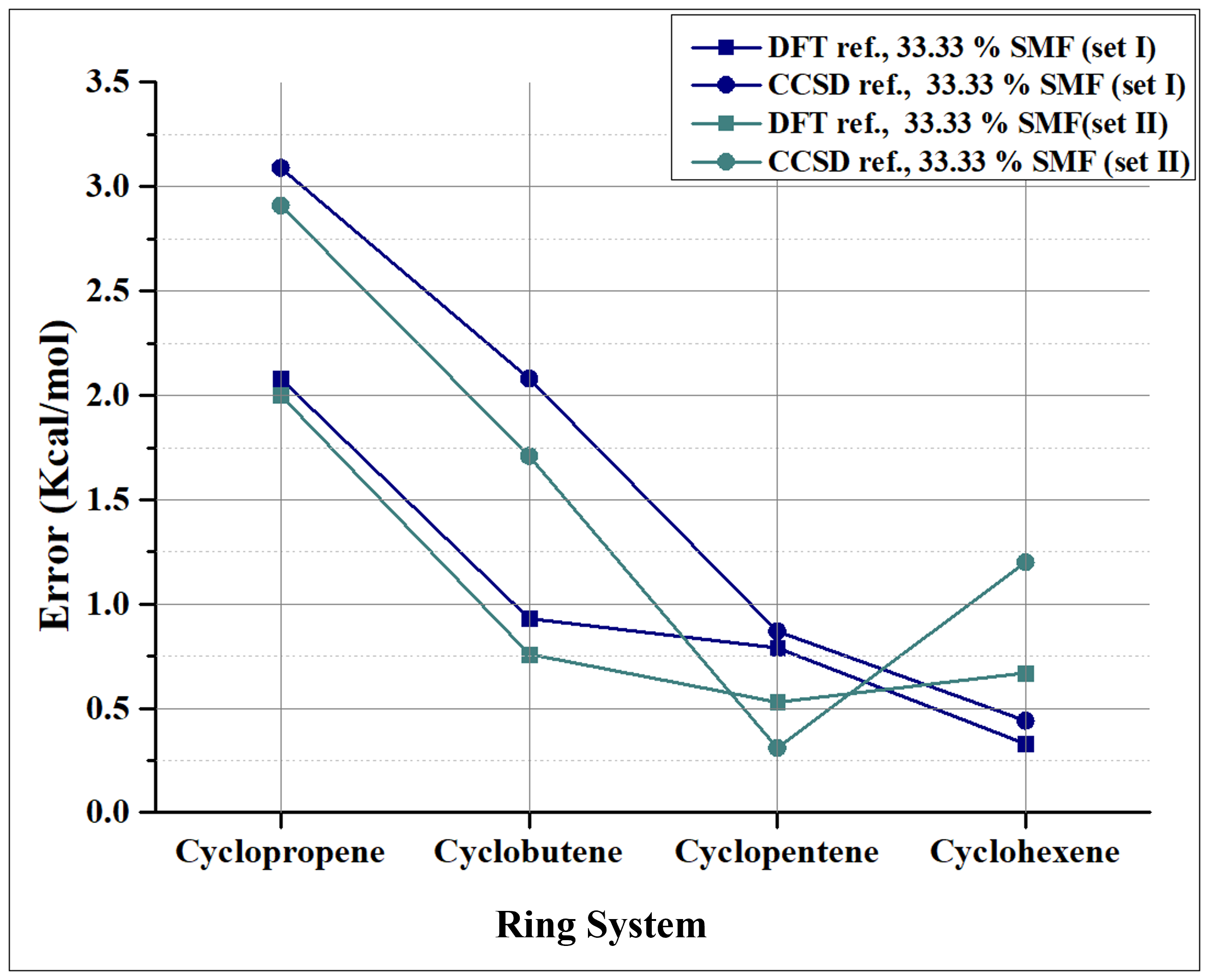}
        \caption{}
    \end{subfigure}
\caption{(a) and (b) Ring strain energies (RSEs) of unsaturated cyclic hydrocarbons (cyclopropene, cyclobutene, cyclopentene, and cyclohexene) computed using Set I and Set II homodesmotic reaction schemes, respectively. Results obtained from symmetry-guided VQE calculations employing an SMF of 33.33\% are compared against DFT and CCSD reference values. (c) Corresponding errors in the VQE-computed RSEs relative to DFT and CCSD references for both homodesmotic schemes. The plots illustrate the performance of the symmetry-guided VQE protocol for unsaturated ring systems and its dependence on the reaction scheme.}
\end{figure*}

\subsection{Unsaturated Cyclic Hydrocarbons}
To further assess the robustness of the symmetry-guided VQE framework, we extended the analysis of ring strain energies to unsaturated cyclic hydrocarbons, namely cyclopropene, cyclobutene, cyclopentene, and cyclohexene, using both Set~I and Set~II homodesmotic reaction schemes. Figure~9 summarizes the RSE values obtained from symmetry-guided VQE calculations employing a common SMF value of $33.33\%$ (restricted to the larger symmetry-consistent active spaces), together with a comparison to DFT and CCSD reference values. Overall, the protocol continues to yield chemically meaningful RSEs for unsaturated rings, with performance that correlates strongly with the degree of ring strain and electronic complexity. The qualitative trends in RSE are consistently reproduced across all systems, and quantitative agreement with both DFT and CCSD is maintained for most cases. \par
For the highly strained unsaturated system cyclopropene, the symmetry-guided VQE approach underestimates the RSE by approximately $2.0$~kcal~mol$^{-1}$ relative to DFT and by $\sim 3.0$~kcal~mol$^{-1}$ relative to CCSD across both homodesmotic reaction schemes. This larger deviation can be attributed to the pronounced ring strain and enhanced $\pi$--$\sigma$ coupling in cyclopropene, which introduce significant near-degeneracy and multireference character. These effects are not fully captured within the compact symmetry-consistent active spaces and the UCCSD ansatz employed here, leading to residual correlation contributions that remain partially unaccounted for. Notably, Set~II homodesmotic reactions offer a modest improvement over Set~I, reflecting more effective error cancellation. \par
For cyclobutene, the protocol yields improved agreement. Deviations relative to DFT remain within chemical accuracy, while comparison to CCSD shows moderate discrepancies, up to $2$~kcal~mol$^{-1}$, that are nevertheless reduced when employing Set~II reaction schemes. This trend mirrors the behavior observed for saturated four-membered rings and underscores the importance of the electronic environment encoded in the reaction design. In contrast, cyclopentene represents a particularly successful case for the present framework. For both Set~I and Set~II reactions, the symmetry-guided VQE results lie well within chemical accuracy relative to both DFT and CCSD. This system demonstrates that, once the extreme strain and strong $\pi$--$\sigma$ coupling characteristic of small unsaturated rings are alleviated, the symmetry-consistent active space selection strategy is capable of capturing the dominant correlation effects governing ring strain with high fidelity. \par
Cyclohexene exhibits similarly robust behavior. The deviations observed are small and slightly negative, indicating mild overestimation of RSE arising from favorable error cancellation rather than a systematic failure of the method. Importantly, these deviations remain within or very close to chemical accuracy for both homodesmotic reaction sets, confirming that the symmetry-guided VQE framework remains reliable for moderately strained unsaturated rings. \par
Overall, the extension of the symmetry-guided VQE framework to unsaturated cyclic hydrocarbons demonstrates its robustness even in the presence of $\pi$-electron correlation. Consistent with the observations for saturated ring systems, the Set~II homodesmotic reaction schemes provide systematically improved agreement with both DFT and CCSD benchmarks, reinforcing their superior error-cancellation characteristics within the VQE paradigm. Taken together, the results for both saturated and unsaturated rings highlight the potential of the symmetry-guided active space selection strategy as a broadly applicable and chemically consistent approach for evaluating virtual properties, such as ring strain energy, using hypothetical reaction schemes within quantum computational chemistry.

\subsection{Symmetry-Matched Fraction and the Role of Intrinsic Molecular Symmetry}

To further rationalize the symmetry-guided active space selection protocol employed in this work, we examined the origin and physical meaning of the symmetry-matched fraction (SMF) by evaluating its value in the full orbital space (6-31G*) for all reactants and products involved in the homodesmotic reactions. In all cases, the highest-order Abelian point group
symmetry intrinsic to each molecule was used. As defined earlier, the SMF corresponds to the fraction of excited configurations that transform according to the same irreducible representation as the Hartree–Fock reference state.

Fig. 9 illustrates the SMF values for representative molecules grouped according to their point-group order. A striking observation emerges: molecules belonging to point groups of the same order exhibit nearly identical SMF values, independent of molecular size or chemical composition. For instance, all species with point-group order 4 yield SMF values
clustered around $\sim 25.6 \%$ [Fig. 9(a)], whereas species with point-group order 2 consistently exhibit SMF values near $\sim 50 \%$ [Fig. 9(b)]. This numerical evidence indicates that SMF is fundamentally governed by molecular symmetry rather than by chemical identity or system size. This observation provides a firm theoretical basis for enforcing a common SMF value across reactants and products in homodesmotic reactions: doing so ensures a symmetry consistent
description of the excitation manifold and, consequently, a balanced treatment of
electron correlation within the VQE framework.


\begin{figure*}[htbp]
    \centering
    \begin{subfigure}{0.45\textwidth}
        \centering
        \includegraphics[width=\linewidth]{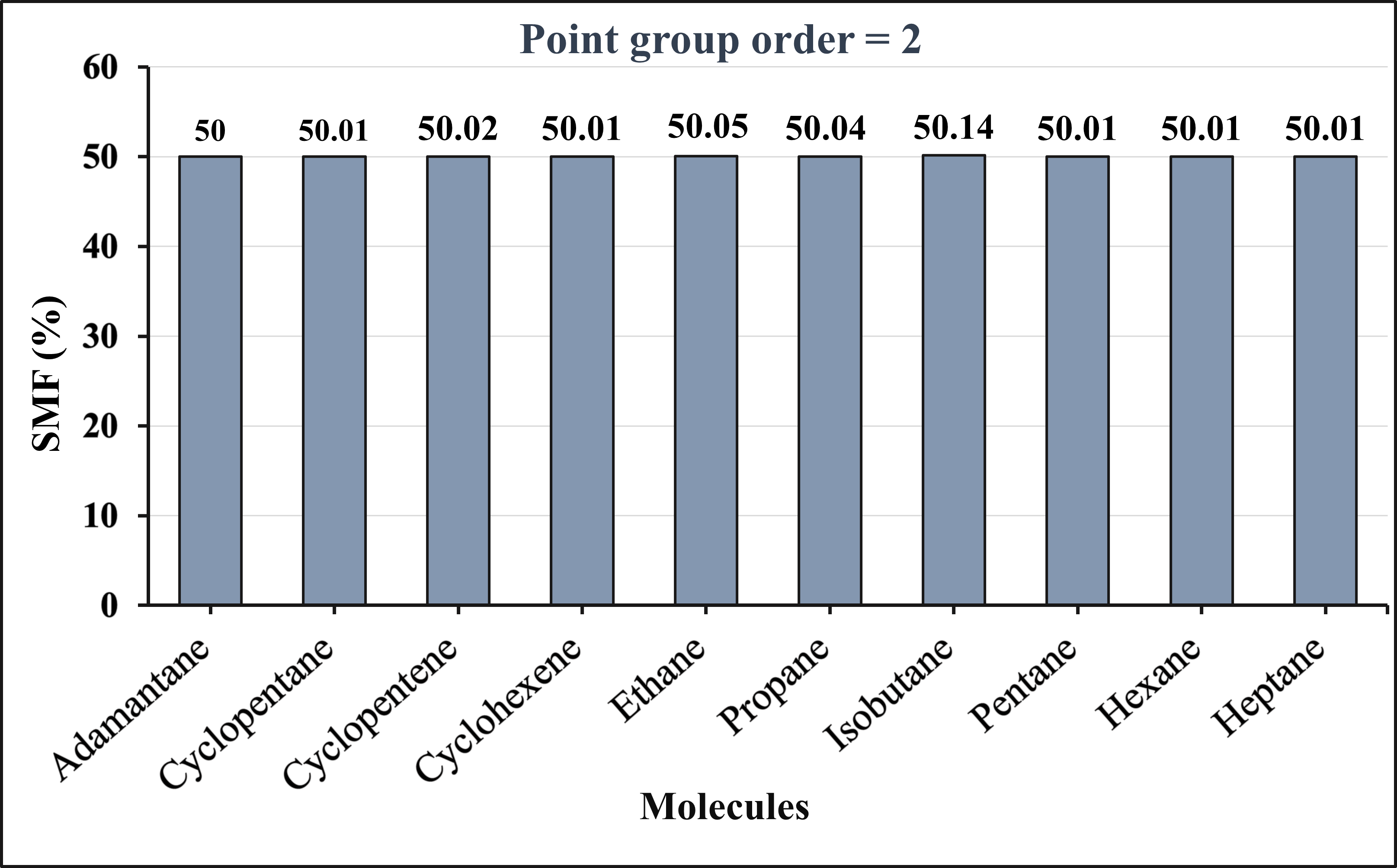}
        \caption{}
    \end{subfigure}
    \hspace{0.04\textwidth}
    \begin{subfigure}{0.45\textwidth}
        \centering
        \includegraphics[width=\linewidth]{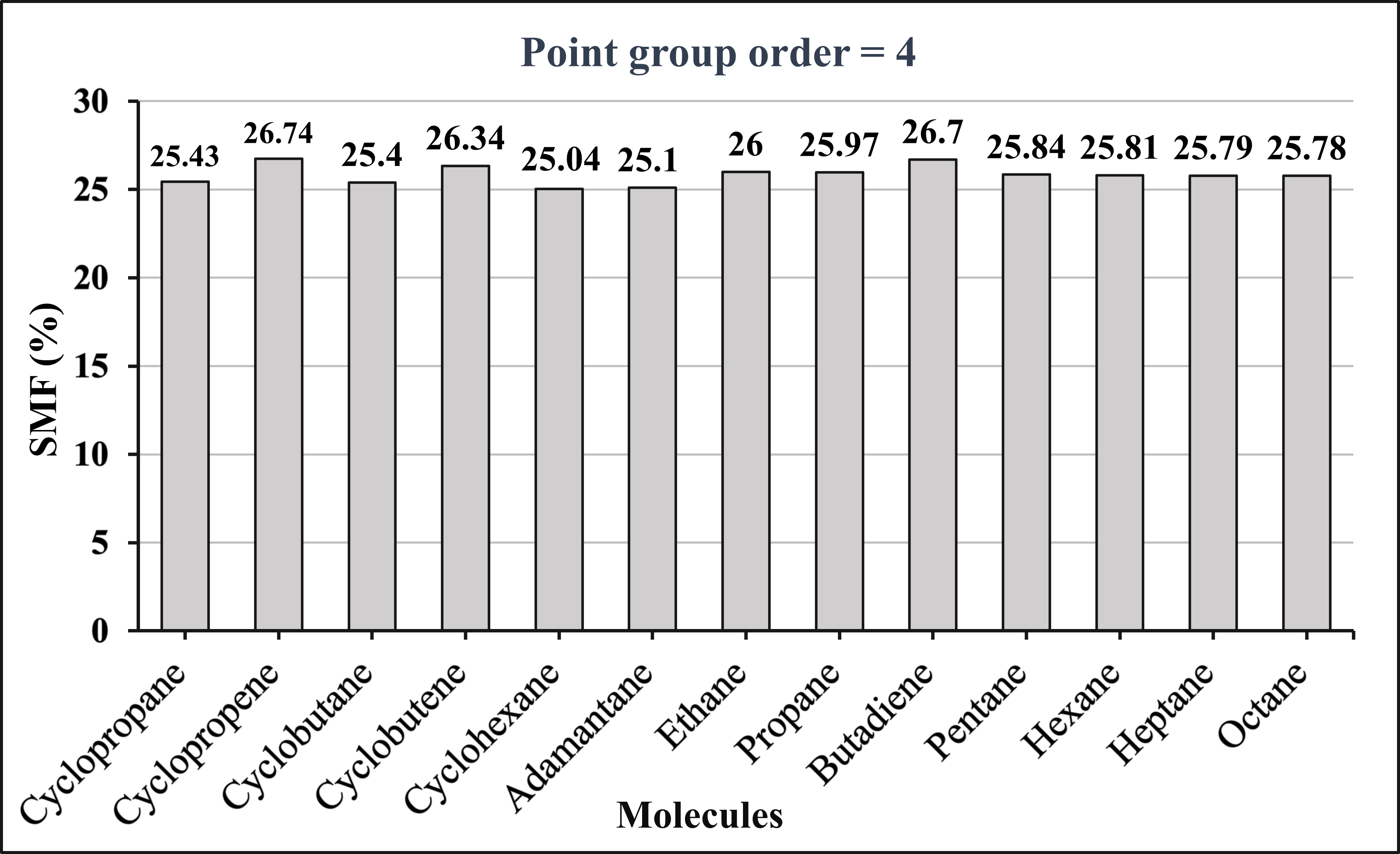}
        \caption{}
    \end{subfigure}
\caption{(a) Symmetry-matched fraction (SMF) values computed from the full orbital space (6-31G*) for representative reaction species belonging to point groups of order 2. (b) Corresponding SMF values for species belonging to point groups of order 4. In each panel, molecules sharing the same symmetry order exhibit very close SMF values with each other. This numerical clustering provides a group-theoretical basis for selecting a common SMF value across all reactants and products in a given homodesmotic reaction scheme}
\end{figure*}

However, an important caveat emerges when symmetry consistency is imposed incorrectly. While SMF correlates strongly with point-group order, artificially enforcing the same pointgroup symmetry across all reaction species, by reducing the intrinsic symmetry of certain molecules, leads to erroneous ring strain energies. This effect is clearly demonstrated for cyclopentane, cyclopentene, cyclohexene and adamantane. Cyclopentane and cyclopentene possess an envelope conformation with an intrinsic highest-order Abelian symmetry of $C_{s}$ (order 2), whereas cyclohexene possesses $C_{2}$ symmetry. Artificially lowering the symmetry of the corresponding reaction species (e.g., ethane, propane, hexane and cis-2-butene, which naturally possess point-group order 4) to enforce a common symmetry order results in severely degraded RSE values, despite achieving a nominally matching SMF (Table I). A similar failure is observed for adamantane, where forcing lower symmetry on highly symmetric species produces unphysical RSE values for both Set I and Set II reaction schemes (Table I).

\begin{table}[h]
\centering
\caption{RSEs obtained when the molecular symmetry of selected reaction species is artificially reduced to enforce a common SMF value across all reactants and products. The resulting RSE values show large deviations and non-physical trends, indicating that forced symmetry lowering disrupts the balanced cancellation of correlation effects in the homodesmotic reaction schemes.}
\renewcommand{\arraystretch}{1.3} 
\setlength{\tabcolsep}{14pt} 
\begin{tabular}{|c|c|c|}
\hline
\multirow{2}{*}{Ring Systems} & \multicolumn{2}{c|}{RSE (kcal/mol)} \\ \cline{2-3}
 & SET I & SET II \\ 
\hline
Cyclopentane & 2.693 & -6.670 \\ \hline
Adamantane & -11.156 & -30.478 \\ \hline
Cyclopentene & -2.500 & -26.600 \\ \hline
Cyclohexene & -3.030 & -35.170 \\ 
\hline
\end{tabular}
\end{table}

In contrast, when the intrinsic highest-order Abelian symmetry of each reaction species is preserved, the computed RSE values recover excellent agreement with both DFT and CCSD values. This behavior can be attributed to the alteration of the degeneracy structure of the electronic Hilbert space upon symmetry lowering, which redistributes molecular orbitals among irreducible representations and modifies the associated correlation contributions. These results establish that SMF consistency must be enforced within each molecule’s intrinsic highest-order symmetry, rather than by artificially forcing all reaction species into a common symmetry order. When this principle is respected, SMF emerges as a robust, symmetry-structured quantity that naturally supports balanced correlation treatment across reactants and products in homodesmotic VQE calculations.


\section{Conclusion}

In this work, we have demonstrated that homodesmotic reaction schemes can be meaningfully and systematically implemented within the Variational Quantum Eigensolver framework to evaluate ring strain energies of cyclic hydrocarbons. By enforcing symmetry-consistent treatment of electron correlation across reactants and products, chemically meaningful reaction energies can be obtained even within the constraints imposed by finite active spaces and NISQ-era quantum algorithms. This study establishes homodesmotic reactions as a viable and powerful construct for reaction-based quantum simulations, extending their long-standing utility in computational chemistry into the quantum computing paradigm. \par
The methodology was validated across a broad range of systems, including small strained rings, moderately sized cyclic hydrocarbons, structurally complex polycyclic systems such as adamantane, and unsaturated rings containing $\pi$ electrons. For all saturated and most unsaturated systems, the symmetry-consistent VQE results show close agreement with density functional theory and coupled-cluster benchmarks, approaching chemical accuracy. The systematic trends in ring strain energies, from highly strained small rings to larger, more flexible frameworks, are faithfully reproduced, confirming that reaction-level error cancellation can be effectively achieved within a quantum algorithmic setting. A consistent observation throughout this study is that Set~II homodesmotic reaction schemes provide improved agreement relative to Set~I schemes across all ring systems considered. This behavior reflects the superior matching of local electronic environments in Set~II reactions, leading to more effective cancellation of residual correlation errors when combined with symmetry-consistent VQE calculations. These findings highlight the importance of reaction design, in addition to active space considerations, for achieving reliable thermochemical predictions on quantum computing platforms. \par
Beyond numerical benchmarking, this work also clarifies the symmetry origin of reaction-consistent correlation treatment in homodesmotic VQE calculations. By analyzing the symmetry-matched fraction, we show that the structure of the excitation manifold is primarily governed by molecular point-group symmetry rather than molecular size or chemical identity. Importantly, we demonstrate that preserving each molecule’s intrinsic highest-order Abelian symmetry is essential; artificial symmetry reduction, even when nominal symmetry-based metrics appear matched, disrupts the degeneracy structure of the electronic Hilbert space and leads to unphysical reaction energetics. These observations establish clear guidelines for the correct application of symmetry-based consistency criteria in reaction-level quantum simulations. \par
For highly strained and electronically complex systems such as cyclopropene, larger deviations are observed, reflecting enhanced $\pi$–$\sigma$ coupling and near-degeneracy effects that are not fully captured within compact active spaces and single-reference ansätze such as UCCSD. These cases delineate the current limits of the approach and point toward natural future extensions involving larger active spaces or more flexible ansätze. Nevertheless, for cyclobutene, cyclopentene, and cyclohexene, the symmetry-consistent VQE framework remains robust, yielding results within or very close to chemical accuracy relative to both DFT and CCSD benchmarks. \par
Overall, this study establishes a general and chemically consistent framework for evaluating virtual thermodynamic properties using quantum algorithms, in which homodesmotic reaction design and symmetry-consistent correlation treatment play central roles. The approach is readily extendable to larger polycyclic systems, heteroatom-containing rings, and additional virtual properties such as aromatic stabilization energies, resonance energies, and hyperconjugative effects. As quantum hardware and algorithms continue to mature, the combination of reaction-based error cancellation and symmetry-aware quantum simulations demonstrated here provides a promising pathway toward scalable and reliable quantum computational chemistry.

\begin{data availability statement}
The data that support the findings of this study are available within the article and its Supporting Information. Additional data and computational details are available from the corresponding author upon reasonable request.

The following materials are available in the Supporting Information:

\begin{itemize}
  \item Optimized molecular geometries for all reactants and products.
  \item Total electronic energies of all reaction species for both Set~I and Set~II homodesmotic reactions.
  \item Highest-order Abelian point-group symmetries employed for each molecule.
  \item All symmetry-consistent active spaces yielding identical SMF values in the group-theoretical analysis used for VQE calculations.
  \item Molecular orbital details, including irreducible representations, for all reaction species from which the active spaces were constructed.
\end{itemize}
\end{data availability statement}

\begin{acknowledgement}    
LR, MS, MT, AK, and MP gratefully acknowledge the Indian Institute of Technology, Jodhpur, for providing the computational facilities essential for completing this work. AK also acknowledges SERB (Grant No. CRG/2022/005979) for funding the project. LR and MS are grateful to the Department of Chemistry, IIT Jodhpur, and the Ministry of Education (MoE) for offering research infrastructure and financial support.
\end{acknowledgement}

\bibliography{references.bib} 
\bibliographystyle{rsc}

\end{document}


\title{Reaction-Level Consistency within the Variational Quantum Eigensolver: Homodesmotic Ring Strain Energies of Cyclic Hydrocarbons}

\author{Lisa Roy}
\affiliation{Quantum Information and Computation Lab, Department of Chemistry,
 Indian Institute of Technology Jodhpur, Rajasthan, India, 342030 
}%
\author{Maitreyee Sarkar}
\affiliation{Quantum Information and Computation Lab, Department of Chemistry,
 Indian Institute of Technology Jodhpur, Rajasthan, India, 342030 
}%
\author{Mitali Tewari}
 \affiliation{Quantum Information and Computation Lab, Department of Chemistry,
 Indian Institute of Technology Jodhpur, Rajasthan, India, 342030 
}%
 \author{Atul Kumar}
 \email{atulk@iitj.ac.in}
\affiliation{Quantum Information and Computation Lab, Department of Chemistry,
 Indian Institute of Technology Jodhpur, Rajasthan, India, 342030 
}%
 \author{Manikandan Paranjothy}
  \email{pmanikandan@iitj.ac.in}

\affiliation{Chemical Dynamics Research Group, Department of Chemistry,
 Indian Institute of Technology Jodhpur, Rajasthan, India, 342030 
}%

\keywords{}
\maketitle

\section{Optimized Geometries and Reference Electronic Energies}
All molecular geometries used in this work were fully optimized at the DFT/B3LYP level using the 6-31G* basis set. The optimized Cartesian coordinates for all reactants and products involved in the homodesmotic reactions are provided in this section.\par

For each optimized structure, single-point electronic energies were evaluated in hartree ($E_{h}$) unit at both the DFT/B3LYP and CCSD levels of theory using the same basis set. These reference energies serve as classical benchmarks for assessing the accuracy of ring strain energy (RSE) calculated by the variational quantum eigensolver (VQE) discussed in the main text.\par
\vspace{10pt}
\begin{center}
\begin{minipage}{0.75\textwidth}
\textbf{(A) Cyclopropane — Optimized Cartesian Co-ordinates (Å)}\\
Electronic energy (DFT/B3LYP) = $-117.89508531974$\\
Electronic energy (CCSD) = $-117.483373008917$
\begin{center}
\begin{verbatim}
 C   0.61328153     0.61328153     0.00000000
 C   0.22447662    -0.83775815     0.00000000
 C  -0.83775815     0.22447662     0.00000000
 H   0.37654920    -1.40530075     0.91083200
 H   0.37654920    -1.40530075    -0.91083200
 H  -1.40530075     0.37654920     0.91083200
 H  -1.40530075     0.37654920    -0.91083200
 H   1.02875155     1.02875155     0.91083200
 H   1.02875155     1.02875155    -0.91083200
\end{verbatim}
\end{center}

\end{minipage}
\end{center}

\newpage
\begin{center}
\begin{minipage}{0.75\textwidth}   
\textbf{(B) Cyclobutane — Optimized Cartesian Co-ordinates (Å)}\\
Electronic energy = $-157.213172627347$\\
Electronic energy (CCSD) = $-156.664956209057$
\begin{center}
\begin{verbatim}
 C  0.76671518     0.76671518     0.12448850
 C -0.76671518     0.76671518    -0.12448850
 C  0.76671518    -0.76671518    -0.12448850
 C -0.76671518    -0.76671518     0.12448850
 H -1.00661158     1.00661158    -1.16634910
 H -1.38976428     1.38976428     0.52569958
 H  1.00661158     1.00661158     1.16634910
 H  1.38976428     1.38976428    -0.52569958
 H  1.38976428    -1.38976428     0.52569958
 H  1.00661158    -1.00661158    -1.16634910
 H -1.00661158    -1.00661158     1.16634910
 H -1.38976428    -1.38976428    -0.52569958
\end{verbatim}
\end{center}

\end{minipage}
\end{center}

\newpage

\begin{center}
\begin{minipage}{0.75\textwidth}   
\textbf{(C) Cyclopentane — Optimized Cartesian Co-ordinates (Å)}\\
Electronic energy = $-196.557079891621$\\
Electronic energy (CCSD) = $-195.873871792808$
\begin{center}
\begin{verbatim}
 C  0.76983700     1.05484568    -0.07450184
 C  1.21391738    -0.40278395     0.20307183
 C -0.79048409     1.03329947    -0.06500233
 C  0.01663170    -1.25795950    -0.24459495
 C -1.19121092    -0.43286682     0.23002802
 H  1.17953581     1.75022875     0.66600892
 H  1.13837716     1.38665468    -1.05220178
 H  2.14895116    -0.66682588    -0.30354808
 H  1.37913307    -0.54672156     1.27920834
 H -1.17847979     1.34255070    -1.04270397
 H -1.21159835     1.72617694     0.67143685
 H  0.03448512    -2.27494137     0.16338235
 H  0.00484676    -1.34363431    -1.34034728
 H -2.12989322    -0.72185640    -0.25604793
 H -1.32935785    -0.57744402     1.31002635

\end{verbatim}
\end{center}

\end{minipage}
\end{center}


\begin{center}
\begin{minipage}{0.75\textwidth}   
\textbf{(D) Cyclohexane — Optimized Cartesian Co-ordinates (Å)}\\
Electronic energy = $-235.880453315122$\\
Electronic energy (CCSD) = $-235.061633080739$

\begin{verbatim}
  C  0.61523334     1.33197468    -0.22949877
  C -0.84564515     1.19895475     0.22947710
  C -1.46075498    -0.13305130    -0.22947350
  C -0.61523334    -1.33197468     0.22949877
  C  1.46075498     0.13305130     0.22947350
  C  0.84564515    -1.19895475    -0.22947710
  H  1.04797788     2.26831200     0.14614644
  H  0.64383126     1.39389763    -1.32781249
  H  1.52790019     0.13929658     1.32782065
  H  2.48823924     0.22672620    -0.14547044
  H  1.44061178    -2.04169076     0.14585272
  H  0.88464361    -1.25424695    -1.32780921
  H -1.04797788    -2.26831200    -0.14614644
  H -0.64383126    -1.39389763     1.32781249
  H -1.52790019    -0.13929658    -1.32782065
  H -2.48823924    -0.22672620     0.14547044
  H -1.44061178     2.04169076    -0.14585272
  H -0.88464361     1.25424695     1.32780921

\end{verbatim}
\end{minipage}
\end{center}

\vspace{8pt}

\begin{center}
\begin{minipage}{0.75\textwidth}
\textbf{(E) Adamantane — Optimized Cartesian Co-ordinates (Å)}\\
Electronic energy (DFT/B3LYP) = $-390.72523957227$\\
Electronic energy (CCSD) = $-389.426347501492$
\begin{center}
\begin{verbatim}
 C -0.71126510     1.37127871    -0.06901305
 C -0.81622960     0.80502677     1.36329672
 C -0.10669337    -0.56441579     1.43534598
 C  1.37975878    -0.39083795     1.05654236
 C  0.77643287     1.54046118    -0.44430459
 C  1.48987732     0.17271245    -0.37644319
 C -0.77696742    -1.54024486     0.44449449
 C -0.67160945    -0.97966017    -0.99002314
 C  0.81622196    -0.80521873    -1.36309385
 C -1.37945816     0.39090998    -1.05682335
 H  0.90499240    -0.42489908    -2.39046117
 H  1.32733199    -1.77794400    -1.33793558
 H -1.14868951    -1.67524233    -1.69350231
 H -1.83198283    -1.68714616     0.71543801
 H -0.29325640    -2.52558258     0.50053049
 H -0.18270024    -0.96518191     2.45512738
 H  1.90046653    -1.35648113     1.12293518
 H  1.87302352     0.28691017     1.76742114
 H  2.54809481     0.29519363    -0.64378375
 H -1.87188409     0.69836420     1.65001640
 H -0.36052551     1.50374270     2.07895634
 H -1.21632332     2.34535846    -0.11821053
 H  1.25954088     2.25171402     0.24035883
 H  0.86461260     1.96144875    -1.45573227
 H -1.32812740     0.79232812    -2.07873021
 H -2.44498025     0.27734764    -0.81229823

\end{verbatim}
\end{center}

\end{minipage}
\end{center}

\begin{center}
\begin{minipage}{0.75\textwidth}
\textbf{(F) Cyclopropene — Optimized Cartesian Co-ordinates (Å)}\\
Electronic energy (DFT/B3LYP) = $-116.619040962101$
Electronic energy (CCSD) = $-116.232325446277$
\begin{center}
\begin{verbatim}
 C -0.00090744     0.84836392    -0.00397410
 C -0.64634752    -0.51589343    -0.01145606
 C  0.64850276    -0.51389066    -0.01234421
 H -0.00275299     1.45622500    -0.91415092
 H -0.00114435     1.44695339     0.91228961
 H -1.57957604    -1.05889613    -0.01431362
 H  1.58347928    -1.05394472    -0.01635300
 \end{verbatim}
\end{center}
\end{minipage}
\end{center}

\vspace{10pt}

\begin{center}
\begin{minipage}{0.75\textwidth}
\textbf{(G) Cyclobutene — Optimized Cartesian Co-ordinates (Å)}\\
Electronic energy (DFT/B3LYP) = $-155.973263820868$
Electronic energy (CCSD) = $-155.449919510845$
\begin{center}
\begin{verbatim}
 C  0.03103601    -0.78577658     0.69186727
 C -0.03103601     0.78577658     0.69186727
 C  0.02662062    -0.66963795    -0.82266026
 C -0.02662062     0.66963795    -0.82266026
 H  0.93899146    -1.21089142     1.13722933
 H -0.84041141    -1.28041114     1.13826605
 H -0.93899146     1.21089142     1.13722933
 H  0.84041141     1.28041114     1.13826605
 H  0.05576362    -1.41981156    -1.60904249
 H -0.05576362     1.41981156    -1.60904249
\end{verbatim}
\end{center}
\end{minipage}
\end{center}

\newpage

\begin{center}
\begin{minipage}{0.75\textwidth}
\textbf{(H) Cyclopentene — Optimized Cartesian Co-ordinates (Å)}\\
Electronic energy (DFT/B3LYP) = $-195.32713997673$
Electronic energy (CCSD) = $-194.6690972$
\begin{center}
\begin{verbatim}
 C  0.256922   -1.207693    0.000000
 C -0.066331   -0.326301    1.237080
 C -0.066331    1.074985    0.667680
 C -0.066331    1.074985   -0.667680
 C -0.066331   -0.326301   -1.237080
 H -0.292335   -2.154414    0.000000
 H  0.662109   -0.460199    2.047158
 H -1.050570   -0.568361    1.667101
 H -0.102623    1.963377    1.292951
 H -0.102623    1.963377   -1.292951
 H  0.662109   -0.460199   -2.047158
 H -1.050570   -0.568361   -1.667101
 H  1.324909   -1.453268    0.000000
\end{verbatim}
\end{center}
\end{minipage}
\end{center}

\begin{center}
\begin{minipage}{0.75\textwidth}
\textbf{(I) Cyclohexene — Optimized Cartesian Co-ordinates (Å)}\\
Electronic energy (DFT/B3LYP) = $-234.648286082659$
Electronic energy (CCSD) = $-233.854160769139$
\begin{center}
\begin{verbatim}
C  0.251500    1.481723    0.047814
C  0.251500   -0.725089   -1.192319
C -0.119231   -0.657807    1.306151
C -0.251500   -1.481723    0.047814
C  0.119231    0.657807    1.306151
C -0.251500    0.725089   -1.192319
H -0.298947    2.425906    0.163460
H  1.350514   -0.723173   -1.192019
H -1.305287   -1.773252   -0.089967
H  0.224347    1.183787    2.254696
H -1.350514    0.723173   -1.192019
H  1.305287    1.773252   -0.089967
H -0.063171   -1.244081   -2.106045
H -0.224347   -1.183787    2.254696
H  0.298947   -2.425906    0.163460
H  0.063171    1.244081   -2.106045

\end{verbatim}
\end{center}
\end{minipage}
\end{center}

\begin{center}
\begin{minipage}{0.75\textwidth}
\textbf{(J) Ethane — Optimized Cartesian Co-ordinates (Å)}\\
Electronic energy (DFT/B3LYP) = $-79.830421384469$\\
Electronic energy (CCSD) = $-79.5274972241149$
\begin{center}
\begin{verbatim}
 C  0.00000000    -0.00000000     0.7653692
 C -0.00000000     0.00000000    -0.76536922
 H -0.72194055     0.72194055     1.16428347
 H  0.98618913     0.26424858     1.16428347
 H -0.26424858    -0.98618913     1.16428347
 H  0.26424858     0.98618913    -1.16428347
 H -0.98618913    -0.26424858    -1.16428347
 H  0.72194055    -0.72194055    -1.16428347
\end{verbatim}
\end{center}
\end{minipage}
\end{center}

\begin{center}
\begin{minipage}{0.75\textwidth}
\textbf{(K) Propane — Optimized Cartesian Co-ordinates (Å)}\\
Electronic energy (DFT/B3LYP) = $-119.144241086761$\\
Electronic energy (CCSD) = $-118.704752170704$
\begin{center}
\begin{verbatim}
 C  1.27734033    -0.00000288    -0.25981551
 C  0.00000000     0.00000000     0.58627637
 C -1.27734033     0.00000288    -0.25981551
 H  2.17596682     0.00395947     0.36788031
 H  1.32071280     0.88257778    -0.91018466
 H  1.32448082    -0.88653804    -0.90449806
 H -0.00000123     0.87767499     1.24686642
 H  0.00000123    -0.87767499     1.24686642
 H -1.32448082     0.88653804    -0.90449806
 H -2.17596682    -0.00395947     0.36788031
 H -1.32071280    -0.88257778    -0.91018466
\end{verbatim}
\end{center}
\end{minipage}
\end{center}

\begin{center}
\begin{minipage}{0.75\textwidth}
\textbf{(L) Isobutane — Optimized Cartesian Co-ordinates (Å)}\\
Electronic energy (DFT/B3LYP) = $-158.458805941477$\\
Electronic energy (CCSD) = $-157.884090664928$
\begin{center}
\begin{verbatim}
 C  0.00000000     0.00000000    -0.37370289
 C  1.03360177     1.03360177     0.09588201
 C -1.41192628     0.37832451     0.09588201
 C  0.37832451    -1.41192628     0.09588201
 H  0.00000000     0.00000000    -1.47421720
 H -2.15893618    -0.33912285    -0.26446245
 H -1.70013215     1.37315710    -0.26446245
 H -1.46779755     0.39329517     1.19244439
 H -0.33912285    -2.15893618    -0.26446245
 H  0.39329517    -1.46779755     1.19244439
 H  1.37315710    -1.70013215    -0.26446245
 H  0.78577909     2.03925501    -0.26446245
 H  2.03925501     0.78577909    -0.26446245
 H  1.07450238     1.07450238     1.19244439

\end{verbatim}
\end{center}
\end{minipage}
\end{center}

\begin{center}
\begin{minipage}{0.75\textwidth}
\textbf{(M) cis-2-butene — Optimized Cartesian Co-ordinates (Å)}\\
Electronic energy (DFT/B3LYP) = $-157.224775200991$\\
Electronic energy (CCSD) = $-156.673143333168$
\begin{center}
\begin{verbatim}
 C  0.66918480     0.00000000     0.66293248
 C -0.66918480     0.00000000     0.66293248
 C  1.59352941     0.00000000    -0.52267396
 C -1.59352941     0.00000000    -0.52267396
 H  1.16927838     0.00000000     1.63230652
 H -1.16927838     0.00000000     1.63230652
 H  1.06145332     0.00000000    -1.47799606
 H  2.25075921     0.88015044    -0.50601485
 H  2.25075921    -0.88015044    -0.50601485
 H -1.06145332     0.00000000    -1.47799606
 H -2.25075921    -0.88015044    -0.50601485
 H -2.25075921     0.88015044    -0.50601485
\end{verbatim}
\end{center}
\end{minipage}
\end{center}

\begin{center}
\begin{minipage}{0.75\textwidth}
\textbf{(N) Pentane — Optimized Cartesian Co-ordinates (Å)}\\
Electronic energy (DFT/B3LYP) = $-197.771755469064$\\
Electronic energy (CCSD) = $-197.059336073665$
\begin{center}
\begin{verbatim}
C  2.560969    0.000756   -0.006845
C  1.284002   -0.001508    0.839631
C  0.000000    0.000000    0.000000
C -1.284002    0.001508    0.839631
C -2.560969   -0.000756   -0.006845
H  3.459125   -0.002172    0.621419
H  2.607897    0.887662   -0.650872
H  2.606712   -0.881463   -0.657375
H  1.282767    0.875094    1.503171
H  1.283253   -0.881102    1.499189
H -0.001008   -0.878599   -0.662708
H  0.001008    0.878599   -0.662708
H -1.283253    0.881102    1.499189
H -1.282767   -0.875094    1.503171
H -2.606712    0.881463   -0.657375
H -3.459125    0.002172    0.621419
H -2.607897   -0.887662   -0.650872
\end{verbatim}
\end{center}
\end{minipage}
\end{center}

\begin{center}
\begin{minipage}{0.75\textwidth}
\textbf{(O) Hexane — Optimized Cartesian Co-ordinates (Å)}\\
Electronic energy (DFT/B3LYP) = $-237.085472147287$\\
Electronic energy (CCSD) = $-236.236633296061$
\begin{center}
\begin{verbatim}
C -0.21984720    -3.22326527    -0.00271266
C  0.54325958    -1.89482970     0.00417255
C -0.37748224    -0.66746490    -0.00167965
C  0.37790439     0.66743759    -0.00066114
C -0.54290419     1.89475728     0.00484893
C  0.22008355     3.22325198    -0.00326701
H  0.46437201    -4.07952276     0.00383791
H -0.85463772    -3.31082228    -0.89318415
H -0.87111403    -3.31049071     0.87582716
H  1.21002887    -1.85046897    -0.86880227
H  1.19670620    -1.85194404     0.88721921
H -1.04232564    -0.70989570     0.87414783
H -1.03485454    -0.71265231    -0.88297366
H  1.03655614     0.71329523    -0.88097064
H  1.04148073     0.70923754     0.87615144
H -1.19592379     1.85237883     0.88823275
H -1.21008152     1.84977464    -0.86778332
H  0.87431613     3.30962020     0.87315828
H -0.46418620     4.07943831     0.00652058
H  0.85184469     3.31180172    -0.89578560

\end{verbatim}
\end{center}
\end{minipage}
\end{center}

\begin{center}
\begin{minipage}{0.75\textwidth}
\textbf{(P) Heptane — Optimized Cartesian Co-ordinates (Å)}\\
Electronic energy (DFT/B3LYP) = $-276.399196573858$\\
Electronic energy (CCSD) = $-275.413928260918$
\begin{center}
\begin{verbatim}
C  3.84465332    -0.00334305    -0.34996308
C  2.56771564     0.00640683     0.49652436
C  1.28380235    -0.00198923    -0.34322946
C  0.00000000     0.00000000     0.49649987
C -1.28380235     0.00198923    -0.34322946
C -2.56771564    -0.00640683     0.49652436
C -3.84465332     0.00334305    -0.34996308
H  4.74266868     0.00612396     0.27836355
H  3.89004844     0.87258775    -1.00898313
H  3.89213874    -0.89638272    -0.98542610
H  2.56753172     0.89154004     1.14862878
H  2.56672385    -0.86443548     1.16762756
H  1.28712408    -0.88457722    -1.00042310
H  1.28288330     0.87234742    -1.01138715
H  0.00128397     0.87849886     1.15902948
H -0.00128397    -0.87849886     1.15902948
H -1.28288330    -0.87234742    -1.01138715
H -1.28712408     0.88457722    -1.00042310
H -2.56672385     0.86443548     1.16762756
H -2.56753172    -0.89154004     1.14862878
H -3.89213874     0.89638272    -0.98542610
H -4.74266868    -0.00612396     0.27836355
H -3.89004844    -0.87258775    -1.00898313


\end{verbatim}
\end{center}
\end{minipage}
\end{center}

\begin{center}
\begin{minipage}{0.75\textwidth}
\textbf{(Q) Octane — Optimized Cartesian Co-ordinates (Å)}\\
Electronic energy = $-315.712913842655$\\
Electronic energy (CCSD) = $-314.591233860341$
\begin{center}
\begin{verbatim}
C -0.44226040    -0.62680211     0.00000000
C  0.44226040     0.62680211     0.00000000
C -0.35207103     1.93914195     0.00000000
C  0.53210701     3.19295171     0.00000000
C -0.26916236     4.49861838     0.00000000
C  0.35207103    -1.93914195     0.00000000
C -0.53210701    -3.19295171     0.00000000
C  0.26916236    -4.49861838     0.00000000
H -1.10436843    -0.60359659     0.87844223
H -1.10436843    -0.60359659    -0.87844223
H  1.10436843     0.60359659    -0.87844223
H  1.10436843     0.60359659     0.87844223
H -1.01427067     1.96359060     0.87848137
H -1.01427067     1.96359060    -0.87848137
H  1.19337279     3.16912274    -0.87797704
H  1.19337279     3.16912274     0.87797704
H  0.39012555     5.37423402     0.00000000
H -0.91445496     4.56784681     0.88452815
H -0.91445496     4.56784681    -0.88452815
H  1.01427067    -1.96359060    -0.87848137
H  1.01427067    -1.96359060     0.87848137
H -1.19337279    -3.16912274     0.87797704
H -1.19337279    -3.16912274    -0.87797704
H -0.39012555    -5.37423402     0.00000000
H  0.91445496    -4.56784681    -0.88452815
H  0.91445496    -4.56784681     0.88452815



\end{verbatim}
\end{center}
\end{minipage}
\end{center}

\section{Point-Group Symmetry}
To enable symmetry-consistent active space selection, point-group symmetries were assigned to all molecules using their optimized geometries. For each system, the highest-order abelian subgroup compatible with the molecular structure was employed.

This choice ensures compatibility with quantum chemistry packages that require abelian symmetry labels while retaining as much symmetry information as possible. The assigned point-group symmetries for all cyclic systems and reference molecules are summarized in Table I.

These symmetry labels were subsequently used to classify molecular orbitals into irreducible representations and to evaluate symmetry-matched fractions (SMF) for active space selection.
\begin{table}[H]
\centering
\captionsetup{justification=centering}
\caption{\textbf{Point-group symmetries}}
\renewcommand{\arraystretch}{1.0}
\setlength{\tabcolsep}{18pt}
\begin{tabular}{|c|c|c|c|}
\hline
\rule{0pt}{20pt} 
\textbf{Molecules} & \makecell{\textbf{Abelian}\\ \textbf{point-group sym.}} &
\textbf{Molecule} & \makecell{\textbf{Abelian}\\ \textbf{point-group sym.}} \\ \hline

Cyclopropane & $C_{2v}$ & Ethane & $C_{2h}$  \\ \hline
Cyclobutane& $C_{2v}$ & Propane & $C_{2v}$ \\ \hline
Cyclopentane& $C_{s}$ & Isobutane & $C_{s}$ \\ \hline
Cyclohexane& $C_{2h}$ & cis-2-butene & $C_{2v}$ \\ \hline
Adamantane& $C_{2v}$ & Pentane & $C_{2v}$ \\ \hline
Cyclopropene& $C_{2v}$ & Hexane & $C_{2h}$ \\ \hline
Cyclobutene& $C_{2v}$ & Heptane & $C_{2v}$ \\ \hline
Cyclopentene& $C_{s}$ & Octane & $C_{2h}$ \\ \hline
Cyclohexene& $C_{2}$ &  &  \\ \hline

\end{tabular}
\end{table}

\section{Symmetry-Matched Fraction (SMF) and Active Space Selection}
The symmetry-matched fraction (SMF) was employed as a quantitative criterion to guide the selection of symmetry-consistent active spaces across all reactants and products participating in a given homodesmotic reaction. For each molecule, multiple candidate active spaces were constructed by varying the number of orbitals and electrons.

Tables 2–18 list representative active spaces that yield identical SMF values (e.g., 33.33\% or 50\%), along with their corresponding VQE ground-state energies. In cases where multiple active spaces exhibited the same SMF, the largest active space (with respect to the number of correlated electrons) was selected for subsequent RSE calculations to improve correlation recovery.

This procedure ensures that all species within a given reaction are treated on an equal symmetry criteria, which is essential for obtaining reliable energy differences.

\begin{table}[H]
\centering
\captionsetup{justification=centering}
\caption{\textbf{Cyclopropane ($C_{2v}$)}}
\renewcommand{\arraystretch}{1.0}
\setlength{\tabcolsep}{18pt}

\begin{tabular}{|c|c|c|c|}
\hline

\multicolumn{2}{|c|}{\rule{0pt}{20pt}\textbf{33.33 \% SMF}} &
\multicolumn{2}{c|}{\rule{0pt}{20pt}\textbf{50 \% SMF}} \\ \hline

\rule{0pt}{20pt} 
\textbf{A.S} & \textbf{VQE G.S Energy ($E_{h}$)} &
\textbf{A.S} & \textbf{VQE G.S Energy ($E_{h}$)}\\ \hline

(2, 2) & -117.057644613782 & (4, 3) & -117.057715465681  \\ \hline
 (2, 4)& -117.057781896548 &  &  \\ \hline
 (2, 5)& -117.057859463175 &  &  \\ \hline
 (2, 7)& -117.061366098853 &  &  \\ \hline
 (6, 4)& -117.057791331871 &  &  \\ \hline

\end{tabular}
\end{table}

\begin{table}[H]
\centering
\captionsetup{justification=centering}
\caption{\textbf{Cyclobutane ($C_{2v}$)}}
\renewcommand{\arraystretch}{1.0}
\setlength{\tabcolsep}{18pt}
\begin{tabular}{|c|c|c|c|}
\hline

\multicolumn{2}{|c|}{\rule{0pt}{20pt}\textbf{33.33 \% SMF}} &
\multicolumn{2}{c|}{\rule{0pt}{20pt}\textbf{50 \% SMF}} \\ \hline

\rule{0pt}{20pt} 
\textbf{A.S} & \textbf{VQE G.S Energy ($E_{h}$)} &
\textbf{A.S} & \textbf{VQE G.S Energy ($E_{h}$)} \\ \hline

(2, 2) & -156.095684427061 & (2, 3) & -156.09577909515  \\ \hline
 (2, 4)& -156.095838267118 &  &  \\ \hline
 (2, 5)& -156.095956827522 &  &  \\ \hline
 (2, 7)& -156.09624483268 &  &  \\ \hline
 (2, 10)& -156.096952828659 &  &  \\ \hline
 (6, 4)& -156.095779741194 &  &  \\ \hline

\end{tabular}
\end{table}

\begin{table}[H]
\centering
\captionsetup{justification=centering}
\caption{\textbf{Cyclopentane ($C_{s}$)}}
\renewcommand{\arraystretch}{1.0}
\setlength{\tabcolsep}{18pt}
\begin{tabular}{|c|c|c|c|}
\hline

\multicolumn{2}{|c|}{\rule{0pt}{20pt}\textbf{33.33 \% SMF}} &
\multicolumn{2}{c|}{\rule{0pt}{20pt}\textbf{50 \% SMF}} \\ \hline

\rule{0pt}{20pt} 
\textbf{A.S} & \textbf{VQE G.S Energy ($E_{h}$)} &
\textbf{A.S} & \textbf{VQE G.S Energy ($E_{h}$)} \\ \hline

(2, 2) & -156.095684427061 & (2, 3) & -195.161893116982  \\ \hline
        &                  & (2, 5) & -195.16200511652 \\ \hline
        &                  & (2, 7) & -195.162076924164 \\ \hline
        &                  & (4, 3) & -195.161822939489 \\ \hline
        &                  & (4, 6) & -195.16228073235 \\ \hline
        &                  & (6, 7) & -195.162425879746 \\ \hline
        &                  & (8, 5) & -195.161898288886 \\ \hline
        &                  & (8, 8) & -195.162582414145 \\ \hline

\end{tabular}
\end{table}

\begin{table}[H]
\centering
\captionsetup{justification=centering}
\caption{\textbf{Cyclohexane ($C_{2h}$)}}
\renewcommand{\arraystretch}{1.0}
\setlength{\tabcolsep}{18pt}
\begin{tabular}{|c|c|c|c|}
\hline

\multicolumn{2}{|c|}{\rule{0pt}{20pt}\textbf{33.33 \% SMF}} &
\multicolumn{2}{c|}{\rule{0pt}{20pt}\textbf{50 \% SMF}} \\ \hline

\rule{0pt}{20pt} 
\textbf{A.S} & \textbf{VQE G.S Energy ($E_{h}$)} &
\textbf{A.S} & \textbf{VQE G.S Energy ($E_{h}$)} \\ \hline

(2, 2) & -156.095684427061 & (4, 3) & -234.205921993365  \\ \hline
(2, 4) & -234.205917467836 &        &                    \\ \hline

\end{tabular}
\end{table}

\begin{table}[H]
\centering
\captionsetup{justification=centering}
\caption{\textbf{Adamantanane ($C_{2v}$)}}
\renewcommand{\arraystretch}{1.0}
\setlength{\tabcolsep}{18pt}
\begin{tabular}{|c|c|c|c|}
\hline

\multicolumn{2}{|c|}{\rule{0pt}{20pt}\textbf{33.33 \% SMF}} &
\multicolumn{2}{c|}{\rule{0pt}{20pt}\textbf{50 \% SMF}} \\ \hline

\rule{0pt}{20pt} 
\textbf{A.S} & \textbf{VQE G.S Energy ($E_{h}$)} &
\textbf{A.S} & \textbf{VQE G.S Energy ($E_{h}$)} \\ \hline

(2, 10) & -388.023703902205 & (2, 3) & -388.023412934743  \\ \hline
(6, 4)  & -388.023405807906 & (4, 3) & -388.023383828651   \\ \hline
(8, 5)  & -388.023413517925 &        &                    \\ \hline
\end{tabular}
\end{table}

\begin{table}[H]
\centering
\captionsetup{justification=centering}
\caption{\textbf{Cyclopropene ($C_{2v}$)}}
\renewcommand{\arraystretch}{1.0}
\setlength{\tabcolsep}{18pt}
\begin{tabular}{|c|c|c|c|}
\hline

\multicolumn{2}{|c|}{\rule{0pt}{20pt}\textbf{33.33 \% SMF}} \\ \hline

\rule{0pt}{20pt} 
\textbf{A.S} & \textbf{VQE G.S Energy ($E_{h}$)} \\ \hline

(2, 2) & -115.8411464 \\ \hline
\end{tabular}
\end{table}

\vspace{1pt}

\begin{table}[H]
\centering
\captionsetup{justification=centering}
\caption{\textbf{Cyclobutene ($C_{2v}$)}}
\renewcommand{\arraystretch}{1.0}
\setlength{\tabcolsep}{18pt}
\begin{tabular}{|c|c|c|c|}
\hline

\multicolumn{2}{|c|}{\rule{0pt}{20pt}\textbf{33.33 \% SMF}} \\ \hline

\rule{0pt}{20pt} 
\textbf{A.S} & \textbf{VQE G.S Energy ($E_{h}$)} \\ \hline

(2, 2) & -154.9140994 \\ \hline
(2, 4) & -154.9141114 \\ \hline
(2, 5) & -154.9142168 \\ \hline
\end{tabular}
\end{table}

\begin{table}[H]
\centering
\captionsetup{justification=centering}
\caption{\textbf{Cyclopentene ($C_{s}$)}}
\renewcommand{\arraystretch}{1.0}
\setlength{\tabcolsep}{18pt}
\begin{tabular}{|c|c|c|c|}
\hline

\multicolumn{2}{|c|}{\rule{0pt}{20pt}\textbf{33.33 \% SMF}} \\ \hline

\rule{0pt}{20pt} 
\textbf{A.S} & \textbf{VQE G.S Energy ($E_{h}$)} \\ \hline

(2, 2) & -193.9885629 \\ \hline

\end{tabular}
\end{table}

\begin{table}[H]
\centering
\captionsetup{justification=centering}
\caption{\textbf{Cyclohexene ($C_{2}$)}}
\renewcommand{\arraystretch}{1.0}
\setlength{\tabcolsep}{18pt}
\begin{tabular}{|c|c|c|c|}
\hline

\multicolumn{2}{|c|}{\rule{0pt}{20pt}\textbf{33.33 \% SMF}} \\ \hline

\rule{0pt}{20pt} 
\textbf{A.S} & \textbf{VQE G.S Energy ($E_{h}$)} \\ \hline

(2, 2) & -233.0286049 \\ \hline

\end{tabular}
\end{table}

\begin{table}[H]
\centering
\captionsetup{justification=centering}
\caption{\textbf{Ethane ($C_{2h}$)}}
\renewcommand{\arraystretch}{1.0}
\setlength{\tabcolsep}{18pt}
\begin{tabular}{|c|c|c|c|}
\hline

\multicolumn{2}{|c|}{\rule{0pt}{20pt}\textbf{33.33 \% SMF}} &
\multicolumn{2}{|c|}{\rule{0pt}{20pt}\textbf{33.33 \% SMF}} \\ \hline

\rule{0pt}{20pt} 
\textbf{A.S} & \textbf{VQE G.S Energy ($E_{h}$)} &
\textbf{A.S} & \textbf{VQE G.S Energy ($E_{h}$)} \\ \hline

(2, 2) & -79.227890223841 & (4, 3) & -79.22797535186 \\ \hline

\end{tabular}
\end{table}

\begin{table}[H]
\centering
\captionsetup{justification=centering}
\caption{\textbf{Propane ($C_{2v}$)}}
\renewcommand{\arraystretch}{1.0}
\setlength{\tabcolsep}{18pt}
\begin{tabular}{|c|c|c|c|}
\hline

\multicolumn{2}{|c|}{\rule{0pt}{20pt}\textbf{33.33 \% SMF}} &
\multicolumn{2}{|c|}{\rule{0pt}{20pt}\textbf{33.33 \% SMF}} \\ \hline

\rule{0pt}{20pt} 
\textbf{A.S} & \textbf{VQE G.S Energy ($E_{h}$)} &
\textbf{A.S} & \textbf{VQE G.S Energy ($E_{h}$)} \\ \hline

(2, 2) & \makecell{(2, 2) \\ -118.262389476592} & (4, 3) & -118.262482027603 \\ \hline
(2, 5) & -118.262896018764 &         &                   \\ \hline
(6, 4) & -118.262535977864 &         &                   \\ \hline

\end{tabular}
\end{table}

\begin{table}[H]
\centering
\captionsetup{justification=centering}
\caption{\textbf{Isobutane ($C_{s}$)}}
\renewcommand{\arraystretch}{1.0}
\setlength{\tabcolsep}{18pt}
\begin{tabular}{|c|c|c|c|}
\hline

\multicolumn{2}{|c|}{\rule{0pt}{20pt}\textbf{50 \% SMF}} \\ \hline

\rule{0pt}{20pt} 
\textbf{A.S} & \textbf{VQE G.S Energy ($E_{h}$)} \\ \hline

(4, 3) & -157.297404662886 \\ \hline
(4, 6) & -157.298164156323 \\ \hline
(8, 5) & -157.297469107271 \\ \hline
(8, 8) & -157.298696058302 \\ \hline

\end{tabular}
\end{table}

\begin{table}[H]
\centering
\captionsetup{justification=centering}
\caption{\textbf{cis-2-butene ($C_{2v}$)}}
\renewcommand{\arraystretch}{1.0}
\setlength{\tabcolsep}{18pt}
\begin{tabular}{|c|c|c|c|}
\hline

\multicolumn{2}{|c|}{\rule{0pt}{20pt}\textbf{50 \% SMF}} \\ \hline

\rule{0pt}{20pt} 
\textbf{A.S} & \textbf{VQE G.S Energy ($E_{h}$)} \\ \hline

(2, 2) & -156.1203142 \\ \hline
(2, 4) & -156.1203393 \\ \hline

\end{tabular}
\end{table}

\begin{table}[H]
\centering
\captionsetup{justification=centering}
\caption{\textbf{Pentane ($C_{2v}$)}}
\renewcommand{\arraystretch}{1.0}
\setlength{\tabcolsep}{18pt}
\begin{tabular}{|c|c|c|c|}
\hline

\multicolumn{2}{|c|}{\rule{0pt}{20pt}\textbf{33.33 \% SMF}} &
\multicolumn{2}{|c|}{\rule{0pt}{20pt}\textbf{33.33 \% SMF}} \\ \hline

\rule{0pt}{20pt} 
\textbf{A.S} & \textbf{VQE G.S Energy ($E_{h}$)} &
\textbf{A.S} & \textbf{VQE G.S Energy ($E_{h}$)} \\ \hline

(2, 2) & -196.3309946 & (2, 3) & -196.331026201365 \\ \hline
(2, 4) & -196.331035013603 &         &                   \\ \hline
(2, 5) & -196.331046293446 &         &                   \\ \hline
(6, 4) & -196.331113073368 &         &                   \\ \hline

\end{tabular}
\end{table}

\begin{table}[H]
\centering
\captionsetup{justification=centering}
\caption{\textbf{Hexane ($C_{2h}$)}}
\renewcommand{\arraystretch}{1.0}
\setlength{\tabcolsep}{18pt}
\begin{tabular}{|c|c|c|c|}
\hline

\multicolumn{2}{|c|}{\rule{0pt}{20pt}\textbf{33.33 \% SMF}} &
\multicolumn{2}{|c|}{\rule{0pt}{20pt}\textbf{33.33 \% SMF}} \\ \hline

\rule{0pt}{20pt} 
\textbf{A.S} & \textbf{VQE G.S Energy ($E_{h}$)} &
\textbf{A.S} & \textbf{VQE G.S Energy ($E_{h}$)} \\ \hline

(2, 4) & -235.365428909493 & (2, 3) & -235.365430391854 \\ \hline
(2, 5) & -235.365439361974 &         &                   \\ \hline
(2, 8) & -235.365505151192 &         &                   \\ \hline
(2, 10) & -235.365586625397 &         &                   \\ \hline

\end{tabular}
\end{table}

\begin{table}[H]
\centering
\captionsetup{justification=centering}
\caption{\textbf{Heptane ($C_{2v}$)}}
\renewcommand{\arraystretch}{1.0}
\setlength{\tabcolsep}{18pt}
\begin{tabular}{|c|c|c|c|}
\hline

\multicolumn{2}{|c|}{\rule{0pt}{20pt}\textbf{33.33 \% SMF}} &
\multicolumn{2}{|c|}{\rule{0pt}{20pt}\textbf{33.33 \% SMF}} \\ \hline

\rule{0pt}{20pt} 
\textbf{A.S} & \textbf{VQE G.S Energy ($E_{h}$)} &
\textbf{A.S} & \textbf{VQE G.S Energy ($E_{h}$)} \\ \hline

(2, 2) & -274.399726792273 & (2, 3) & -274.399739263643 \\ \hline
(2, 4) & -274.399754833975 &         &                   \\ \hline
(2, 5) & -274.39976040572 &         &                   \\ \hline
(2, 10) & -274.399845859321 &         &                   \\ \hline
(6, 4) & -274.399807258758 &         &                   \\ \hline

\end{tabular}
\end{table}

\begin{table}[H]
\centering
\captionsetup{justification=centering}
\caption{\textbf{Octane ($C_{2h}$)}}
\renewcommand{\arraystretch}{1.0}
\setlength{\tabcolsep}{18pt}
\begin{tabular}{|c|c|c|c|}
\hline

\multicolumn{2}{|c|}{\rule{0pt}{20pt}\textbf{33.33 \% SMF}} &
\multicolumn{2}{|c|}{\rule{0pt}{20pt}\textbf{33.33 \% SMF}} \\ \hline

\rule{0pt}{20pt} 
\textbf{A.S} & \textbf{VQE G.S Energy ($E_{h}$)} &
\textbf{A.S} & \textbf{VQE G.S Energy ($E_{h}$)} \\ \hline

(2, 10) & -313.434147043079 & (2, 3) & -313.434073118054 \\ \hline
(6, 4) & -313.434132682269 &  (4, 3) & -313.434115199662  \\ \hline

\end{tabular}
\end{table}
\section{Ring Strain Energy Calculations}
Ring strain energies (RSEs) were computed using homodesmotic reaction schemes following two distinct sets of reactions (Set I and Set II), as detailed in the main text. This section reports RSE values obtained using DFT/B3LYP, CCSD, VQE with a uniform minimal active space of (2,2), and VQE with symmetry-guided active spaces selected based on a common SMF criterion.
Tables 19–36 summarize the computed RSEs for saturated and unsaturated ring systems.
\vspace{10pt}
\begin{figure*}[htbp]
\centering
\includegraphics[width=0.6\linewidth]{sat_rings/cyclopropane_set_I.png}
\end{figure*}

\vspace{0.5pt}
\begin{table}[H]
\centering
\caption{\textbf{Cyclopropane (Set I)}}
\renewcommand{\arraystretch}{1.0}
\setlength{\tabcolsep}{4.5pt}
\begin{tabular}{|c|c|c|c|c|c|}
\hline

\rule{0pt}{20pt} 
\textbf{Methods} & \textbf{Cyclopropane} &
\textbf{Ethane} & \textbf{Pentane} & \makecell{\textbf{RSE }\\ \textbf{($E_{h}$)}} & \makecell{\textbf{RSE }\\ \textbf{(kcal/mol)}} \\ \hline

 \makecell{DFT/ \\ B3LYP} & -117.8950853 & -79.83042138 & -197.7717555 & 0.04624882 & 29.0215 \\ \hline

CCSD & -117.483373 & -79.52749722 & -197.0593361 & 0.048465841 & 30.4127 \\ \hline

\makecell{VQE \\ (2, 2)} & \makecell{(2, 2) \\ -117.0576446} & \makecell{(2, 2) \\ -79.22789022} & \makecell{(2, 2) \\ -196.3309946} & 0.045459762 & 28.5264 \\ \hline

\makecell{VQE/ \\ 33.33\% SMF} & \makecell{(6, 4) \\ -117.0577913} & \makecell{(2, 2) \\ -79.22789022} & \makecell{(6, 4) \\ -196.3311131} & 0.045431518 & 28.5086 \\ \hline

\end{tabular}
\end{table}


\newpage
\begin{figure*}[htbp]
\centering
\includegraphics[width=0.6\linewidth]{sat_rings/cyclopropane_set_II.png}
\end{figure*}

\vspace{0.5pt}
\begin{table}[H]
\centering
\caption{\textbf{Cyclopropane (Set II)}}
\renewcommand{\arraystretch}{1.0}
\setlength{\tabcolsep}{4.5pt}
\begin{tabular}{|c|c|c|c|c|c|}
\hline

\rule{0pt}{20pt} 
\textbf{Methods} & \textbf{Cyclopropane} &
\textbf{Ethane} & \textbf{Propane} & \makecell{\textbf{RSE }\\ \textbf{($E_{h}$)}} &  \makecell{\textbf{RSE }\\ \textbf{(kcal/mol)}} \\ \hline

 \makecell{DFT/ \\ B3LYP} & -117.8950853 & -79.83042138 & -119.1442411 & 0.046373842 & 29.1 \\ \hline

CCSD & -117.483373 & -79.52749722 & -118.7047522 & 0.048391831 & 30.3663 \\ \hline

\makecell{VQE \\ (2, 2)} & \makecell{(2, 2) \\ -117.0576446} & \makecell{(2, 2) \\ -79.22789022} & \makecell{(2, 2) \\ -118.2623895
} & 0.045853144 & 28.7732 \\ \hline

\makecell{VQE/ \\ 33.33\% SMF} & \makecell{(6, 4) \\ -117.0577913} & \makecell{(2, 2) \\ -79.22789022} & \makecell{(6, 4) \\ -118.262536
} & 0.04614593 & 28.9569 \\ \hline

\end{tabular}
\end{table}

\newpage
\begin{figure*}[htbp]
\centering
\includegraphics[width=0.6\linewidth]{sat_rings/cyclobutane_set_I.png}
\end{figure*}

\vspace{0.5pt}
\begin{table}[H]
\centering
\caption{\textbf{Cyclobutane (Set I)}}
\renewcommand{\arraystretch}{1.0}
\setlength{\tabcolsep}{4.5pt}
\begin{tabular}{|c|c|c|c|c|c|}
\hline

\rule{0pt}{20pt} 
\textbf{Methods} & \textbf{Cyclobutane} &
\textbf{Ethane} & \textbf{Hexane} & \makecell{\textbf{RSE }\\ \textbf{($E_{h}$)}} & \makecell{\textbf{RSE }\\ \textbf{(kcal/mol)}} \\ \hline

 \makecell{DFT/ \\ B3LYP} & -157.2131726 & -79.83042138 & -237.0854721 & 0.041878135 & 26.2789\\ \hline

CCSD & -156.6649562 & -79.52749722 & -236.2366333 & 0.044179863 & 27.7232\\ \hline

\makecell{VQE \\ (2, 2)} & \makecell{(2, 2) \\ -156.0956844} & \makecell{(2, 2) \\ -79.22789022} & \makecell{(2, 2) \\ -235.3654289} & 0.041854259 & 26.2639\\ \hline

\makecell{VQE/ \\ 33.33\% SMF} & \makecell{(6, 4) \\ -156.0957797} & \makecell{(2, 2) \\ -79.22789022} & \makecell{(6, 4) \\ -235.3655866
} & 0.04191666 & 26.3030\\ \hline

\end{tabular}
\end{table}


\begin{figure*}[htbp]
\centering
\includegraphics[width=0.6\linewidth]{sat_rings/cyclobutane_set_II.png}
\end{figure*}

\vspace{0.5pt}
\begin{table}[H]
\centering
\caption{\textbf{Cyclobutane (Set II)}}
\renewcommand{\arraystretch}{1.0}
\setlength{\tabcolsep}{4.5pt}
\begin{tabular}{|c|c|c|c|c|c|}
\hline

\rule{0pt}{20pt} 
\textbf{Methods} & \textbf{Cyclobutane} &
\textbf{Ethane} & \textbf{Propane} & \makecell{\textbf{RSE }\\ \textbf{($E_{h}$)}} & \makecell{\textbf{RSE }\\ \textbf{(kcal/mol)}}\\ \hline

 \makecell{DFT/ \\ B3LYP} & -157.2131726 & -79.83042138 & -119.1442411 & 0.042106182 & 26.4220\\ \hline

CCSD & -156.6649562 & -79.52749722 & -118.7047522 & 0.044063577 & 27.6502\\ \hline

\makecell{VQE \\ (2, 2)} & \makecell{(2, 2) \\ -156.0956844} & \makecell{(2, 2) \\ -79.22789022} & \makecell{(2, 2) \\ -118.2623895
} & 0.042312584 & 26.5515\\ \hline

\makecell{VQE/ \\ 33.33\% SMF} & \makecell{(6, 4) \\ -156.0957797} & \makecell{(2, 2) \\ -79.22789022} & \makecell{(6, 4) \\ -118.262536
} & 0.042803275 & 26.8594\\ \hline

\end{tabular}
\end{table}


\newpage
\begin{figure*}[htbp]
\centering
\includegraphics[width=0.6\linewidth]{sat_rings/cyclopentane_set_I.png}
\end{figure*}

\vspace{0.5pt}
\begin{table}[H]
\centering
\caption{\textbf{Cyclopentane (Set I)}}
\renewcommand{\arraystretch}{1.0}
\setlength{\tabcolsep}{4.5pt}
\begin{tabular}{|c|c|c|c|c|c|}
\hline

\rule{0pt}{20pt} 
\textbf{Methods} & \textbf{Cyclopentane} &
\textbf{Ethane} & \textbf{Heptane} & \makecell{\textbf{RSE }\\ \textbf{($E_{h}$)}} & \makecell{\textbf{RSE }\\ \textbf{(kcal/mol)}} \\ \hline

 \makecell{DFT/ \\ B3LYP} & -196.5570799 & -79.83042138 & -276.3991966 & 0.011695298 & 7.3389\\ \hline

CCSD & -195.8738718 & -79.52749722 & -275.4139283 & 0.012559244 & 7.8810\\ \hline

\makecell{VQE \\ (2, 2)} & \makecell{(2, 2) \\ -195.1617634} & \makecell{(2, 2) \\ -79.22789022} & \makecell{(2, 2) \\ -274.3997268
} & 0.010073177 & 6.3210\\ \hline

\makecell{VQE/ \\ 33.33\% SMF} & \makecell{(6, 4) \\ -195.1617634} & \makecell{(2, 2) \\ -79.22789022} & \makecell{(6, 4) \\ -274.3998073
} & 0.010153643 & 6.3715\\ \hline

\end{tabular}
\end{table}


\begin{figure*}[htbp]
\centering
\includegraphics[width=0.6\linewidth]{sat_rings/cyclopentane_set_II.png}
\end{figure*}

\vspace{0.5pt}
\begin{table}[H]
\centering
\caption{\textbf{Cyclopentane (Set II)}}
\renewcommand{\arraystretch}{1.0}
\setlength{\tabcolsep}{4.5pt}
\begin{tabular}{|c|c|c|c|c|c|}
\hline

\rule{0pt}{20pt} 
\textbf{Methods} & \textbf{Cyclopentane} &
\textbf{Ethane} & \textbf{Propane} & \makecell{\textbf{RSE }\\ \textbf{($E_{h}$)}} & \makecell{\textbf{RSE }\\ \textbf{(kcal/mol)}} \\ \hline

 \makecell{DFT/ \\ B3LYP} & -196.5570799 & -79.83042138 & -119.1442411 & 0.01201862 & 7.5417\\ \hline

CCSD & -195.8738718 & -79.52749722 & -118.7047522 & 0.01240294 & 7.7829 \\ \hline

\makecell{VQE \\ (2, 2)} & \makecell{(2, 2) \\ -195.1617634} & \makecell{(2, 2) \\ -79.22789022} & \makecell{(2, 2) \\ -118.2623895
} & 0.010732872 & 6.7349\\ \hline

\makecell{VQE/ \\ 33.33\% SMF} & \makecell{(6, 4) \\ -195.1617634} & \makecell{(2, 2) \\ -79.22789022} & \makecell{(6, 4) \\ -118.262536
} & 0.011465378 & 7.1946\\ \hline

\end{tabular}
\end{table}


\newpage
\begin{figure*}[htbp]
\centering
\includegraphics[width=0.6\linewidth]{sat_rings/cyclohexane_set_I.png}
\end{figure*}

\vspace{0.5pt}
\begin{table}[H]
\centering
\caption{\textbf{Cyclohexane (Set I)}}
\renewcommand{\arraystretch}{1.0}
\setlength{\tabcolsep}{4.5pt}
\begin{tabular}{|c|c|c|c|c|c|}
\hline

\rule{0pt}{20pt} 
\textbf{Methods} & \textbf{Cyclohexane} &
\textbf{Ethane} & \textbf{Octane} & \makecell{\textbf{RSE }\\ \textbf{($E_{h}$)}} & \makecell{\textbf{RSE }\\ \textbf{(kcal/mol)}} \\ \hline

 \makecell{DFT/ \\ B3LYP} & -235.8804533 & -79.83042138 & -315.7129131 & 0.002038413 & 1.2791\\ \hline

CCSD & -235.0616331 & -79.52749722 & -314.5912339 & 0.002103555 & 1.32 \\ \hline

\makecell{VQE \\ (2, 2)} & \makecell{(2, 2) \\ -234.2058849} & \makecell{(2, 2) \\ -79.22789022} & \makecell{(2, 2) \\ -313.4340659
} & 0.000290789 & 0.1824\\ \hline

\makecell{VQE/ \\ 33.33\% SMF} & \makecell{(2, 4) \\ -234.2059175} & \makecell{(2, 2) \\ -79.22789022} & \makecell{(6, 4) \\ -313.4341327
} & 0.000324991 & 0.2039\\ \hline

\end{tabular}
\end{table}


\begin{figure*}[htbp]
\centering
\includegraphics[width=0.6\linewidth]{sat_rings/cyclohexane_set_II.png}
\end{figure*}

\vspace{0.5pt}
\begin{table}[H]
\centering
\caption{\textbf{Cyclohexane (Set II)}}
\renewcommand{\arraystretch}{1.0}
\setlength{\tabcolsep}{4.5pt}
\begin{tabular}{|c|c|c|c|c|c|}
\hline

\rule{0pt}{20pt} 
\textbf{Methods} & \textbf{Cyclohexane} &
\textbf{Ethane} & \textbf{Propane} & \makecell{\textbf{RSE }\\ \textbf{($E_{h}$)}} & \makecell{\textbf{RSE }\\ \textbf{(kcal/mol)}} \\ \hline

 \makecell{DFT/ \\ B3LYP} & -235.8804533 & -79.83042138 & -119.1442411 & 0.002464899 & 1.5467\\ \hline

CCSD & -235.0616331 & -79.52749722 & -118.7047522 & 0.001896599 & 1.1901\\ \hline

\makecell{VQE \\ (2, 2)} & \makecell{(2, 2) \\ -234.2058849} & \makecell{(2, 2) \\ -79.22789022} & \makecell{(2, 2) \\ -118.2623895
} & 0.001110645 & 0.6969\\ \hline

\makecell{VQE/ \\ 33.33\% SMF} & \makecell{(2, 4) \\ -234.2059175} & \makecell{(2, 2) \\ -79.22789022} & \makecell{(6, 4) \\ -118.262536
} & 0.001957056 & 1.2280\\ \hline

\end{tabular}
\end{table}


\newpage
\begin{figure*}[htbp]
\centering
\includegraphics[width=0.8\linewidth]{sat_rings/adamantane_set_I.png}
\end{figure*}

\vspace{0.5pt}
\begin{table}[H]
\centering
\caption{\textbf{Adamantane (Set I)}}
\renewcommand{\arraystretch}{1.0}
\setlength{\tabcolsep}{4.5pt}
\begin{tabular}{|c|c|c|c|c|c|c|}
\hline

\rule{0pt}{20pt} 
\textbf{Molecules} & \textbf{Adamantane} &
\textbf{Ethane} & \textbf{Pentane} & \textbf{Isobutane} & \textbf{Hexane} & \makecell{\textbf{RSE }} \\ \hline

 \makecell{DFT/ \\ B3LYP} & -390.7252396 & -79.83042138 & -197.7717555 & -158.4588059 & -237.0854721 & \makecell{0.009755955 $E_{h}$ \\ \hline 6.1219 kcal/mol} \\ \hline

CCSD & -389.4263475 & -79.52749722 & -197.0593361 & -157.8840907 & -236.2366333 & \makecell{0.008815148 $E_{h}$ \\ \hline 5.5315 kcal/mol}\\ \hline

\makecell{VQE \\ (2, 2)} & \makecell{(2, 2) \\ -388.0233615} & \makecell{(2, 2) \\ -79.227890223} & \makecell{(2, 2) \\ -196.3309946
} & \makecell{(2, 2) \\ -157.2973597} & \makecell{(2, 2) \\ -235.3654289} & \makecell{0.005341726 $E_{h}$\\ \hline 3.3519 kcal/mol} \\ \hline

\makecell{VQE \\ 50\% SMF} & \makecell{(4, 3) \\ -388.0233838} & \makecell{(4, 3) \\ -79.22797535} & \makecell{(2, 3) \\ -196.3310262
} & \makecell{(8, 8) \\ -157.2986961} & \makecell{(2, 3) \\ -235.3654304} & \makecell{0.009973436 $E_{h}$ \\ \hline 6.2584 kcal/mol} \\ \hline

\end{tabular}
\end{table}


\begin{figure*}[htbp]
\centering
\includegraphics[width=0.6\linewidth]{sat_rings/adamantane_set_II.png}
\end{figure*}

\vspace{0.5pt}
\begin{table}[H]
\centering
\caption{\textbf{Adamantane (Set II)}}
\renewcommand{\arraystretch}{1.0}
\setlength{\tabcolsep}{4.5pt}
\begin{tabular}{|c|c|c|c|c|c|c|}
\hline

\rule{0pt}{20pt} 
\textbf{Molecules} & \textbf{Adamantane} &
\textbf{Ethane} & \textbf{Isobutane} & \textbf{Propane} & \makecell{\textbf{RSE }\\ \textbf{(Eh)}} & \makecell{\textbf{RSE }\\ \textbf{(kcal/mol)}} \\ \hline

 \makecell{DFT/ \\ B3LYP} & -390.7252396 & -79.83042138 & -158.4588059 & -119.14424111 & 0.010374101 & 6.5098\\ \hline

CCSD & -389.4263475 & -79.52749722 & -157.8840907 & -118.7047522 & 0.008561493 & 5.3724\\ \hline

\makecell{VQE \\ (2, 2)} & \makecell{(2, 2) \\ -388.0233615} & \makecell{(2, 2) \\ -79.227890223} & \makecell{(2, 2) \\ -157.2973597} & \makecell{(2, 2) \\ -118.2623895} & 0.005731385 & 3.5964\\ \hline

\makecell{VQE \\ 50\% SMF} & \makecell{(4, 3) \\ -388.0233838} & \makecell{(4, 3) \\ -79.22797535} & \makecell{(8, 8) \\ -157.2986961} & \makecell{(4, 3) \\ -118.262482} & 0.010588348 & 6.6442\\ \hline

\end{tabular}
\end{table}


\newpage
\begin{figure*}[htbp]
\centering
\includegraphics[width=0.8\linewidth]{unsaturated_rings/Reactions/cyclopropene_set_I.png}
\end{figure*}

\vspace{0.5pt}
\begin{table}[H]
\centering
\caption{\textbf{Cyclopropene (Set I)}}
\renewcommand{\arraystretch}{1.0}
\setlength{\tabcolsep}{4.5pt}
\begin{tabular}{|c|c|c|c|c|c|c|}
\hline

\rule{0pt}{20pt} 
\textbf{Methods} & \textbf{Cyclopropene} &
\textbf{Ethane} & \textbf{Pentane} & \makecell{\textbf{cis-2 }\\ \textbf{-butene}} & \textbf{Hexane} & \makecell{\textbf{RSE }\\ \textbf{(Eh)}} \\ \hline

 \makecell{DFT/ \\ B3LYP} & -116.619041 & -79.83042138 & -197.7717555 & -157.2247752 & -237.0854721 & \makecell{0.089029532 $E_{h}$ \\ \hline 55.8668 kcal/mol} \\ \hline

CCSD & -116.2323254 & -79.52749722 & -197.0593361 & -156.6731433 & -236.2366333 & \makecell{0.090617885 $E_{h}$ \\ \hline 56.8635 kcal/mol}\\ \hline

\makecell{VQE \\ (2, 2)} & \makecell{(2, 2) \\ -115.8411464} & \makecell{(2, 2) \\ -79.227890223} & \makecell{(2, 2) \\ -196.3309946
} & \makecell{(2, 2) \\ -156.1203142} & \makecell{(2, 2) \\ -235.3654289} & \makecell{0.085711886 $E_{h}$\\ \hline 53.7849 kcal/mol} \\ \hline

\makecell{VQE \\ 33.33\% \\ SMF} & \makecell{(4, 3) \\ -115.8411464} & \makecell{(4, 3) \\ -79.22797535} & \makecell{(6, 4) \\ -196.3311131
} & \makecell{(2, 4) \\ -156.1203393} & \makecell{(2, 10) \\ -235.3655866} & \makecell{0.085776228 $E_{h}$ \\ \hline 53.8253 kcal/mol} \\ \hline

\end{tabular}
\end{table}


\begin{figure*}[htbp]
\centering
\includegraphics[width=0.6\linewidth]{unsaturated_rings/Reactions/cyclopropene_set_II.png}
\end{figure*}

\vspace{0.5pt}
\begin{table}[H]
\centering
\caption{\textbf{Cyclopropene (Set II)}}
\renewcommand{\arraystretch}{1.0}
\setlength{\tabcolsep}{4.5pt}
\begin{tabular}{|c|c|c|c|c|c|c|}
\hline

\rule{0pt}{20pt} 
\textbf{Methods} & \textbf{Cyclopropene} &
\textbf{Ethane} & \makecell{\textbf{cis-2 }\\ \textbf{-butene}} & \textbf{Propane} & \makecell{\textbf{RSE }\\ \textbf{($E_{h}$)}} & \makecell{\textbf{RSE }\\ \textbf{(kcal/mol)}} \\ \hline

 \makecell{DFT/ \\ B3LYP} & -116.619041 & -79.83042138 & -157.2247752
 & -119.14424111 & 0.089132557 & 55.9314\\ \hline

CCSD & -116.2323254 & -79.52749722 & -156.6731433 & -118.7047522 & 0.090575609 & 56.8370\\ \hline

\makecell{VQE \\ (2, 2)} & \makecell{(2, 2) \\ -115.8411464} & \makecell{(2, 2) \\ -79.227890223} & \makecell{(2, 2) \\ -156.1203142
} & \makecell{(2, 2) \\ -118.2623895} & 0.085776829 & 53.8257\\ \hline

\makecell{VQE \\ 33.33\% \\ SMF} & \makecell{(4, 3) \\ -115.8411464} & \makecell{(4, 3) \\ -79.22797535} & \makecell{(2, 4) \\ -156.1203393
} & \makecell{(6, 4) \\ -118.262536} & 0.08594843 & 53.9334\\ \hline

\end{tabular}
\end{table}


\newpage
\begin{figure*}[htbp]
\centering
\includegraphics[width=0.8\linewidth]{unsaturated_rings/Reactions/cyclobutene_set_I.png}
\end{figure*}

\vspace{0.5pt}
\begin{table}[H]
\centering
\caption{\textbf{Cyclobutene (Set I)}}
\renewcommand{\arraystretch}{1.0}
\setlength{\tabcolsep}{4.5pt}
\begin{tabular}{|c|c|c|c|c|c|c|}
\hline

\rule{0pt}{20pt} 
\textbf{Methods} & \textbf{Cyclobutene} &
\textbf{Ethane} & \textbf{Pentane} & \makecell{\textbf{cis-2 }\\ \textbf{-butene}} & \textbf{Hexane} & \makecell{\textbf{RSE }} \\ \hline

 \makecell{DFT/ \\ B3LYP} & -155.9732638 & -79.83042138 & -197.7717555 & -157.2247752 & -237.0854721 & \makecell{0.048523352
 $E_{h}$ \\ \hline 30.4488 kcal/mol} \\ \hline

CCSD & -155.4499195 & -79.52749722 & -197.0593361 & -156.6731433 & -236.2366333 & \makecell{0.050321043 $E_{h}$ \\ \hline 31.5769 kcal/mol}\\ \hline

\makecell{VQE \\ (2, 2)} & \makecell{(2, 2) \\ -154.9140994} & \makecell{(2, 2) \\ -79.227890223} & \makecell{(2, 2) \\ -196.3309946
} & \makecell{(2, 2) \\ -156.1203142} & \makecell{(2, 2) \\ -235.3654289} & \makecell{0.047193195 $E_{h}$\\ \hline 29.6141 kcal/mol} \\ \hline

\makecell{VQE \\ 33.33\% \\ SMF} & \makecell{(2, 5) \\ -154.9142168} & \makecell{(4, 3) \\ -79.22797535} & \makecell{(6, 4) \\ -196.3311131
} & \makecell{(2, 4) \\ -156.1203393} & \makecell{(2, 10) \\ -235.3655866} & \makecell{0.04717938 $E_{h}$ \\ \hline 29.6054 kcal/mol} \\ \hline

\end{tabular}
\end{table}


\begin{figure*}[htbp]
\centering
\includegraphics[width=0.6\linewidth]{unsaturated_rings/Reactions/cyclobutene_set_II.png}
\end{figure*}

\vspace{0.5pt}
\begin{table}[H]
\centering
\caption{\textbf{Cyclobutene (Set II)}}
\renewcommand{\arraystretch}{1.0}
\setlength{\tabcolsep}{4.5pt}
\begin{tabular}{|c|c|c|c|c|c|c|}
\hline

\rule{0pt}{20pt} 
\textbf{Methods} & \textbf{Cyclobutene} &
\textbf{Ethane} & \makecell{\textbf{cis-2 }\\ \textbf{-butene}} & \textbf{Propane} & \makecell{\textbf{RSE }\\ \textbf{($E_{h}$)}} & \makecell{\textbf{RSE }\\ \textbf{(kcal/mol)}} \\ \hline

 \makecell{DFT/ \\ B3LYP} & -155.9732638 & -79.83042138 & -157.2247752 & -119.14424111 & 0.0487294 & 30.5781\\ \hline

CCSD & -155.4499195 & -79.52749722 & -156.6731433 & -118.7047522 & 0.050236491 & 31.5238\\ \hline

\makecell{VQE \\ (2, 2)} & \makecell{(2, 2) \\ -154.9140994} & \makecell{(2, 2) \\ -79.227890223} & \makecell{(2, 2) \\ -156.1203142
} & \makecell{(2, 2) \\ -118.2623895} & 0.047323082 & 29.6956\\ \hline

\makecell{VQE \\ 33.33\% \\ SMF} & \makecell{(2, 5) \\ -154.9142168
} & \makecell{(4, 3) \\ -79.22797535} & \makecell{(2, 4) \\ -156.1203393} & \makecell{(6, 4) \\ -118.262536} & 0.047523784 & 29.8216\\ \hline

\end{tabular}
\end{table}


\newpage
\begin{figure*}[htbp]
\centering
\includegraphics[width=0.8\linewidth]{unsaturated_rings/Reactions/cyclopentene_set_I.png}
\end{figure*}

\vspace{0.5pt}
\begin{table}[H]
\centering
\caption{\textbf{Cyclopentene (Set I)}}
\renewcommand{\arraystretch}{1.0}
\setlength{\tabcolsep}{4.5pt}
\begin{tabular}{|c|c|c|c|c|c|c|}
\hline

\rule{0pt}{20pt} 
\textbf{Methods} & \textbf{Cyclopentene} &
\textbf{Ethane} & \textbf{Pentane} & \makecell{\textbf{cis-2 }\\ \textbf{-butene}} & \textbf{Hexane} & \makecell{\textbf{RSE }} \\ \hline

 \makecell{DFT/ \\ B3LYP} & -195.32714 & -79.83042138 & -197.7717555 & -157.2247752 & -237.0854721 & \makecell{0.008363874
 $E_{h}$ \\ \hline 5.2484 kcal/mol} \\ \hline

CCSD & -194.6690972 & -79.52749722 & -197.0593361 & -156.6731433 & -236.2366333 & \makecell{0.008440576 $E_{h}$ \\ \hline 5.2965 kcal/mol}\\ \hline

\makecell{VQE \\ (2, 2)} & \makecell{(2, 2) \\ -193.9885629} & \makecell{(2, 2) \\ -79.227890223} & \makecell{(2, 2) \\ -196.3309946
} & \makecell{(2, 2) \\ -156.1203142} & \makecell{(2, 2) \\ -235.3654289} & \makecell{0.007164005 $E_{h}$\\ \hline 4.4954 kcal/mol} \\ \hline

\makecell{VQE \\ 33.33\% \\ SMF} & \makecell{(2, 2) \\ -154.9142168} & \makecell{(4, 3) \\ -79.22797535} & \makecell{(6, 4) \\ -196.3311131
} & \makecell{(2, 2) \\ -193.9885629} & \makecell{(2, 10) \\ -235.3655866} & \makecell{0.007306832 $E_{h}$ \\ \hline 4.5851 kcal/mol} \\ \hline

\end{tabular}
\end{table}


\begin{figure*}[htbp]
\centering
\includegraphics[width=0.6\linewidth]{unsaturated_rings/Reactions/cyclopentene_set_II.png}
\end{figure*}

\vspace{0.5pt}
\begin{table}[H]
\centering
\caption{\textbf{Cyclopentene (Set II)}}
\renewcommand{\arraystretch}{1.0}
\setlength{\tabcolsep}{4.5pt}
\begin{tabular}{|c|c|c|c|c|c|c|}
\hline

\rule{0pt}{20pt} 
\textbf{Methods} & \textbf{Cyclopentene} &
\textbf{Ethane} & \makecell{\textbf{cis-2 }\\ \textbf{-butene}} & \textbf{Propane} & \makecell{\textbf{RSE }\\ \textbf{($E_{h}$)}} & \makecell{\textbf{RSE }\\ \textbf{(kcal/mol)}} \\ \hline

 \makecell{DFT/ \\ B3LYP} & -195.32714 & -79.83042138 & -157.2247752 & -119.14424111 & 0.008672947 & 5.4423\\ \hline

CCSD & -194.6690972 & -79.52749722 & -156.6731433 & -118.7047522 & 0.008313749 & 5.2169 \\ \hline

\makecell{VQE \\ (2, 2)} & \makecell{(2, 2) \\ -193.9885629} & \makecell{(2, 2) \\ -79.227890223} & \makecell{(2, 2) \\ -156.1203142
} & \makecell{(2, 2) \\ -118.2623895} & 0.007358834 & 4.61773\\ \hline

\makecell{VQE \\ 33.33\% \\ SMF} & \makecell{(2, 2) \\ -193.9885629
} & \makecell{(4, 3) \\ -79.22797535} & \makecell{(2, 4) \\ -156.1203393} & \makecell{(6, 4) \\ -118.262536} & 0.007823438 & 4.9092\\ \hline

\end{tabular}
\end{table}


\newpage
\begin{figure*}[htbp]
\centering
\includegraphics[width=0.8\linewidth]{unsaturated_rings/Reactions/cyclohexene_set_I.png}
\end{figure*}

\vspace{0.5pt}
\begin{table}[H]
\centering
\caption{\textbf{Cyclohexene (Set I)}}
\renewcommand{\arraystretch}{1.0}
\setlength{\tabcolsep}{4.5pt}
\begin{tabular}{|c|c|c|c|c|c|c|}
\hline

\rule{0pt}{20pt} 
\textbf{Methods} & \textbf{Cyclohexene} &
\textbf{Ethane} & \textbf{Pentane} & \makecell{\textbf{cis-2 }\\ \textbf{-butene}} & \textbf{Hexane} & \makecell{\textbf{RSE }} \\ \hline

 \makecell{DFT/ \\ B3LYP} & -234.6482861 & -79.83042138 & -197.7717555 & -157.2247752 & -237.0854721 & \makecell{0.000934446
 $E_{h}$ \\ \hline 0.5863 kcal/mol} \\ \hline

CCSD & -233.8541608 & -79.52749722 & -197.0593361 & -156.6731433 & -236.2366333 & \makecell{0.000674229 $E_{h}$ \\ \hline 0.4230 kcal/mol}\\ \hline

\makecell{VQE \\ (2, 2)} & \makecell{(2, 2) \\ -233.0286049} & \makecell{(2, 2) \\ -79.227890223} & \makecell{(2, 2) \\ -196.3309946
} & \makecell{(2, 2) \\ -156.1203142} & \makecell{(2, 2) \\ -235.3654289} & \makecell{0.001556314 $E_{h}$\\ \hline 0.9766 kcal/mol} \\ \hline

\makecell{VQE \\ 33.33\% \\ SMF} & \makecell{(2, 2) \\ -233.0286049} & \makecell{(4, 3) \\ -79.22797535} & \makecell{(6, 4) \\ -196.3311131} & \makecell{(2, 2) \\ -193.9885629} & \makecell{(2, 10) \\ -235.3655866} & \makecell{0.001738384 $E_{h}$ \\ \hline 1.0908 kcal/mol} \\ \hline

\end{tabular}
\end{table}


\begin{figure*}[htbp]
\centering
\includegraphics[width=0.6\linewidth]{unsaturated_rings/Reactions/cyclohexene_set_II.png}
\end{figure*}

\vspace{0.5pt}
\begin{table}[H]
\centering
\caption{\textbf{Cyclohexene (Set II)}}
\renewcommand{\arraystretch}{1.0}
\setlength{\tabcolsep}{4.5pt}
\begin{tabular}{|c|c|c|c|c|c|c|}
\hline

\rule{0pt}{20pt} 
\textbf{Methods} & \textbf{Cyclohexene} &
\textbf{Ethane} & \makecell{\textbf{cis-2 }\\ \textbf{-butene}} & \textbf{Propane} & \makecell{\textbf{RSE }\\ \textbf{($E_{h}$)}} & \makecell{\textbf{RSE }\\ \textbf{(kcal/mol)}} \\ \hline

 \makecell{DFT/ \\ B3LYP} & -234.6482861 & -79.83042138 & -157.2247752 & -119.14424111 & 0.0013465437 & 0.8449\\ \hline

CCSD & -233.8541608 & -79.52749722 & -156.6731433 & -118.7047522 & 0.000505126 & 0.3169 \\ \hline

\makecell{VQE \\ (2, 2)} & \makecell{(2, 2) \\ -233.0286049} & \makecell{(2, 2) \\ -79.227890223} & \makecell{(2, 2) \\ -156.1203142
} & \makecell{(2, 2) \\ -118.2623895} & 0.001816087 & 1.1396\\ \hline

\makecell{VQE \\ 33.33\% \\ SMF} & \makecell{(2, 5) \\ -233.0286049
} & \makecell{(4, 3) \\ -79.22797535} & \makecell{(2, 4) \\ -156.1203393} & \makecell{(6, 4) \\ -118.262536} & 0.002427192 & 1.5230\\ \hline

\end{tabular}
\end{table}

\section{Molecular Orbital Symmetries}
To facilitate symmetry-guided active space construction, the irreducible representations of molecular orbitals were determined from DFT/B3LYP calculations using the same optimized geometries and basis set employed throughout this work. For each molecule, orbital energies, occupations, and symmetry labels are reported in Tables 37–53 .
\vspace{14pt}
\begin{table}[H]
\centering
\captionsetup{justification=centering}
\caption{\textbf{Cyclopropane ($C_{2v}$)}}
\renewcommand{\arraystretch}{1.0}
\setlength{\tabcolsep}{18pt}
\begin{tabular}{|c|c|c|c|}
\hline
\rule{0pt}{20pt} 
\textbf{MO Index} & \textbf{Energy ($E_{h}$)} &
\textbf{Occupation} & \textbf{Symmetry Irrep.} \\ \hline

8 & 0.4.612559 & 2.0 & $a_{1}$ \\ \hline
9 & 0.3.595667 & 2.0 & $b_{1}$ \\ \hline
10& -0.3.595663 & 2.0 & $a_{2}$  \\ \hline
11& -0.2.879821 & 2.0 & $a_{1}$ \\ \hline
12& -0.2.879817 & 2.0 & $b_{2}$ \\ \hline
13& -0.1.053475 & 0.0 & $a_{1}$ \\ \hline
14& -0.1.103289 & 0.0 & $b_{1}$ \\ \hline
15& 0.1.578402 & 0.0 & $b_{2}$ \\ \hline
16& 0.1.578407 & 0.0 & $a_{1}$ \\ \hline
17& 0.1.587890 & 0.0 & $b_{2}$ \\ \hline
18& 0.2.331767 & 0.0 & $b_{1}$ \\ \hline
19& 0.2.331771 & 0.0 & $a_{2}$ \\ \hline
20& 0.2.638234 & 0.0 & $a_{1}$ \\ \hline
21& 0.2.638237 & 0.0 & $b_{2}$ \\ \hline
22& 0.4.747883 & 0.0 & $a_{1}$ \\ \hline
23& 0.5.497686 & 0.0 & $b_{1}$ \\ \hline

\end{tabular}
\end{table}

\newpage
\begin{table}[H]
\centering
\captionsetup{justification=centering}
\caption{\textbf{Cyclobutane ($C_{2v}$)}}
\renewcommand{\arraystretch}{1.0}
\setlength{\tabcolsep}{18pt}
\begin{tabular}{|c|c|c|c|}
\hline
\rule{0pt}{20pt} 
\textbf{MO Index} & \textbf{Energy ($E_{h}$)} &
\textbf{Occupation} & \textbf{Symmetry Irrep.} \\ \hline

12 & -0.3789239 & 2.0 & $b_{1}$  \\ \hline
13& -0.3432059 & 2.0 & $a_{2}$ \\ \hline
14& -0.3104972 & 2.0 & $a_{1}$ \\ \hline
15& -0.2986533 & 2.0 & $b_{2}$ \\ \hline
16& -0.2986533 & 2.0 & $b_{1}$ \\ \hline
17& 0.9673501 & 0.0 & $a_{1}$ \\ \hline
18& 0.1027169 & 0.0 & $a_{1}$ \\ \hline
19& 0.1375360 & 0.0 & $b_{2}$ \\ \hline
20& 0.1375360 & 0.0 & $b_{1}$ \\ \hline
21& 0.1826816 & 0.0 & $a_{1}$ \\ \hline
22& 0.1901736 & 0.0 & $b_{2}$ \\ \hline
23& 0.1901736 & 0.0 & $b_{1}$ \\ \hline
24& 0.2075811 & 0.0 & $b_{2}$ \\ \hline
25& 0.2075811 & 0.0 & $b_{1}$ \\ \hline

\end{tabular}
\end{table}

\newpage
\begin{table}[H]
\centering
\captionsetup{justification=centering}
\caption{\textbf{Cyclopentane ($C_{s}$)}}
\renewcommand{\arraystretch}{1.0}
\setlength{\tabcolsep}{18pt}
\begin{tabular}{|c|c|c|c|}
\hline
\rule{0pt}{20pt} 
\textbf{MO Index} & \textbf{Energy ($E_{h}$)} &
\textbf{Occupation} & \textbf{Symmetry Irrep.} \\ \hline

16& -0.3305889 & 2.0 & $a'$  \\ \hline
17& -0.3264048 & 2.0 & $a'$ \\ \hline
18& -0.3225831 & 2.0 & $a''$ \\ \hline
19& -0.3174673 & 2.0 & $a'$ \\ \hline
20& -0.3092582 & 2.0 & $a''$ \\ \hline
21& 0.8539591 & 0.0 & $a'$ \\ \hline
22& 0.1006851 & 0.0 & $a'$ \\ \hline
23& 0.1475302 & 0.0 & $a'$ \\ \hline
24& 0.1476220 & 0.0 & $a''$ \\ \hline
25& 0.1719134 & 0.0 & $a'$ \\ \hline
26& 0.1783948 & 0.0 & $a''$ \\ \hline
27& 0.1791338 & 0.0 & $a'$ \\ \hline
28& 0.1875959 & 0.0 & $a'$ \\ \hline
29& 0.1899393 & 0.0 & $a''$ \\ \hline

\end{tabular}
\end{table}

\newpage
\begin{table}[H]
\centering
\captionsetup{justification=centering}
\caption{\textbf{Cyclohexane ($C_{2h}$)}}
\renewcommand{\arraystretch}{1.0}
\setlength{\tabcolsep}{18pt}
\begin{tabular}{|c|c|c|c|}
\hline
\rule{0pt}{20pt} 
\textbf{MO Index} & \textbf{Energy ($E_{h}$)} &
\textbf{Occupation} & \textbf{Symmetry Irrep.} \\ \hline

20 & -0.3348652 & 2.0 & $b_{u}$  \\ \hline
21& -0.348240 & 2.0 & $a_{u}$ \\ \hline
22& -0.3098703 & 2.0 & $a_{g}$ \\ \hline
23& -0.2919409 & 2.0 & $a_{g}$ \\ \hline
24& -0.2918530 & 2.0 & $b_{g}$ \\ \hline
25& 0.8777587 & 0.0 & $a_{g}$ \\ \hline
26& 0.1131145 & 0.0 & $b_{u}$ \\ \hline
27& 0.1362218 & 0.0 & $b_{u}$ \\ \hline
28& 0.1362866 & 0.0 & $a_{u}$ \\ \hline
29& 0.1652217 & 0.0 & $b_{u}$ \\ \hline
30& 0.1741130 & 0.0 & $b_{g}$ \\ \hline
31& 0.1741452 & 0.0 & $a_{g}$ \\ \hline
32& 0.1744722 & 0.0 & $a_{g}$ \\ \hline
33& 0.1867262 & 0.0 & $a_{g}$ \\ \hline

\end{tabular}
\end{table}

\newpage
\begin{table}[H]
\centering
\captionsetup{justification=centering}
\caption{\textbf{Adamantane ($C_{2v}$)}}
\renewcommand{\arraystretch}{1.0}
\setlength{\tabcolsep}{18pt}
\begin{tabular}{|c|c|c|c|}
\hline
\rule{0pt}{20pt} 
\textbf{MO Index} & \textbf{Energy ($E_{h}$)} &
\textbf{Occupation} & \textbf{Symmetry Irrep.} \\ \hline

34& -0.3173258 & 2.0 & $b_{1}$  \\ \hline
35& -0.3172563 & 2.0 & $b_{2}$ \\ \hline
36& -0.2733644 & 2.0 & $b_{2}$ \\ \hline
37& -0.2733588 & 2.0 & $b_{1}$ \\ \hline
38& -0.2732725 & 2.0 & $a_{1}$ \\ \hline
39& 0.6936682 & 0.0 & $a_{1}$ \\ \hline
40& 0.1298254 & 0.0 & $b_{1}$ \\ \hline
41& 0.1298572 & 0.0 & $a_{1}$ \\ \hline
42& 0.1298629 & 0.0 & $b_{2}$ \\ \hline
43& 0.1360074 & 0.0 & $a_{1}$ \\ \hline
44& 0.1360915 & 0.0 & $b_{2}$ \\ \hline
45& 0.1361323 & 0.0 & $b_{1}$ \\ \hline
46& 0.1515498 & 0.0 & $a_{1}$ \\ \hline
47& 0.1677760 & 0.0 & $a_{2}$ \\ \hline

\end{tabular}
\end{table}

\newpage
\begin{table}[H]
\centering
\captionsetup{justification=centering}
\caption{\textbf{Cyclopropene ($C_{2v}$)}}
\renewcommand{\arraystretch}{1.0}
\setlength{\tabcolsep}{18pt}
\begin{tabular}{|c|c|c|c|}
\hline
\rule{0pt}{20pt} 
\textbf{MO Index} & \textbf{Energy ($E_{h}$)} &
\textbf{Occupation} & \textbf{Symmetry Irrep.} \\ \hline

7 & -0.5007447 & 2.0 & $a_{1}$  \\ \hline
8& -0.4293838 & 2.0 & $b_{1}$ \\ \hline
9& -0.3428927 & 2.0 & $a_{1}$ \\ \hline
10& -0.2903605 & 2.0 & $b_{2}$ \\ \hline
11& -0.2478430 & 2.0 & $b_{1}$ \\ \hline
12& 0.1724116 & 0.0 & $a_{2}$ \\ \hline
13& 0.1242538 & 0.0 & $b_{2}$ \\ \hline
14& 0.1275673 & 0.0 & $a_{1}$ \\ \hline
15& 0.1471494 & 0.0 & $a_{1}$ \\ \hline
16& 0.1848895 & 0.0 & $b_{2}$ \\ \hline
17& 0.1917303 & 0.0 & $b_{1}$ \\ \hline
18& 0.2389962 & 0.0 & $a_{1}$ \\ \hline
19& 0.3405598 & 0.0 & $b_{2}$ \\ \hline
20& 0.4957064 & 0.0 & $a_{1}$ \\ \hline

\end{tabular}
\end{table}

\newpage
\begin{table}[H]
\centering
\captionsetup{justification=centering}
\caption{\textbf{Cyclobutene ($C_{2v}$)}}
\renewcommand{\arraystretch}{1.0}
\setlength{\tabcolsep}{18pt}
\begin{tabular}{|c|c|c|c|}
\hline
\rule{0pt}{20pt} 
\textbf{MO Index} & \textbf{Energy ($E_{h}$)} &
\textbf{Occupation} & \textbf{Symmetry Irrep.} \\ \hline

11& -0.3739475 & 2.0 & $a_{1}$  \\ \hline
12& -0.3574676 & 2.0 & $a_{2}$ \\ \hline
13& -0.3230840 & 2.0 & $a_{1}$ \\ \hline
14& -0.3135043 & 2.0 & $b_{2}$ \\ \hline
15& -0.2434532 & 2.0 & $b_{1}$ \\ \hline
16& 0.2735712 & 0.0 & $a_{2}$ \\ \hline
17& 0.1161045 & 0.0 & $a_{1}$ \\ \hline
18& 0.1396150 & 0.0 & $a_{1}$ \\ \hline
19& 0.1447575 & 0.0 & $b_{1}$ \\ \hline
20& 0.1483449 & 0.0 & $b_{2}$ \\ \hline
21& 0.1668853 & 0.0 & $b_{2}$ \\ \hline
22& 0.1865544 & 0.0 & $a_{1}$ \\ \hline
23& 0.2089280 & 0.0 & $b_{2}$ \\ \hline
24& 0.2313675 & 0.0 & $a_{2}$ \\ \hline

\end{tabular}
\end{table}

\newpage
\begin{table}[H]
\centering
\captionsetup{justification=centering}
\caption{\textbf{Cyclopentene ($C_{s}$)}}
\renewcommand{\arraystretch}{1.0}
\setlength{\tabcolsep}{18pt}
\begin{tabular}{|c|c|c|c|}
\hline
\rule{0pt}{20pt} 
\textbf{MO Index} & \textbf{Energy ($E_{h}$)} &
\textbf{Occupation} & \textbf{Symmetry Irrep.} \\ \hline

15& -0.3421174 & 2.0 & $a'$ \\ \hline
16& -0.3414832 & 2.0 & $a''$  \\ \hline
17& -0.3373874 & 2.0 & $a'$ \\ \hline
18& -0.3328453 & 2.0 & $a''$ \\ \hline
19& -0.2328254 & 2.0 & $a'$ \\ \hline
20& -0.3312223 & 0.0 & $a''$ \\ \hline
21& 0.9927131 & 0.0 & $a'$ \\ \hline
22& 0.1252173 & 0.0 & $a'$ \\ \hline
23& 0.1488224 & 0.0 & $a'$ \\ \hline
24& 0.1531221 & 0.0 & $a''$ \\ \hline
25& 0.1724665 & 0.0 & $a'$ \\ \hline
26& 0.1751117 & 0.0 & $a''$ \\ \hline
27& 0.1934649 & 0.0 & $a''$ \\ \hline
28& 0.1942655 & 0.0 & $a'$ \\ \hline

\end{tabular}
\end{table}

\newpage
\begin{table}[H]
\centering
\captionsetup{justification=centering}
\caption{\textbf{Cyclohexene ($C_{2}$)}}
\renewcommand{\arraystretch}{1.0}
\setlength{\tabcolsep}{18pt}
\begin{tabular}{|c|c|c|c|}
\hline
\rule{0pt}{20pt} 
\textbf{MO Index} & \textbf{Energy ($E_{h}$)} &
\textbf{Occupation} & \textbf{Symmetry Irrep.} \\ \hline

20& -0.3225074 & 2.0 & $a$  \\ \hline
22& -0.3108892 & 2.0 & $a$ \\ \hline
23& -0.2992470 & 2.0 & $b$ \\ \hline
24& -0.2334454 & 2.0 & $b$ \\ \hline
25& -0.3418307 & 2.0 & $a$ \\ \hline
26& 0.9208232 & 0.0 & $a$ \\ \hline
27& 0.1307619 & 0.0 & $b$ \\ \hline
28& 0.1369928 & 0.0 & $b$ \\ \hline
29& 0.1414337 & 0.0 & $a$ \\ \hline
30& 0.1681633 & 0.0 & $b$ \\ \hline
31& 0.1690318 & 0.0 & $a$ \\ \hline
32& 0.1723182 & 0.0 & $a$ \\ \hline
33& 0.1876243 & 0.0 & $b$ \\ \hline
34& 0.1956346 & 0.0 & $a$ \\ \hline

\end{tabular}
\end{table}

\newpage
\begin{table}[H]
\centering
\captionsetup{justification=centering}
\caption{\textbf{Ethane ($C_{2h}$)}}
\renewcommand{\arraystretch}{1.0}
\setlength{\tabcolsep}{18pt}
\begin{tabular}{|c|c|c|c|}
\hline
\rule{0pt}{20pt} 
\textbf{MO Index} & \textbf{Energy ($E_{h}$)} &
\textbf{Occupation} & \textbf{Symmetry Irrep.} \\ \hline

5& -0.4300871 & 2.0 & $b_{u}$  \\ \hline
6& -0.4300868 & 2.0 & $a_{u}$ \\ \hline
7& -0.36276463 & 2.0 & $a_{g}$ \\ \hline
8& -0.3400984 & 2.0 & $a_{g}$ \\ \hline
9& -0.3400984 & 2.0 & $b_{g}$ \\ \hline
10& 0.1047701 & 0.0 & $a_{g}$ \\ \hline
11& 0.1560553 & 0.0 & $b_{u}$ \\ \hline
12& 0.1637176 & 0.0 & $a_{u}$ \\ \hline
13& 0.1637178 & 0.0 & $b_{u}$ \\ \hline
14& 0.1902123 & 0.0 & $a_{g}$ \\ \hline
15& 0.1902125 & 0.0 & $b_{g}$ \\ \hline
16& 0.2424501 & 0.0 & $b_{u}$ \\ \hline
17& 0.5131700 & 0.0 & $a_{g}$ \\ \hline
18& 0.5437522 & 0.0 & $b_{u}$ \\ \hline

\end{tabular}
\end{table}

\newpage
\begin{table}[H]
\centering
\captionsetup{justification=centering}
\caption{\textbf{Propane ($C_{2v}$)}}
\renewcommand{\arraystretch}{1.0}
\setlength{\tabcolsep}{18pt}
\begin{tabular}{|c|c|c|c|}
\hline
\rule{0pt}{20pt} 
\textbf{MO Index} & \textbf{Energy ($E_{h}$)} &
\textbf{Occupation} & \textbf{Symmetry Irrep.} \\ \hline

10& -0.3981662 & 2.0 & $a_{2}$  \\ \hline
11& -0.3934170 & 2.0 & $b_{1}$ \\ \hline
12& -0.3603363 & 2.0 & $a_{1}$ \\ \hline
13& -0.3344306 & 2.0 & $b_{2}$ \\ \hline
14& -0.1576648 & 2.0 & $a_{1}$ \\ \hline
15& 0.9309762 & 0.0 & $b_{1}$ \\ \hline
16& 0.1057658 & 0.0 & $a_{1}$ \\ \hline
17& 0.1380316 & 0.0 & $b_{2}$ \\ \hline
18& 0.1404120 & 0.0 & $b_{1}$ \\ \hline
19& 0.1545550 & 0.0 & $a_{1}$ \\ \hline
20& 0.1771186 & 0.0 & $a_{2}$ \\ \hline
21& 0.1825377 & 0.0 & $b_{2}$ \\ \hline
22& 0.2077311 & 0.0 & $a_{1}$ \\ \hline
23& 0.2330171 & 0.0 & $b_{1}$ \\ \hline

\end{tabular}
\end{table}

\newpage
\begin{table}[H]
\centering
\captionsetup{justification=centering}
\caption{\textbf{Isobutane ($C_{s}$)}}
\renewcommand{\arraystretch}{1.0}
\setlength{\tabcolsep}{18pt}
\begin{tabular}{|c|c|c|c|}
\hline
\rule{0pt}{20pt} 
\textbf{MO Index} & \textbf{Energy ($E_{h}$)} &
\textbf{Occupation} & \textbf{Symmetry Irrep.} \\ \hline

13& 0.3.785722 & 2.0 & $a'$ \\ \hline
14& 0.3.589584 & 2.0 & $a''$ \\ \hline
15& -0.3.223921 & 2.0 & $a'$ \\ \hline
16& -0.3.223920 & 2.0 & $a''$  \\ \hline
17& -0.3.177430 & 2.0 & $a'$ \\ \hline
18& -0.8.605392 & 0.0 & $a'$ \\ \hline
19& -0.1.371217 & 0.0 & $a'$ \\ \hline
20& -0.1.544118 & 0.0 & $a''$ \\ \hline
21& 0.1.544124 & 0.0 & $a'$ \\ \hline
22& 0.1.585785 & 0.0 & $a'$ \\ \hline
23& 0.1.585788 & 0.0 & $a''$ \\ \hline
24& 0.1.759932 & 0.0 & $a'$ \\ \hline
25& 0.1.967860 & 0.0 & $a''$ \\ \hline
26& 0.1.967865 & 0.0 & $a'$ \\ \hline

\end{tabular}
\end{table}

\newpage
\begin{table}[H]
\centering
\captionsetup{justification=centering}
\caption{\textbf{cis-2-butene ($C_{2v}$)}}
\renewcommand{\arraystretch}{1.0}
\setlength{\tabcolsep}{18pt}
\begin{tabular}{|c|c|c|c|}
\hline
\rule{0pt}{20pt} 
\textbf{MO Index} & \textbf{Energy ($E_{h}$)} &
\textbf{Occupation} & \textbf{Symmetry Irrep.} \\ \hline

12& -0.3967572 & 2.0 & $a_{2}$ \\ \hline
13& -0.3805681 & 2.0 & $b_{1}$ \\ \hline
14& -0.3576487 & 2.0 & $a_{1}$ \\ \hline
15& -0.3270648 & 2.0 & $b_{1}$ \\ \hline
16& 0.2345955 & 2.0 & $b_{2}$ \\ \hline
17& 0.3684579 & 0.0 & $a_{2}$ \\ \hline
18& 0.1028931 & 0.0 & $a_{1}$ \\ \hline
19& 0.1200412 & 0.0 & $a_{1}$ \\ \hline
20& 0.1431323 & 0.0 & $b_{1}$ \\ \hline
21& 0.1686074 & 0.0 & $b_{2}$ \\ \hline
22& 0.1760362 & 0.0 & $a_{1}$ \\ \hline
23& 0.1899734 & 0.0 & $a_{2}$ \\ \hline
24& 0.1904207 & 0.0 & $b_{1}$ \\ \hline

\end{tabular}
\end{table}

\newpage
\begin{table}[H]
\centering
\captionsetup{justification=centering}
\caption{\textbf{Pentane ($C_{2v}$)}}
\renewcommand{\arraystretch}{1.0}
\setlength{\tabcolsep}{18pt}
\begin{tabular}{|c|c|c|c|}
\hline
\rule{0pt}{20pt} 
\textbf{MO Index} & \textbf{Energy ($E_{h}$)} &
\textbf{Occupation} & \textbf{Symmetry Irrep.} \\ \hline

17& -0.3484585 & 2.0 & $b_{1}$  \\ \hline
18& -0.3316241 & 2.0 & $a_{2}$ \\ \hline
19& -0.3218434 & 2.0 & $a_{1}$ \\ \hline
20& -0.3172332 & 2.0 & $b_{2}$ \\ \hline
21& -0.3107255 & 2.0 & $b_{1}$ \\ \hline
22& 0.9371722 & 0.0 & $a_{1}$ \\ \hline
23& 0.1111909 & 0.0 & $b_{1}$ \\ \hline
24& 0.1282134 & 0.0 & $b_{2}$ \\ \hline
25& 0.1357248 & 0.0 & $a_{1}$ \\ \hline
26& 0.1534412 & 0.0 & $a_{1}$ \\ \hline
27& 0.1746415 & 0.0 & $b_{1}$ \\ \hline
28& 0.1748441 & 0.0 & $a_{2}$ \\ \hline
29& 0.1786405 & 0.0 & $a_{1}$ \\ \hline
30& 0.1975477 & 0.0 & $b_{2}$ \\ \hline

\end{tabular}
\end{table}

\newpage
\begin{table}[H]
\centering
\captionsetup{justification=centering}
\caption{\textbf{Hexane ($C_{2h}$)}}
\renewcommand{\arraystretch}{1.0}
\setlength{\tabcolsep}{18pt}
\begin{tabular}{|c|c|c|c|}
\hline
\rule{0pt}{20pt} 
\textbf{MO Index} & \textbf{Energy ($E_{h}$)} &
\textbf{Occupation} & \textbf{Symmetry Irrep.} \\ \hline

21& -0.3367311 & 2.0 & $b_{u}$ \\ \hline
22& -0.3237155 & 2.0 & $a_{u}$ \\ \hline
23& -0.3202144 & 2.0 & $a_{g}$ \\ \hline
24& -0.3166229 & 2.0 & $b_{g}$ \\ \hline
25& 0.3042931 & 2.0 & $a_{g}$ \\ \hline
26& 0.9277298 & 0.0 & $a_{g}$ \\ \hline
27& 0.1045822 & 0.0 & $b_{u}$ \\ \hline
28& 0.1230903 & 0.0 & $a_{u}$ \\ \hline
29& 0.1323686 & 0.0 & $b_{u}$ \\ \hline
30& 0.1369442 & 0.0 & $a_{g}$ \\ \hline
31& 0.1628419 & 0.0 & $b_{g}$ \\ \hline
32& 0.1685872 & 0.0 & $a_{g}$ \\ \hline
33& 0.1694078 & 0.0 & $b_{u}$ \\ \hline

\end{tabular}
\end{table}

\newpage
\begin{table}[H]
\centering
\captionsetup{justification=centering}
\caption{\textbf{Heptane ($C_{2v}$)}}
\renewcommand{\arraystretch}{1.0}
\setlength{\tabcolsep}{18pt}
\begin{tabular}{|c|c|c|c|}
\hline
\rule{0pt}{20pt} 
\textbf{MO Index} & \textbf{Energy ($E_{h}$)} &
\textbf{Occupation} & \textbf{Symmetry Irrep.} \\ \hline

25& -0.3311559 & 2.0 & $b_{1}$  \\ \hline
26& -0.3197094 & 2.0 & $a_{2}$ \\ \hline
27& -0.3185487 & 2.0 & $a_{1}$ \\ \hline
28& -0.3163461 & 2.0 & $b_{2}$ \\ \hline
29& -0.2994859 & 2.0 & $b_{1}$ \\ \hline
30& 0.9197429 & 0.0 & $a_{1}$ \\ \hline
31& 0.1016322 & 0.0 & $b_{1}$ \\ \hline
32& 0.1197137 & 0.0 & $b_{2}$ \\ \hline
33& 0.1226148 & 0.0 & $a_{1}$ \\ \hline
34& 0.1299200 & 0.0 & $a_{1}$ \\ \hline
35& 0.1528544 & 0.0 & $a_{2}$ \\ \hline
36& 0.1549060 & 0.0 & $b_{1}$ \\ \hline
37& 0.1700154 & 0.0 & $b_{1}$ \\ \hline
38& 0.1730565 & 0.0 & $a_{1}$ \\ \hline

\end{tabular}
\end{table}

\newpage
\begin{table}[H]
\centering
\captionsetup{justification=centering}
\caption{\textbf{Octane ($C_{2h}$)}}
\renewcommand{\arraystretch}{1.0}
\setlength{\tabcolsep}{18pt}
\begin{tabular}{|c|c|c|c|}
\hline
\rule{0pt}{20pt} 
\textbf{MO Index} & \textbf{Energy ($E_{h}$)} &
\textbf{Occupation} & \textbf{Symmetry Irrep.} \\ \hline

29& -0.3262071 & 2.0 & $b_{u}$ \\ \hline
30& -0.3179687 & 2.0 & $a_{g}$ \\ \hline
31& -0.3176338 & 2.0 & $a_{u}$ \\ \hline
32& -0.3162469 & 2.0 & $b_{g}$ \\ \hline
33& 0.2957062 & 2.0 & $a_{g}$ \\ \hline
34& 0.9137391 & 0.0 & $a_{g}$ \\ \hline
35& 0.9953405 & 0.0 & $b_{u}$ \\ \hline
36& 0.1150390 & 0.0 & $a_{g}$ \\ \hline
37& 0.1173179 & 0.0 & $a_{u}$ \\ \hline
38& 0.1267486 & 0.0 & $b_{u}$ \\ \hline
39& 0.1448314 & 0.0 & $b_{g}$ \\ \hline
40& 0.1456144 & 0.0 & $b_{u}$ \\ \hline
41& 0.1530669 & 0.0 & $a_{g}$ \\ \hline

\end{tabular}
\end{table}
